\documentclass[letter,11pt]{article}%
\usepackage{caption}
\usepackage{amsmath,bm}
\usepackage{amsfonts}
\usepackage{amssymb}
\usepackage{lscape}
\usepackage{float}
\usepackage{graphicx}
\usepackage{rotating}
\usepackage[]{subfigure}
\usepackage{caption}
\usepackage{subcaption}
\usepackage[letterpaper,margin=1.25in]{geometry}
\usepackage[colorlinks=true,citecolor=blue, linkcolor=black]{hyperref}
\usepackage[authoryear]{natbib}
\usepackage{color}
\usepackage{colortbl}
\usepackage{appendix}
\usepackage{paralist}
\usepackage{threeparttable}
\usepackage{multirow}
\usepackage{amsmath}
\usepackage{tabularx}
\usepackage{algorithm}
\usepackage{algpseudocode}
\usepackage{pifont}%
\usepackage{array}
\newcolumntype{P}[1]{>{\centering\arraybackslash}p{#1}}
\usepackage{titlecaps}
\Addlcwords{for a an and with in of to the for}
\usepackage{caption} 

\providecommand{\U}[1]{\protect\rule{.1in}{.1in}}

\RequirePackage{hypernat}
\newtheorem{theorem}{Theorem}

\newtheorem{assumption}{Assumption}

\newtheorem{corollary}[theorem]{Corollary}

\newtheorem{lemma}[theorem]{Lemma}

\newtheorem{proposition}[theorem]{Proposition}

\newtheorem{definition}{Definition}

\newtheorem{remark}{Remark}

\catcode`~=11
\newcommand{\urltilde}{\kern -.15em\lower .7ex\hbox{~}\kern .04em}
\catcode`~=13
\def \@seccntformat#1{\csname the#1\endcsname.\quad}
\makeatother
\numberwithin{equation}{section}

\DeclareCaptionLabelFormat{table}{TABLE~#2}
\begin{document}
	
	\title{Debiased Machine Learning for Unobserved Heterogeneity: High-Dimensional Panels and Measurement Error Models\thanks{We thank Stéphane Bonhomme, Jesús Carro, Clément de Chaisemartin, Koen Jochmans and Martin Weidner for their valuable comments. Research funded by Ministerio de Ciencia e
			Innovaci\'{o}n grant PID2021-127794NB-I00 and Comunidad de Madrid, grants
			EPUC3M11 (VPRICIT) and H2019/HUM-589.}}
	\author{Facundo Arga\~{n}araz\thanks{Department of Economics. E-mail:
			\href{mailto:facundo.arganaraz@sciencespo.fr}{facundo.arganaraz@sciencespo.fr}.}\\\textit{Sciences Po}
		\and Juan Carlos Escanciano\thanks{Corresponding author: Department of Economics.
			E-mail: \href{mailto:jescanci@indiana.edu}{jescanci@econ.uc3m.es}. }\\\textit{Universidad Carlos III de Madrid}}
	\date{\today}
	\maketitle
	
	\begin{abstract}
		Developing robust inference for models with nonparametric Unobserved Heterogeneity (UH) is both important and challenging. We propose novel Debiased Machine Learning (DML) procedures for valid inference on functionals of UH, allowing for partial identification of multivariate target and high-dimensional nuisance parameters. Our main contribution is a full characterization of \textit{all} relevant Neyman-orthogonal moments in models with nonparametric UH, where \textit{relevance} means informativeness about the parameter of interest. Under additional support conditions, orthogonal moments are globally robust to the distribution of the UH. They may still involve other high-dimensional nuisance parameters, but their local robustness reduces regularization bias and enables valid DML inference. We apply these results to: (i) common parameters, average marginal effects, and variances of UH in panel data models with high-dimensional controls; (ii) moments of the common factor in the Kotlarski model with a factor loading; and (iii) smooth functionals of teacher value-added. Monte Carlo simulations show substantial efficiency gains from using efficient orthogonal moments relative to ad-hoc choices. We illustrate the practical value of our approach by showing that existing estimates of the average and variance effects of maternal smoking on child birth weight are robust.\vspace{2mm}
		
		\begin{description}
			\item[Keywords:] Debiased Inference; Machine Learning; Unobserved Heterogeneity.
			
			\item[\emph{JEL classification:}] C14; C31; C33; C35.\newpage
			
		\end{description}
	\end{abstract}
	
	\section{Introduction}
	
	One of the most important insights of empirical economics is the pervasiveness of heterogeneity in observed economic behavior \citep[][]{heckman2001micro}. In particular, Unobserved Heterogeneity (UH) plays a central role in bridging economic modeling with empirical data, with significant implications for policy \citep[][]{Browning_Carro_2007}. Structural analysis often focuses on functionals of the nonparametric UH, which are key summary objects of interest in many empirical applications. These structural parameters may include average treatment effects, average marginal effects, variances, counterfactual outcomes, linear projections of UH on covariates, welfare measures, bounds, and other downstream functionals. Many of these parameters can be expressed as functionals of the unknown distribution of UH and other parameters of the model.

This paper introduces a systematic approach to conduct inferences for functionals of the UH based on Neyman-orthogonal moments \cite[][]{neyman1959}.\footnote{Throughout the paper, we use
	indistinguishably the terms Neyman-orthogonal, orthogonal, debiased, and Locally Robust (LR) moments.} Neyman-orthogonal moments for a target functional are robust to local deviations of the nuisance parameters of the target at the truth. The application of locally robust inference to contexts with nonparametric UH is naturally important for (at least) four reasons. First, the distribution of UH is typically not nonparametrically identified, e.g., in short panel data models with discrete outcomes. This implies that nuisance parameters of target functionals are often not identified in these settings, so obtaining moments that are locally robust to them is important. Second, even when the distribution of UH is point-identified and estimated, plug-in estimators may suffer from large regularization biases---so much so that they render inference on the target functional invalid. Third, nonparametric estimators of UH may have very slow (e.g., logarithmic) rates of convergence, invalidating critical assumptions in semiparametric estimation theory.\footnote{Plug-in estimators with these properties are available from a large literature on nonparametric estimation of the distribution of UH in \cite{arellano2012identifying}, \cite{bonhomme2010generalized}, \cite{carroll1988optimal}, 
	\cite{heckman1984method},
	\cite{horowitz1996semiparametric}, 
	\cite{kato2021robust},
	\cite{kiefer1956consistency}, \cite{koenker2014convex}, \cite{li1998nonparametric}, and \cite{szekely2000identifiability}, among many others.} Orthogonal moments are free from the distribution of the UH \textit{if} an additional support condition holds, which overcomes the problem of the regularization bias and slow rate of convergence of nonparametric estimators of the distribution of the UH. Fourth, efficient moments are locally robust moments. It is thus important to provide inference methods in models with UH that are asymptotically valid (provable), robust to the UH and additional nuisance parameters, and informative (relevant) about the target parameter. We show that our Debiased Machine Learning (DML) for models with UH (DML-UH in short) is a systematic approach to achieving these goals.

As the main result of this paper, we provide necessary and sufficient conditions for the existence and relevance of orthogonal moments for general smooth functionals in models with nonparametric UH. The precise sense of smoothness of the functional is defined in Section \ref{Smoothfunctional}. Relevant Neyman-orthogonal moments, a concept we introduce in this paper, are informative about the target functional even when its Fisher information matrix is singular but not zero (e.g., in some partially identified cases). We thus characterize \textit{all} relevant Neyman-orthogonal moments in models with nonparametric UH. Our necessary and sufficient conditions are constructive, providing new tools for identification, estimation, and inference on functionals of UH in a broad range of settings.


Based on the necessary and sufficient conditions for orthogonality, we propose a three-step algorithm for their construction. Specifically, we show that relevant orthogonal moments must solve a system of two functional equations. The first is an integral equation that is free of the distribution of UH when certain additional support conditions hold. This means that locally robust moments are globally robust to the distribution of UH. In some cases, this first equation admits a closed-form solution; more generally, it can be solved numerically or via functional approximation. Such support conditions and similar equations to our first one have notably appeared in the functional differencing literature, see the important work by \cite{bonhomme2012functional}, and more recently \cite{honore2024moment}. Yet, ours is the first paper to underscore the role played by support conditions to obtain global robustness. We show that similar arguments invoked by such literature apply more broadly, e.g., to semiparametric/nonparametric conditional likelihoods given the UH. Thus, the first equation characterizing orthogonal moments generalizes estimating equations appearing in the functional differencing literature.

More importantly, we show that this first equation alone does not give an orthogonal moment for general functionals. For example, existing moments for average marginal effects (e.g., \cite{chamberlain1992efficiency}, \cite{aguirregabiria2024identification}) that solve the first equation are not orthogonal in general. This lack of robustness may be particularly problematic when additional high-dimensional nuisance parameters are present, as in the high-dimensional random coefficient model we consider in this paper. Orthogonal moments must also satisfy a second functional equation to ensure orthogonality with respect to other nuisance parameters different from the UH. We develop a three-step algorithm that finds a solution to both equations simultaneously, and hence delivers an orthogonal moment. 

Once a relevant orthogonal moment has been constructed using our method, we introduce a score-type test based on it. We derive its asymptotic properties and show how it can be used for valid inference. Importantly, our characterization and inference results allow for multivariate partially identified target parameters and the use of machine learning methods to estimate high-dimensional nuisance components, incorporating cross-fitting to mitigate overfitting bias, as in \cite{chernozhukov2018double}.\footnote{While our characterization of local robustness allows for non-identified high-dimensional nuisance parameters, the implementation of our inference procedures requires the orthogonal moment to be identified. This situation is analogous to that considered in \cite{santos2011instrumental}, \cite{escanciano2021optimal}, and, more recently, in \cite{bennett2023source}. As in these references, regularization of the nuisance parameters helps in identifying one element of the nuisance parameter's identified set.} We also present a number of novel results on identification and LR estimation.

We illustrate our three-step construction and inferences with three distinct applications involving nonparametric UH. These specific applications build on our general characterization of relevant orthogonal moments. The first example application develops DML inferences in a random coefficient panel data model with high-dimensional exogenous controls and a finite number of periods. Such specification may be well-motivated for several reasons, such as the availability of large data sets that contain a myriad of controls, concerns about strong functional form assumptions, or to make causal statements more credible.\footnote{In Section \ref{sectioneventstudy} of the Supplementary Appendix, we build on \cite{borusyak2024revisiting} and argue that our procedure complements the literature on event-study designs by allowing for flexible modeling of parallel trends assumptions. We are not aware of inference procedures---using LR moments---that allow for high-dimensional regressors in this context, which might be of independent interest.} We build on the important work by \cite{chamberlain1992efficiency} and \cite{arellano2012identifying} to generalize their results to a setting with high-dimensional covariates and robust inferences. We construct in Section~\ref{sectionhighdimensionalpanel} novel orthogonal moments and develop DML-UH procedures for estimating linear combinations of common parameters, average marginal effects, and variances of the UH. We are currently developing an R package, \texttt{dmlpanel.R}, that implements all the procedures described in this section.\footnote{A preliminary version of this function can be obtained from \texttt{https://github.com/argafacu/dmlpanel}.} 

A second application is to a canonical measurement error model--the Kotlarski model with a factor loading. The identification of this model has been recently investigated in \cite{lewbel2022kotlarski}, who also provides a rich historical perspective. We develop in Section~\ref{Kotlarski} the first debiased inferences for non-central moments of the common factor in this setting. This model has been used, for example, in studies of income dynamics \citep[][]{bonhomme2010generalized, horowitz1996semiparametric}, the formation of cognitive and noncognitive skills in childhood \citep[][]{cunha2010estimating}, studies on the impact of grade retention \citep[][]{fruehwirth2016timing}, the intergenerational mobility literature \citep[][]{nybom2016heterogeneous, haider2006life}, panel data  \citep[][]{evdokimov2010identification,lewbel2022kotlarski}, and the triangular two-equation system without instruments \citep[][]{lewbel2024identification}. Inference on this model is notoriously difficult, see \cite{kato2021robust}, as the conditional likelihood of the data given the common factor contains two high-dimensional nuisance parameters (the idiosyncratic errors' distributions). Our orthogonal moments are free from these nuisance parameters.

 The third example application that we work on belongs to the literature on the quality of institutions, such as hospitals, doctors, nurses, teachers, schools, neighborhoods, and firms \citep[][]{kane2008estimating, fletcher2014estimating, yakusheva2014nurse, chetty2014measuring, angrist2016interpreting, angrist2017leveraging, finkelstein2017adjusting, doyle2019evaluating, kline2022systemic, chetty2018impacts, abowd1999high}. These studies typically employ linear random coefficient models with multidimensional UH, drawing on “value-added” estimation methods originally developed in the teacher value-added literature \citep[see][for a review]{koedel2015value}, and focus on functionals of UH using plug-in procedures; see \cite{bonhomme2024estimating} for discussion of this framework.

 We consider the standard teacher value-added model with a nonparametric distribution of “value-added” or random effects, as in, e.g., \cite{gilraine2020new}. We make two main contributions within this framework in Section~\ref{TAV}. First, we use the necessity of our conditions for orthogonality to show the nonexistence of orthogonal moments for prominent policy parameters in the teacher value-added literature. This result has important policy implications, as it shows the non-robustness of certain counterfactual measures that inform teacher hiring and retention decisions. Second, we develop robust inference methods for smooth functionals of “value-added”, leveraging Neyman-orthogonality. Notably, our approach to constructing orthogonal moments does not require estimating the distribution of UH, and therefore, it provides a robust alternative to commonly employed regularized plug-in methods. 

There is nothing fundamentally special about the previous settings. Our results can be applied to other contexts involving some form of UH or latent variables, where Neyman-orthogonality principles for general smooth functionals (and possibly under more general model assumptions) are not yet known. Examples include, among many others, panel data models with censored, truncated, or missing data \citep[see, e.g.,][]{honore1992trimmed}; nonparametric panel data with interactive fixed effects \citep[see, e.g.,][]{freyberger2018non}; non-classical measurement error models with instrumental variables \citep[see, e.g.,][]{hu2008instrumental}; mixed logit models \citep[see, e.g.,][]{rafi2024nonparametric}; production function models with firm specific technologies \citep[see][]{duffy2004capital, MairesseGriliches1990, dobbelaere2013panel}, structural models of firm investment and capital formation \citep[extending, e.g.,][]{cooper2006nature}; duration models \citep[e.g., as in][]{heckman1984identifiability, heckman1984method, horowitz1999semiparametric}, including multiple-spell duration models with individual heterogeneity in different parameters \citep[][]{van2001duration, alvarez2016decomposing}; auctions with nonparametric UH \citep[see, e.g.,][]{krasnokutskaya2011identification}; linear and nonlinear random coefficient models, \citep[e.g., as in][]{lewbel2017unobserved, masten2018random}; individual-specific income profile/earning dynamic models \citep[][]{lillard1979components, guvenen2009empirical, arellano2017earnings}; and learning models \citep[see, e.g.,][]{bunting2024heterogeneity} . 

The most common approach to nonparametric inference in these applications is to use a nonparametric regularized maximum likelihood (sieve) approach followed by a plug-in principle. However, this approach has several important limitations relative to the methods proposed here. First, it requires fully nonparametric identification, which may entail strong assumptions. Second, it can be computationally intensive, as it needs to solve a high-dimensional nonlinear optimization problem. Finally, it can be very sensitive to the choice of the sieve basis, since estimating these nuisance parameters is an ill-posed inverse problem. Nevertheless, sieve methods could be combined with our orthogonal moments to provide DML inferences with desirable properties. We illustrate this point in Monte Carlo simulations, where we show the large efficiency gains from using DML-UH with efficient moments (which debias a plug-in moment). This application emphasizes the practical importance of relevant orthogonal moments. 

An additional important setting where our results may be useful is the analysis of heterogeneous treatment effects under flexible specifications. In our empirical application, we study heterogeneity in the effect of smoking during pregnancy on child birth weight, an outcome that has received considerable attention in economics due to its strong correlation with later-life outcomes such as health, education, and labor market performance \citep[see][]{hack1995long, corman1995effects, corman1998effect, currie1999impact}. Our approach extends \cite{abrevaya2006estimating} by allowing for heterogeneous slopes, and builds on \cite{arellano2012identifying} by incorporating high-dimensional controls and constructing novel moments that are locally robust to the covariance structure of the idiosyncratic errors. Our empirical findings are broadly consistent with those in \cite{arellano2012identifying} and show, in particular, that their conclusions remain robust to flexible specifications of the control variables. Yet, our results suggest slightly more variability of the treatment effects of smoking across mothers than previously determined. 


The rest of the paper is organized as follows. Section~\ref{literaturereview} reviews the related literature. Section~\ref{examplesandfunctionals} introduces the examples, target functionals, and a heuristic summary of the main results. Section~\ref{SecExistence} establishes conditions for the existence and relevance of orthogonal moments in models with nonparametric UH. Section~\ref{Construction} proposes a three-step method for constructing relevant debiased moments. Section~\ref{sectiontest} introduces a test statistic for DML inference, and Section~\ref{sectionExamples} applies the general results to our examples. Section~\ref{MC} presents Monte Carlo simulations. Section~\ref{empiricalapplication} reports an empirical application to the effect of smoking on birth weight. Finally, Section~\ref{Conclusions} concludes. The Supplementary Appendix contains proofs, asymptotic results, regularity conditions, and additional discussion.

	\section{Literature Review}
	\label{literaturereview}
	
	This paper contributes to two different strands of the literature---the literature on DML and the literature on models with UH or latent variables. We explain our contributions relative to these two literatures.
	
	While the literature on DML has been growing rapidly in the last years, the work on orthogonal moments for models with UH is relatively scarce, and more so for general functionals. The most recent literature on DML has shown that inferences based on LR moments have several
	advantages over plug-in approaches. Most notably, LR moments are useful to
	reduce model selection and regularization biases when Machine Learning (ML) estimators are used
	as first steps \citep[see][]{chernozhukov2018double}, they lead to optimal inferences \citep[see][]{lee2024locallyregularefficienttests}, and they provide valid confidence
	intervals for structural parameters under more general conditions than plug-in
	methods \citep[see][]{chernozhukov2022locally}. 
	
	In this paper, we contribute to the DML literature by studying functionals of the UH, a setting for which existing DML methods remain underdeveloped. In the context of linear panel data models with additive and scalar UH, our work complements recent DML inference methods proposed by \cite{clarke2025double}, \cite{klosin2022estimating}, and \cite{semenova2023inference}. Like some of these papers, we build on the estimation framework of \cite{belloni2016inference}. However, we depart from this literature in the specific application to high-dimensional panel data by allowing for multivariate UH and considering general functionals of UH (e.g., average marginal effects and variances).
	
	Moreover, our approach is conceptually distinct from the existing DML literature, which requires nuisance and target parameters to be identified from a starting GMM moment condition, see, e.g., \cite{chernozhukov2022locally}. Instead, we adopt a score operator approach to characterize Neyman-orthogonal moments, which complements the identification results of \cite{escanciano2022semiparametric} and the operator-based identification and orthogonality results for causal models in \cite{navjeevan2023identification}. In particular, our paper is the first to characterize all (relevant) Neyman-orthogonal moments in models with UH. 
	
	Our paper also relates to the important literature on robust inference in the presence of UH in panel data settings with a finite number of time periods or repeated measures, and an unrestricted distribution of individual effects given covariates (i.e., a fixed effects approach).\footnote{See the differencing methods proposed in, e.g., \cite{andersen1970asymptotic}, \cite{bonhomme2012functional}, \cite{chamberlain1984panel}, \cite{honore1992trimmed}, \cite{honore2023momentconditionsdynamicpanel}, and \cite{mundlak1978pooling}. There is also an extensive literature on debiasing for large $T$ panel data, see \cite{arellano2007understanding} for a survey, where $T$ denotes the number of periods. Our DML inferences are valid for fixed $T$.}  In the linear random coefficient model with strictly exogenous covariates, \cite{chamberlain1992efficiency} introduced a GMM estimator for the common parameters and average marginal effects that does not rely on the unknown distribution of UH. \cite{graham2012identification} further examined the irregular identification and estimation of these parameters.\footnote{In this paper, we focus on regular identification for target parameters, in the general sense that the information matrix for them is non-zero but possibly of reduced rank; the irregular case—characterized here by an identically zero information matrix—is left for future research.}
	
	Building on \cite{chamberlain1992efficiency}, \cite{arellano2012identifying} used covariance and additional independence restrictions on the idiosyncratic errors to identify variances or even the full distribution of UH. They proposed an estimator for the distribution of UH and reported plug-in estimates for functionals such as quantiles in their empirical application. 
	
	In this paper, we show that the moments commonly used to estimate average marginal effects, variances, and other functionals of the UH are generally not orthogonal. We derive novel locally robust (Neyman-orthogonal) moments for these parameters—when they exist—and provide valid asymptotic inference procedures based on them.
	
	This paper also relates to the important literature on functional differencing (\cite{bonhomme2012functional}). Under parametric assumptions on the conditional distribution of the dependent variables given the exogenous covariates and the UH—a setting we refer henceforth to as conditionally parametric—\cite{bonhomme2012functional} developed a systematic approach, known as functional differencing, to construct moment conditions for common parameters that are free from the distribution of individual effects. This method has seen several successful applications, including recent work by \cite{honore2024moment} and \cite{dobronyi2024identificationdynamicpanellogit} in the context of dynamic logit models, and \cite{bonhommeDano2024functional} in network models. 
	
	We show that when our method is applied to: (i) the common parameters as the functional of interest; (ii) conditionally parametric models; and (iii) under additional support conditions, it yields functional differencing as a special case. Thus, ours is the first paper to link the literature of functional differencing with the DML literature. This link is useful, as it enables the application of the full suite of theoretical and computational tools developed in the DML literature to functional differencing. In particular, this link provides a formal framework for the theoretical justification of cross-fitted efficient implementations of functional differencing. We illustrate the substantial efficiency gains of employing cross-fitted efficient functional differencing in a Monte Carlo simulation study.
	
	As \cite{bonhommeDano2024functional} succinctly write: “\textit{The scope of functional differencing is not limited to panel data, and the approach can be applied to any setting with a parametric conditional distribution involving latent variables”.} Our paper shows that the seminal contribution by \cite{bonhomme2012functional} on functional differencing has much broader applicability, beyond conditionally parametric models. However, the moments it produces must be modified to achieve orthogonality for general functionals. This is our main contribution relative to the literature of functional differencing. We fully characterize these additional restrictions that functional differencing moments must satisfy through score operators and their adjoints.\footnote{The first version of this paper, \cite{arganaraz2023existence}, has characterization results on general models, beyond models with UH.}

	Recently, \cite{bonhomme2024neyman} has extended the concept of Neyman (first-order) orthogonality to higher-order orthogonal moments within the conditionally parametric panel data and large $T$ setting. In contrast, our analysis is restricted to first-order orthogonal moments, but applies to more general models and fixed $T$, including high-dimensional panel data models and the Kotlarski model. In this sense, our results are complementary to those of \cite{bonhomme2024neyman}.
	
	More broadly, our paper contributes to the recent literature on how to systematically construct Neyman-orthogonal moments. Our approach complements the influence function and GMM-based approach of \cite{chernozhukov2022locally} by considering a likelihood approach for models with nonparametric UH. The construction of orthogonal moments is the fundamental step in the implementation of popular DML inferences, see \cite{chernozhukov2018double}. We propose novel orthogonal moments and DML inferences in panel data models with high-dimensional covariates, the Kotlarski model and the teacher value-added models. These applications demonstrate the utility of the general characterization of orthogonal moments in this paper.

	\section{Examples, Functionals and Heuristics}
	\label{examplesandfunctionals}
	\subsection{A High-Dimensional Linear Random Coefficient Model}
	\label{rcexample}
	Consider the following linear Random Coefficient model
	\begin{equation}
		\label{lineareqY}
		Y_{i}=W_{i}\beta_{0}+V_{i}\alpha_{i}+\varepsilon_{i},
	\end{equation}
	where $Y_{i}=(Y_{i1},...,Y_{iT})$ is a $T\times1$ vector of dependent
	variables, for a finite and fixed $T,$ $X_{i}=(W_{i},V_{i})$ are strictly
	exogenous covariates, for $T\times p$ and $T\times q$ matrices $W_{i}\ $and
	$V_{i},$ respectively, $\alpha_{i}$ is a $q\times1$ vector of individual heterogeneity (UH), and $\varepsilon_{i}$ are idiosyncratic
	(unobserved) errors.\footnote{Other names for $\alpha_{i}$ include fixed
		effects, individual-specific effects, incidental parameters, etc.} The data
	observation is $Z_{i}=(Y_{i},X_{i})$ for a large number of units $i=1,...,n.$ We assume that ${(Z_{i},\alpha_{i}):i=1,...,n}$ are independent and identically distributed (iid), with $Z_{i}$ following the distribution $\mathbb{P}_0$.
	To simplify notation, we drop the reference to $i$ when is convenient. This
	model is parametrized by $\lambda_{0}=\left(  \theta_{0},\eta_{0}\right)$ where $\theta_{0}=(\beta_{0},f_{\left.  \varepsilon\right\vert
		\alpha,X})\in\mathbb{R}^{p}\times\mathcal{F}%
	_{\mathcal{\varepsilon}}$ and $\eta_{0}=f_{\left.  \alpha\right\vert X},$ and
	where henceforth, for generic random vectors $U_{1}$ and $U_{2},$ $f_{\left.
		U_{1}\right\vert U_{2}}$ and $f_{U_{1}}$ denote, respectively, the conditional
	density of $U_{1}$ given $U_{2}$ and the marginal density of $U_{1}$. The set $\mathcal{F}_\varepsilon$ of error distributions might be large, i.e., we are not necessarily imposing parametric assumptions on $f_{\left.  \varepsilon\right\vert
		\alpha,X}$. We will be concerned with situations where $p$, the dimension of $W$, might be large, specifically, larger than the sample size. This is consistent with modern applications of panel data that may contain a myriad of time-varying characteristics or with flexible specifications of a low-dimensional set of controls that can be well approximated by a dictionary given by the columns of $W$.\footnote{We note that our results can be extended to nonlinear  (in covariates and common parameters) models, as in Equation (4.1) of \cite{chamberlain1992efficiency}, but with high-dimensional regressors; however, this extension is beyond the scope of the current paper.}
	
	In the more classical setting with $p$ fixed, \cite{chamberlain1992efficiency} proposed GMM estimators for $\beta_{0}$ and $\mathbb{E}%
	\left[  \alpha\right]  $ when the following strict
	exogeneity condition on $f_{\left.  \varepsilon\right\vert \alpha,X}\ $holds,
	\begin{equation}
		\mathbb{E}\left[  \left.  \varepsilon\right\vert \alpha,X\right]  =0,\text{
			almost surely (a.s.)}\label{exo}%
	\end{equation}
	He also considered the key assumption that $\mathbb{P}_{0}\left(
	rank(V^{\prime}V)=q\right)  =1$ and $q<T$. This is a type of non-surjectivity condition which plays a key role for the existence of orthogonal moments, see the first version of this paper \cite{arganaraz2023existence} for further discussion on existence. Under additional covariance restrictions (independence conditions of UH) of the idiosyncratic errors $\varepsilon$ given the exogenous regressors, \cite{arellano2012identifying} showed identification of the variance of $\alpha$ (and its distribution, respectively). All these results consider a fixed $p$. 
	
	We present in Section \ref{sectionExamples} novel orthogonal moments, DML inferences, and estimators for linear combinations of $\beta_{0}$, average marginal effects, and quadratic forms of the UH (e.g., variances) in this model with high-dimensional regressors, i.e., when $p$ is potentially larger than the sample size. 
	
	\subsection{Nonlinear Panel Data Models and Functional Differencing}
	\label{nonlinearexample}
	A strand of the literature on panel data has considered a model for $f_{Y|X}$ given by\footnote{Henceforth, for the sake of exposition, we drop the region of integration, as it will be clear from the context.}
	\begin{equation}
		f_{\lambda_{0}}\left(  z\right)  =\int f_{Y|\alpha,X}\left(  y|\alpha
		,x;\theta_{0}\right)  \eta_{0}(\alpha|x)d\alpha,\label{mixt}%
	\end{equation}
	where $f_{Y|\alpha,X}\left(  y|\alpha,x;\theta_{0}\right)$ is the conditional
	density of $Y$ given $(\alpha,X),$ which is assumed to be known up the finite-dimensional parameter $\theta_{0}$. This setting includes static and dynamic models, as $X$ may include exogeneous variables and initial conditions. We refer to this class of models as conditionally parametric models. It is in this setting where the recent and important literature on functional differencing has been developed, \citep[see, e.g.,][among others]{bonhomme2011,bonhomme2012functional,honore2024moment,dobronyi2021identificationdynamicpanellogit,davezies2021identification,bonhomme2024neyman}.
	
	We show that functional differencing is a special case of our approach for the common parameters as the functional of interest, when $\theta_{0}$ is a finite-dimensional parameter. Our characterization of orthogonal moments applies to more general models and other functionals, such as average marginal effects. We note that the moments resulting from functional differencing are generally not locally robust for general functionals (e.g., for average marginal effects). We thus provide novel orthogonal moments for those. We also develop DML-UH inferences allowing for partial identification of the multivariate target and nuisance parameters, and provide a theoretical framework for efficient implementations of functional differencing. We illustrate in a small Monte Carlo exercise the efficiency gains of optimal implementations of functional differencing and the importance of relevant orthogonal moments in practice.   
	
	\subsection{Kotlarski Model with a Factor Loading}
	\label{Kotlarskicexample}
	\bigskip \noindent This example illustrates that we allow for nonparametric specifications of $f_{\left.
		Y\right\vert \alpha}$. Specifically, suppose the observed data $Z=(Y_{1},Y_{2})$ satisfies%
	\begin{align}
		Y_{1} &  =\alpha+\varepsilon_{1},\label{K1}\\
		Y_{2} &  =\beta_{0}\alpha+\varepsilon_{2},\label{K2}%
	\end{align}
	where $\alpha$ is a common unobserved factor, independent of the bivariate error $\varepsilon=(\varepsilon_{1},\varepsilon_{2})$, which also has independent components with zero mean and densities $f_{\varepsilon_1}$ and $f_{\varepsilon_2}$, respectively. The scalar parameter
	$\beta_{0}$ is the unknown factor loading, with $\beta_{0}\neq0$. The density of the data is
	given by (\ref{mixt}) with $X$ empty, $Z=Y=(Y_{1},Y_{2})$, and
	\[
	f_{Y/\alpha}(y_{1},y_{2};\theta_{0})=f_{\varepsilon_{1}}(y_{1}-\alpha)f_{\varepsilon_{2}}(y_{2}-\beta_{0}\alpha).
	\]
	Thus, here $\theta_{0}=(\beta_{0},f_{\varepsilon_{1}},f_{\varepsilon_{2}})$ is infinite-dimensional. This model is called the Kotlarski model with a factor loading in \cite{lewbel2022kotlarski}, and it generalizes the classical Kotlarski model first used in
	econometrics by \cite{li1998nonparametric}. See \cite{rao1983nonparametric} for a description
	of the basic Kotlarski's lemma and \cite{evdokimov2012some} for extensions. The identification of the model above has been
	studied in detail by \cite{lewbel2022kotlarski}. 
    
    	The Kotlarski model has several prominent applications in economics. It has been used to study income dynamics, where $Y_1$ and $Y_2$ denote individual earnings at two different periods and $\alpha$ represents a persistent component of earnings \citep[][]{bonhomme2010generalized, horowitz1996semiparametric}. \cite{cunha2010estimating} analyze the formation of cognitive and noncognitive skills in childhood, where $Y_1$ and $Y_2$ are different measurements, e.g., test scores and $\alpha$ is unobserved skill.\footnote{\cite{cunha2010estimating} do not necessarily assume independence between $\varepsilon_1$ and $\varepsilon_2$.} \cite{fruehwirth2016timing} focus on the treatment effect of grade retention (or grade repetition), where $Y_1$ is an observed measure that is involved in the retention decision and $Y_2$ is some outcome of interest, e.g., test score, and $\alpha$ is an unobservable common ``cause"  that determines the effect of grade retention and whether a student is retained.  Other interesting applications have appeared to measure the importance of information sets while making optimal decisions \citep[][]{navarro2017identifying} and in the intergenerational mobility literature \citep[][]{nybom2016heterogeneous, haider2006life}. \cite{lewbel2022kotlarski} discusses further applications of the model (e.g., to panel data), and \cite{lewbel2024identification} has extended these applications to a triangular two-equation system without instruments.\footnote{Note that some of the aforementioned applications control for strictly exogenous regressors that appear in \eqref{K1}-\eqref{K2}. For simplicity, we focus on the case where these regressors are absent from the analysis. Our results could be extended to one-step or two-step methods that account for exogenous covariates.}
	
	The Kotlarski model is representative of a large class of semiparametric models with a similar mathematical structure for which our results apply. To illustrate, consider the general class of non-classical measurement error models put forward by \cite{hu2008instrumental}. In these models $\alpha$ is a latent (unobserved) regressor, with a contaminated counterpart $Y_2$ and a dependent variable $Y_1$, so  $Y=(Y_1,Y_2)$. The vector $X$ is a set of instrumental variables. Under their conditions, the model is a semiparametric mixture model as above, but with 
	\[
	f_{Y/\alpha}(y_{1},y_{2};\theta_{0})=f_{Y_1/\alpha}(y_{1};\beta_{0},\alpha)f_{Y_2/\alpha}(y_{2};\alpha),
	\]
	where $\theta_{0}=(\beta_{0},f_{Y_2/\alpha})$ is again infinite-dimensional. \cite{hu2008instrumental} proposed a joint sieve method for $\theta_{0}$ and $\eta_{0}$. Other examples with a similar mathematical structure include nonparametric panel data with interactive fixed effects in \cite{freyberger2018non} and learning models in \cite{bunting2024heterogeneity}, among many others. 
    
    In this application, we obtain debiased inferences for the $k$-order moment of the common factor and the parameter $\beta_{0}$ that do not require estimating the distribution of UH and other high-dimensional nuisance parameters. Thus, our methods in this application overcome the aforementioned limitations of plug-in functionals. 
	
	\subsection{Teacher Value-Added}
	\label{TVAcexample}
	In this model, $ij$ indexes student-teacher pairs (teacher $i$, student $j$), $t=1,\dots, T$ denotes periods, $Y_{ijt}$ captures $j$-student's test scores of a particular teacher $i$ at time $t$, which follows the model 
	$$
	Y_{ijt}=X_{ijt}^{\prime}\beta_{0}+\alpha_{i}+\varepsilon_{ijt},
	$$
	where $X_{ijt}$ is a $q-$dimensional vector of observed characteristics of the student and the teacher. The literature assumes that the entries of $\varepsilon$ are iid normal (across $i$ and $T$) with unknown and common variance and that the linear projection of $Y_{ijt}$ on $X_{ijt}$ identifies $\beta_{0}$. Recently, \cite{gilraine2020new} has proposed a nonparametric maximum likelihood estimation method for the distribution of UH from the equation that averages over $j$ and $t$, denoted by     
	$$
	Y_i = \alpha_i + \varepsilon_i.
	$$
	We propose LR moments for smooth (analytic) functionals of the teacher value-added, which are free from its distribution, and also show that LR moments do not exist for some empirically relevant policy parameters, see Section \ref{TAV}. These results abstract from the estimation of $\beta_{0}$ and the selection of controls $X_{ijt}$, as the results from the high-dimensional random coefficient model apply here to provide robust inferences to the selection of controls. 
	
	\subsection{Functionals of Interest}
	
	Researchers working with the previous models might be interested in various parameters, which typically depend on the UH, i.e., functionals of UH. We will denote a generic parameter by $\psi_0 \equiv \psi(\lambda_0)$, with $\lambda_0 = (\theta_0,\eta_0)$. Some examples of these functionals include:
	\begin{enumerate}
		\item[i)] \textit{Components of common parameters:} $\psi_0 = C_1^{\prime}\beta_0$, where $C_1$ is a $p \times p_1$ known matrix of coefficients, with rank $p_1$. An example is a single coefficient of $\beta_0$.
		
		\item[ii)] \textit{Expectations of linear combinations  of UH:} $\psi_0 = \mathbb{E}\left[C_2^{\prime}\alpha\right]$, where $C_2$ is a $q \times q_2$ known matrix of coefficients, with rank $q_2$. 
		
		\item[iii)] \textit{Linear projections of UH on covariates:} $\psi_0 = \left(\mathbb{E}\left[X^{\prime}X\right]\right)^{-1} \mathbb{E}\left[X^{\prime}\alpha\right]$, where $X$ is some relevant matrix of covariates. These functionals represent linear projections, e.g., from regressing neighborhood effects on average income in the neighborhood as in \cite{chetty2014measuring}. 
		
		\item[iv)] \textit{Moments that are quadratic in UH:} $\psi_0 = \mathbb{E}\left[\alpha^{\prime}\Omega \alpha\right]$, where $\Omega$ is a $q \times q$ known matrix. A leading situation in applications is to report the variance of UH; e.g., \cite{abowd1999high} measure variances of workers and firm effects along with their covariances. As an example, we can take $\psi_0 = \mathbb{E}\left[\frac{1}{q}\sum^q_{j=1}\left(\alpha_j - \frac{1}{q}\sum^q_{j^{\prime} = 1}\alpha_{j^{\prime}}\right)^2\right]$, but other weighting schemes are possible. 
		
		\item[v)]  \textit{CDFs and quantiles of UH:} In several studies, the cumulative distribution functions (CDFs) of UH is a central part. For instance, \cite{kline2022systemic} assess how concentrated racial discrimination is among employers in the US when it comes to hiring. To determine this, examination of the CDF of firm effects plays a key role. In this case, we might look at $\psi_0 = \mathbb{E}\left[\frac{1}{q} \sum^q_{j=1} \bm{1}\left\{\alpha_j \leq \bar{\alpha}\right\}\right]$, where $\bar{\alpha}$ is fixed and $\bm{1}\left\{A\right\}$ takes value one if the event $A$ occurs and zero otherwise. Similarly, it is common in applications to report quantiles of the distribution of UH. For example, \cite{aaronson2007teachers} compute the 10th, 25th, 50th, 75th, and 90th quantiles of teacher fixed effects on math scores in ninth grade. 
		\item[vi)] \textit{Policy parameters:} We can also consider parameters resulting from counterfactual exercises. A policy recommendation in the teacher value-added literature, which has been considered quite seriously, suggests replacing poor-quality teachers (i.e., teachers in the lowest part of the value-added distribution) with mean-quality ones; see \cite{hanusheck09}, \cite{hanushek2011economic}, and \cite{chetty2014measuring}. The marginal gain in test scores resulting from replacing the bottom $\varphi$\% (typically, 5\%), in the standard model of teacher value-added introduced above, is the functional $\psi(\lambda_0) = -\mathbb{E}\left[\alpha \bm{1}\left\{ \alpha < F^{-1}_\alpha(\varphi)\right\}\right]$,
		where $F_\alpha$ is the CDF of $\alpha$. Another recommendation might involve releasing the bottom $\varphi$\%. In this case, the researcher might be interested in $\psi(\lambda_0) = \mathbb{E}\left[\alpha \bm{1}\left\{\alpha \geq F^{-1}_\alpha(\varphi)\right\}\right]$.
	\end{enumerate}
	
	\bigskip Most of the previous quantities are special cases of the following generic functional 
	\begin{equation}
		\label{generalfunct}
		\psi\left(\lambda_0\right) = \mathbb{E}\left[r\left(\alpha,X,\theta_0\right)\right],
	\end{equation}
	where $r$ is known up to $\theta_0$. Our results apply to more general functionals of UH than in (\ref{generalfunct}), but we focus for the most part on this generic functional for simplicity of exposition.  

\subsection{A Heuristic Summary of the Main Results}
    
	The main result of this paper can be heuristically stated as follows. For a general functional $\psi_{0}=\mathbb{E}\left[r\left(\alpha,X,\theta_0\right)\right]$, and for a model with nonparametric UH, the set of relevant Neyman-orthogonal moments for $\psi_{0}$ is characterized as the set of non-zero moment functions $g(Z,\rho_{0},\psi_{0})$ with mean zero and finite variance, depending on additional nuisance parameters $\rho_{0}$ (which may include $\theta_0$ as well as other parameters different from $\psi_0$), satisfying the following two equations:
\begin{equation}
			\label{eq1}
\mathbb{E}\left[  \left.  g(Z,\rho_{0},\psi_{0})\right\vert \alpha,X\right]
	=r\left(\alpha,X,\theta_0\right)-\psi_{0},\text{ a.s.},
\end{equation}
and 
\begin{equation}
			\label{eq2}
\frac{\partial\mathbb{E}\left[  g(Z,\rho_{0},\psi_{0})\right]  }{\partial\rho}=0.
\end{equation}
	The first equation (\ref{eq1}) depends on the distribution of UH only through its support. This equation makes the moment $\mathbb{E}\left[  g(Z,\psi_{0},\rho_{0})\right]$ locally robust to the distribution of UH. In fact, under some commonly employed additional support conditions, the solutions to this equation will be globally robust to the distribution of UH, i.e., $\rho_{0}$ will not contain $\eta_0$. 
    
    The second equation is a heuristic and commonly used locally robust condition for the orthogonal moment with respect to the additional nuisance parameters. We provide a formal (and corrected) version of (\ref{eq2}) in terms of pathwise derivatives, or equivalently, in terms of a certain score operator and the derivative of the functional $\psi(\lambda_0)$. 
    
    The main result shows that these two equations are necessary and sufficient for a relevant orthogonal moment, up to scale, with the necessary part being the most difficult part to establish. We then propose a three-step procedure to construct a moment function $g$ that satisfies these two equations. We illustrate the wide applicability of our three-step procedure in our examples and their finite sample performance through a Monte Carlo and an empirical application. 

    The next section provides a formal framework and regularity conditions to prove the main result. Applied researchers can skip this and the next sections and move directly to the examples in Section~\ref{sectionExamples}, where the general results are applied.

    \section{Existence and Relevance of Debiased Moments}
	\label{SecExistence}
	
	This section formally introduces the model, regularity conditions, and the main theoretical result of the paper: the necessary and sufficient conditions for relevant orthogonal moments in models with nonparametric UH. We introduce some notation that will be used throughout the paper. 
	
	\bigskip \noindent \textbf{Notation:} Let
	$f_{\lambda}$ denote the density of $\mathbb{P}_{\lambda}$ with respect to (wrt) a $\sigma$-finite measure $\mu$, and let $f_{0}\equiv f_{\lambda_0}$ be the density pertaining to $\mathbb{P}_{0}$. Let $L_{2}\equiv L_{2}(f_{0})$ denote the
	Hilbert space of $\mathbb{P}_{0}-$square integrable measurable functions with
	inner product $\langle g_{1},g_{2}\rangle=\mathbb{E}\left[  g_{1}\left(  Z\right)  g_{2}\left(  Z\right)  \right]
	$ and norm $\left\Vert g\right\Vert_{2} ^{2}=\mathbb{E}\left[
	g^{2}\left(  Z\right)  \right]  $. The set $L^{0}_{2}\equiv L^{0}_{2}(f_{0})$
	is the subspace of zero-mean functions in $L_2,$ i.e., $g$ with
	$\mathbb{E}\left[  g\left(  Z\right)  \right]  =0$ and $\mathbb{E}\left[
	g^{2}\left(  Z\right)  \right]  <\infty.$ More generally, $L_{2}(f)$ and $L_{2}^{0}(f)$ are defined analogously for any density $f.$ For a vector $\mathbf{v}$ in $\mathbb{R}^{p}$, we denote $\left|\mathbf{v}\right|_1$ and $\|\mathbf{v}\|$ the  $l_1$ and $l_2$ norms, respectively. If $A$ is a matrix, let $\left|\left|A\right|\right|$ be its spectral norm, $vec(A)$ its vectorization, let $A^{\prime}$ be the transpose of $A$, and let $A^{\dagger}$ denote the generalized Moore-Penrose inverse of $A$. Henceforth, $I_{r}$ denotes the identity matrix of order $r\ $($r\times	r$ identity matrix). For any linear operator $S$, let $\mathcal{R}(S)$ and $\mathcal{N}\left(S\right)$ denote its range and null space, respectively. 
    For any linear space $V$, let $\overline{V}$ be its closure and $\Pi_{\overline{V}}$ be the orthogonal projection operator onto $\overline{V}$ (the topology will be clear from the context). For any subset $B$ of a generic Hilbert space $\mathcal{H}$, let $B^{\perp}$ be the orthogonal complement of $B$ in $\mathcal{H}$. 
    Henceforth, $C$ is a generic positive constant that can change from expression to expression.

	\subsection{Model, Regularity and Scores}
	The model we consider has a density with respect to $\mu=\mu_{Y}\times v_{X},$ where $\mu_{Y}$ is
		a $\sigma$-finite measure and $v_{X}$ is the probability measure of $X$, given by
	\begin{equation}
		f_{\lambda_{0}}\left(  z\right)  =\int f_{Y|\alpha,X}\left(  y|\alpha
		,x;\theta_{0}\right)  \eta_{0}(\alpha|x)d\alpha,\label{mixt2}%
	\end{equation}
	where $f_{Y|\alpha,X}\left(  y|\alpha,x;\theta_{0}\right)$ is the conditional
	density of $Y$ given $(\alpha,X)$, which is known up to the parameter $\theta_0$, and $\eta_{0}(\alpha|x)$ is the conditional density of the UH given the exogenous regressors $X$, which is unknown and unrestricted (i.e., we consider a fixed effects approach). The model is indexed by the parameter $\lambda_{0}=\left(\theta_{0},\eta_{0}\right) \in \Lambda\equiv\Theta\times\Xi$, where $\Theta$ and $\Xi$ are subsets of possibly infinite-dimensional spaces. For simplicity of presentation, we consider the UH to be continuously distributed, but this can be trivially relaxed. Additionally, we note that the dimension of the UH is unrestricted for the moment.
	
	The observed data is an iid sample
	$Z_{i}=(Y_{i},X_{i})$, $i=1,...,n,$ from  $\mathbb{P}_{0}\in\mathcal{P}=\{d\mathbb{P}_\lambda/d\mu =f_{\lambda} \text{ in } (\ref{mixt2}):\lambda=(\theta
	,\eta)\in\Lambda\}$. The parameters that
	generated the data are denoted by $\lambda_{0}=\left(  \theta_{0},\eta
	_{0}\right)  \in\Lambda,$ i.e., $\mathbb{P}_{0}=\mathbb{P}_{\theta_{0},\eta
		_{0}}.$ However, the model may be partially identified in the sense that
	the equation $\mathbb{P}_{0}=\mathbb{P}_{\lambda}$ may have more than one
	solution in $ \lambda \in \Lambda.$ 
	
	Our first assumption is that the model in (\ref{mixt2}) is regular, as defined in Section \ref{regpaneldatagral} of the Supplementary Appendix. Regularity includes the concept of Differentiability in Quadratic Mean (DQM), which is a standard smoothness assumption in the literature, see \cite{newey1990semiparametric}. 
	
	\begin{assumption}[Regularity of the Model]
		\label{regularity}
		The model in (\ref{mixt2}) is regular. 
	\end{assumption}
	
	To explain what regularity entails, we need to introduce some notation that will also be useful for subsequent developments. To model deviations from $\theta_0$ and $\eta_0$, we introduce the Hilbert space $\mathbf{H}=\mathcal{H}_{\theta}\times \mathcal{H}_{\eta} $, which has the inner product $\langle(\delta_{1},b_{1}%
	),(\delta_{2},b_{2})\rangle_{\mathbf{H}}:=\langle\delta_{1},\delta_{2}%
	\rangle_{\mathcal{H}_{\theta}}+\langle b_{1},b_{2}\rangle_{\mathcal{H}_{\eta}%
	}$, and where $\mathcal{H}_{\theta}$ and $\mathcal{H}_{\eta}\ $are Hilbert
	spaces endowed with the inner products $\langle\cdot,\cdot\rangle
	_{\mathcal{H}_{\theta}}$ and $\langle\cdot,\cdot\rangle_{\mathcal{H}_{\eta}},$
	respectively$.$ Since we are considering a fixed effects approach, the space of deviations from $\eta_0$ is unrestricted and given by $\mathcal{H}_{\eta}=L_2(\eta_0\times f_X)$, where $f_X$ is the density of $X$ (with respect to some dominating measure). The inner product associated to $\mathcal{H}_{\eta}$ is then simply $\langle b_{1},b_{2}\rangle_{\mathcal{H}_{\eta}%
	}=\mathbb{E}\left[  b_{1}(\alpha,X)b_{2}(\alpha,X)\right]$. 
	
	We will then consider paths $\tau\in\lbrack0,\epsilon)\rightarrow\lambda_{\tau}=(\theta_{\tau},\eta_{\tau})\in \Lambda$ such that $d\log\eta_{\tau}(\alpha)/d\tau= b\in\mathcal{H}_{\eta}$ and $d\theta_{\tau}/d\tau= \delta\in\mathcal{H}_{\theta}$, where henceforth $d/d\tau$ is the derivative from the right (i.e., for nonnegative values of $\tau$) at $\tau=0$. To achieve this, it will often suffice to consider $\theta_{\tau}=\theta_0+\tau \delta$ and $\eta_{\tau}=(1+\tau b)\eta_{0}$. Note that $b$ has the interpretation of the (conditional) score for $\eta_0$. If $b$ is bounded with $\int b(\alpha,x) \eta_0(\alpha|x) d\alpha = 0$ a.s., then $\eta_{\tau}$ will be a density for a small enough $\tau$. 
	
	Under regularity, scores are well-defined in a quadratic mean sense, but for simplicity of notation and exposition, in the main text, we will express them pointwise, as in
	\[
	S_{\theta} \delta = \frac{d}{d\tau} \log f_{\theta_{\tau}, \eta_{0}}(z), 
	\]
    \[
	S_{\eta} b = \frac{d}{d\tau} \log f_{\theta_{0}, \eta_{\tau}}(z).
	\]
	Since these derivatives may not exist for all deviations $h=(\delta,b)$ in $\mathbf{H}=\mathcal{H}_{\theta}\times
		\mathcal{H}_{\eta}$, we consider a subset $\Delta(\lambda
	_{0})=B(\theta_{0})\times B(\eta_{0})\subseteq\mathbf{H,}$ with the property that $B(\eta_{0})$ is dense in $\mathcal{H}_{\eta}\cap L^{0}_2(\eta_0)$ (consistent with the fixed effects approach). Thus, an implication of regularity and the chain rule is that the score of the model corresponding to the path $\lambda_{\tau}$ is 
	\begin{equation}
		S_{\lambda}h:=S_{\theta}\delta+S_{\eta}b,\text{ }h=(\delta,b)\in
		\Delta(\lambda_{0})\subset\mathbf{H}.\label{scoresemi}%
	\end{equation}
	Moreover, from a direct calculation with the likelihood of the model, in Section \ref{regpaneldatagral} of the Supplementary Appendix, we show that the scores for $\theta$ and $\eta$, have the expressions given, respectively, by
	\[
	S_{\theta} \delta (z)= \mathbb{E}\left[  \left.  \ell_\theta(Y\vert\alpha,X,\theta_0	)[\delta]\right\vert Z=z\right], 
	\]
    \[
	S_{\eta} b (z)= \mathbb{E}\left[  \left.  b(\alpha,X	)\right\vert Z=z\right],
	\]
	where $\ell_\theta(Y\vert\alpha,X,\theta_0	)[\delta]$ is the conditional score operator for $\theta$ at $\theta_0$, defined formally in Section \ref{regpaneldatagral} of the Supplementary Appendix. These formulas for the scores, and in particular that for $S_\eta$, will have fundamental consequences for the characterization of orthogonal moments. The score operators $S_{\theta}$ and $S_{\eta}$ transform \textquotedblleft
	scores\textquotedblright\ of the parameter spaces in $\Delta(\lambda
	_{0})\subseteq\mathbf{H,}$ into scores of the model in $L_{2}$.\footnote{See, e.g., \cite{begun1983information} for an introduction to score operators.}
	
	To illustrate the calculation of $S_{\theta}\delta$ for a simple case, consider conditionally parametric models where $\theta$ is finite-dimensional (say $p$-dimensional). In this case, $\mathcal{H}_{\theta}$ is $\mathbb{R}^p$, the inner product $\langle\cdot,\cdot\rangle
	_{\mathcal{H}_{\theta}}$ is the standard Euclidean inner product, and $S_{\theta}\delta(z)=s_\theta(z)'\delta$, where $s_\theta(z)=\partial \log f_{\theta_0,\eta_0}/\partial\theta$ is the classical score for $\theta$ at $\theta_0$. Here, $\ell_\theta(y\vert\alpha,x,\theta_0	)[\delta]=(\partial \log f_{Y|\alpha,X}\left(  y|\alpha
	,x;\theta_{0}\right)/\partial\theta)'\delta$ is the conditional score in the ``direction'' $\delta\in\mathbb{R}^p$. Further discussion on regularity is given in Section \ref{regpaneldatagral} of the Supplementary Appendix.
	
	\subsection{Relevant orthogonal moments}
	
	Inference on $\psi(\lambda_{0})$ is based on a non-zero vector of moment
	functions $g\in L_2$ such
	that, for all $\lambda_0\in\Lambda$,
	\begin{equation}
		\mathbb{E}\left[  g\left(  Z,\lambda_{0}\right)  \right]  =0,
		\label{eq:moment}%
	\end{equation}
	where $Z$ is distributed as $Z_{i}\ $and $\mathbb{E}$ denotes expectation
	under $\mathbb{P}_{0}$, i.e., under $\lambda_0$. 
	
	Our goal is to construct relevant orthogonal moments for inference on $\psi(\lambda_{0})$. To explain what relevant orthogonal moments for $\psi(\lambda_{0})$ are,
	let again $\lambda_{\tau}$ denote a path in $\Lambda,$ i.e., a mapping $\tau
	\in\lbrack0,\epsilon)\rightarrow\lambda_{\tau}\in\Lambda,$ with
	$\epsilon>0,$ such that $\lambda_{\tau}$ equals $\lambda_{0}$ at $\tau=0$
	and where other regularity conditions given in Section \ref{regpaneldatagral} of the Supplementary Appendix are satisfied. 
	
	In addition to the regularity of the model, we also require regularity conditions on the moments to be able to interchange derivatives and expectations. This will be a maintained assumption throughout for all the moments we consider.  Let $\mathbb{E}_{\tau}$ denote
	the expectation under the path $\mathbb{P}_{\lambda_{\tau}}\equiv\mathbb{P}_{\tau}.$
	
	\begin{assumption}[Regularity of moments]
		\label{RM}
		$g\in
		L^{0}_{2}$ satisfies (\ref{eq:moment}) and for all paths $\mathbb{P}_{\tau}\equiv\mathbb{P}_{\lambda_{\tau}}$ with scores $s=S_{\lambda_{0}}h$, $h\in
		\Delta(\lambda_{0})\subset\mathbf{H},$ fulfilling DQM, the derivatives
		$d\mathbb{E}[g(Z,\lambda_{\tau})]/d\tau$ and $d\mathbb{E}_{\tau}%
		[g(Z,\lambda_{0})]/d\tau$ are well-defined, and moreover,
		\begin{equation}
			\frac{d}{d\tau}\mathbb{E}_{\tau}\left[  g(Z,\lambda_{0})\right]
			=\mathbb{E}\left[  g(Z,\lambda_{0})s(Z)\right]  . \label{GI}%
		\end{equation}
	\end{assumption}

	\begin{remark}
		These regularity conditions are again standard in the literature, see, e.g., \cite{newey1990semiparametric}. Condition (\ref{GI}) holds, for example,
		when $g$ is bounded or more generally, if
		\[
		g(Z,\theta,\eta)=m(Z)-%
		{\textstyle\int}
		m(z)f_{\theta,\eta}(z)d\mu(z),
		\]
		provided that for all paths in Assumption \ref{RM},
		\[
		\lim\sup_{\tau\downarrow0}%
		{\textstyle\int}
		m^{2}(z)f_{\theta_\tau,\eta_\tau}(z)d\mu(z)<\infty;
		\]
		see \cite{ibragimovkhasminskii1981}, Lemma 7.2, p. 67. 
	\end{remark}

	Then, we give a formal definition of an orthogonal moment function for $\psi(\lambda)$ at $\psi_{0}$.

    \begin{definition}[Orthogonal Moment Function]
        An orthogonal moment function for $\psi(\lambda)$ at $\psi_{0}$ $\equiv\psi(\lambda_{0})$ is a (regular) moment function $g$
	satisfying (\ref{eq:moment}) and the locally robust property
	\begin{equation*}
		\frac{d}{d\tau}\mathbb{E}\left[  g(Z,\lambda_{\tau})\right]  =0,\label{LR2}%
	\end{equation*}
	for all paths $\tau\in\lbrack0,\epsilon)\rightarrow\lambda_{\tau}\in
	\Lambda$ with scores $h\in\Delta(\lambda
	_{0})\subseteq\mathbf{H,}$ such that $\psi(\lambda_{\tau})=\psi_{0}\ $for all $\tau\in
	\lbrack0,\epsilon)$. 
    \end{definition} 
	We will show below that the lack of
	sensitivity of the moment $\mathbb{E}\left[  g(Z,\lambda_{\tau})\right]  $ for
	all restricted paths fixing the parameter of interest, i.e., with $\psi
	(\lambda_{\tau})=\psi_{0},$ is the fundamental notion to extend DML to general functionals. The DML literature has considered moments with a partition of the parameters as $\lambda_0=(\psi_0,\gamma_0)$, for some $\gamma_0$, but such partition is not possible for most functionals $\psi(\lambda_0)$.

    We now define relevant orthogonal moments.
    \begin{definition}
    The moment $\mathbb{E}\left[g(Z,\lambda_{\tau})\right]$ will be relevant for $\psi(\lambda)$ at $\psi_{0}$ if 
	\begin{equation*}
		\frac{d}{d\tau}\mathbb{E}\left[  g(Z,\lambda_{\tau})\right]  \neq0,\label{LR2a}%
	\end{equation*}
	for some path $\tau\in\lbrack0,\epsilon)\rightarrow\lambda_{\tau}\in
	\Lambda$ such that $\psi(\lambda_{\tau})\neq\psi_{0}\ $for all $\tau\in
	\lbrack0,\epsilon)$. 
    \end{definition}
	
    Note that the concept of relevance is in a weak sense, as the non-zero derivative has to hold for some path for which $\psi(\lambda_{\tau})\neq\psi_{0}$ and not for all such paths. Relevant orthogonal moments give rise to score tests with a non-trivial local power function. Further discussion on local power analysis and relevance is given in Section \ref{LocalPower} of the Supplementary Appendix. 
	
	\subsection{Smooth Functionals}
	\label{Smoothfunctional}
	After discussing the regularity of the model and the moments, this section introduces smoothness assumptions on the functionals considered. For simplicity of exposition, we focus on a scalar functional 
	\begin{equation}
		\label{functionalUH}
		\psi(\lambda_{0})=\mathbb{E}\left[
		r(\alpha,X,\theta_{0})\right].
	\end{equation}
	Extensions to vector functionals are trivial.\footnote{The scalar case is without loss of generality. The case $\psi(\lambda_{0})\in\mathbb{R}^{d_{\psi}}$, with $d_{\psi}>1$, can be analyzed by considering one-dimensional projections $C'\psi(\lambda_{0})$, for a selected basis of $C\in\mathbb{R}^{d_{\psi}}$. In fact, our results also hold for $d_{\psi}=\infty$, but we focus on a finite-dimensional target parameter for simplicity of exposition.} Also, extensions to nonlinear functionals of UH are considered in Section \ref{TAV}, and are permitted in our theory. Examples of empirically relevant nonlinear functionals are quantiles of the distribution of the UH. 
	
	\begin{assumption}[Smoothness of the Functional]
		\label{functsmooth}
		The pathwise derivative of the mapping $\theta \mapsto \mathbb{E}\left[
		r(\alpha,X,\theta)\right]$ exists at $\theta_0$, and has the representation $d\mathbb{E}\left[
		r(\alpha,X,\theta_{\tau})\right]/d\tau=\langle r_\theta,\delta%
		\rangle_{\mathcal{H}_{\theta}}$, for all $\delta\in \overline{B(\theta
			_{0})}$ and some $r_\theta\in \overline{B(\theta
			_{0})}$. Moreover, $\mathbb{E}\left[r^2(\alpha,X,\theta_0)\right] < \infty$. 
	\end{assumption}
	
	This assumption is weaker than differentiability of the functional in the sense of \cite{van1991differentiable}, since in addition to Assumption \ref{functsmooth}, differentiability requires a special representation for $r$ and $r_\theta$; see Remark \ref{remarkDiff} for further discussion. 
    
    To illustrate this assumption, in the case of a finite-dimensional $\theta$, it requires the existence of the standard derivative $r_{\theta}=\partial \mathbb{E}\left[
	r(\alpha,X,\theta_{0})\right]/\partial \theta$. 
	
    \subsection{Main result: existence of relevant orthogonal moments}

   	Under the regularity of the model (Assumption \ref{regularity}), the adjoint operator of $S_{\theta}$ is well-defined: there exists a continuous linear operator $S_{\theta}^{\ast}%
	:L_2\rightarrow\overline{B(\theta_{0})}$ satisfying for all $\delta\in\overline{B(\theta_{0})}\subset{\mathcal{H}_{\theta}}$ and all
	$g\in L_2,$%
	\begin{equation}
		\mathbb{E}\left[  S_{\theta}\delta(Z)g(Z)\right]  =\langle \delta,S_{\theta
		}^{\ast}g\rangle_{\mathcal{H}_{\theta}}.
		\label{adjoint0}
	\end{equation}
	To illustrate, when $\theta$ is finite-dimensional, since $S_{\theta}\delta(z)=s_\theta(z)'\delta$, with $s_\theta(z)=\partial \log f_{\theta_0,\eta_0}/\partial\theta$, we have that $\mathbb{E}\left[\delta's_\theta(Z) g(Z)\right]  =\delta'\mathbb{E}\left[s_\theta(Z) g(Z)\right]$, and hence $S_{\theta}^{\ast}g=\mathbb{E}\left[s_\theta(Z) g(Z)\right]$. Therefore, $S^{*}_\theta$ is continuous when $\mathbb{E}\left[\|s_\theta(Z)\|^2 \right]<\infty$, which is a standard assumption in the literature. Furthermore, by the information equality, for a regular $g\in L^{0}_2$, 
	\begin{equation}
		S^{*}_\theta g=-\mathbb{E}\left[  \frac{\partial g(Z,\lambda_{0})}{\partial\theta^{\prime}%
		}\right].
	\end{equation}
For future reference, the adjoint operator for $\eta$ is given by 
        \begin{equation*}
			S^{*}_{\eta}g=\mathbb{E}\left[  \left.  g(Z,\lambda_{0})\right\vert \alpha,X\right]
			,\text{ a.s}.
		\end{equation*}
		 	
We are ready to give the main theoretical result of this paper. Define the $\eta$-influence function
	\begin{equation}
		\label{r_eta}
		r_\eta(\alpha,X,\psi_0,\theta_{0}):=r(\alpha,X,\theta_{0})-\psi_0,
	\end{equation}
 and recall $r_{\theta}$ has been defined in Assumption \ref{functsmooth}.
	\begin{theorem}
		\label{LRmomentUHresult}
		Let Assumptions \ref{regularity} to \ref{functsmooth} hold. Then, an orthogonal moment function $0\neq g$ for $\psi(\lambda_0)$ in model \eqref{mixt2} exists iff $0\neq g\in L^0_2$ solves: 
		\begin{equation}
			\label{02s}
			\mathbb{E}\left[  \left.  g(Z,\lambda_{0})\right\vert \alpha,X\right]
			=cr_\eta(\alpha,X,\theta_{0}),\text{ a.s},
		\end{equation}
		and 
		\begin{equation}
			\label{01s}
			S_{\theta}^{\ast}g=cr_{\theta},
		\end{equation}
		where $c$ is a constant that may depend on $\lambda_0$ and the functional $\psi$. Moreover, the orthogonal moment $g$ is relevant iff $c\neq0$.
	\end{theorem}
    \begin{remark}
        Equation (\ref{01s}) is equivalent to
        \begin{equation}
			\label{Cor01s'}
			\frac{d\mathbb{E}\left[  g(Z,\theta_{\tau},\eta_{0})\right]  }{d\tau}=0, \quad \textit{for all } \delta\in B(\theta_0) \textit{ with } \langle r_\theta,\delta%
		\rangle_{\mathcal{H}_{\theta}}=0.
		\end{equation}
        Under (\ref{02s}), equation (\ref{01s}) is equivalent to
        \begin{equation}
			\label{Cor01s''}
			\frac{d\mathbb{E}\left[  g(Z,\lambda_{\tau})\right]  }{d\tau}=0, \quad \textit{for all paths with }  \frac{d\psi(\lambda_{\tau})  }{d\tau}=0.
        \end{equation}
        This is the precise version of the heuristic condition in (\ref{eq2}).
    \end{remark}
    
    \begin{remark}
        The main conclusion of Theorem \ref{LRmomentUHresult} can be concisely stated as follows: the set of relevant orthogonal moments is the set of solutions of $S^{*}_{\lambda}g=r_{\psi}$ and scalar multiples of those, where $S^{*}_{\lambda}=(S^{*}_{\theta},S^{*}_{\eta})$ and $r_{\psi}=(r_{\theta},r_{\eta})$. The existence of a relevant orthogonal moment then requires $r_{\psi}\in\mathcal{R}(S^{*}_{\lambda})$, and in this case, a solution is given by 
        \begin{equation*}
        g=(S^{*}_{\lambda}))^{\dagger}r_{\psi}.    
        \end{equation*}
        In general, it will be difficult to implement this (optimal) solution, and one has to resort to numerical methods to solve the integral equation $S^{*}_{\lambda}g=r_{\psi}$, see \cite{engl1996regularization}. We propose a sequential three-step method to find a solution in Section \ref{Construction}.    
    \end{remark}
    \begin{remark}[Multivariate Functionals]
    \label{RemMultFunc}
For a multivariate functional $\psi(\lambda_{0})\in\mathbb{R}^{d_{\psi}}$ with Riesz representer $r_{\psi}=(r_{\psi}^{(1)},...,r_{\psi}^{(d_{\psi})})$, an orthogonal (real-valued) moment function $g$ exists iff $S_{\lambda}^{\ast}g=c^{\prime
}r_{\psi},$ for some $c\in\mathbb{R}^{d_{\psi}},$ where we define $c^{\prime}r_{\psi}:=\sum
_{j=1}^{d_{\psi}}c_{j}r_{\psi}^{(j)}\in\mathbf{H},$ where
$c=(c_{1},...,c_{d_{\psi}})^{\prime}\in\mathbb{R}^{d_{\psi}} $. Additionally, relevance is equivalent to $c\neq0.$ To illustrate, consider the Kotlarski model and the bivariate functional $\psi(\lambda_{0})=(\mathbb{E}\left[ \alpha\right],\mathbb{E}\left[ \alpha^2\right])\in\mathbb{R}^{2}$. For this example, $r_{\psi}^{(1)}=(0,\alpha-\mathbb{E}\left[ \alpha\right])$ and $r_{\psi}^{(2)}=(0,\alpha^2-\mathbb{E}\left[ \alpha^2\right])$, and the characterization of orthogonality is
        \begin{equation*}
			\mathbb{E}\left[  \left.  g(Z,\lambda_{0})\right\vert \alpha\right]
			=c_1(\alpha-\mathbb{E}\left[ \alpha\right]))+c_2(\alpha^2-\mathbb{E}\left[ \alpha^2\right]),\text{ a.s},
		\end{equation*}
and $S^{*}_\theta g=0$. We find such an orthogonal moment in Section \ref{Kotlarski}. 		
    \end{remark}
We give a corollary of this theorem, which may be very useful for identification and for a simpler verification of the orthogonality conditions. 
    \begin{corollary}
    \label{CorLRmomentUHresult}
		Under the assumptions of Theorem \ref{LRmomentUHresult}, the following two conditions are necessary and sufficient for the moment function $g=m_0-\psi_0$ to be a relevant orthogonal moment for $\psi(\lambda_0)$ with $c=1$ in the model \eqref{mixt2}: 
		\begin{equation}
			\label{Cor02s}
			\mathbb{E}\left[  \left.  m_0(Z,\lambda_{0})\right\vert \alpha,X\right]
			=r(\alpha,X,\theta_{0}),\text{ a.s},
		\end{equation}
		and 
		\begin{equation}
			\label{Cor01s}
			\frac{d\mathbb{E}\left[  m_0(Z,\theta_{\tau},\eta_{0})\right]  }{d\tau}=0, \quad \textit{for all } \delta\in B(\theta_0).
		\end{equation}
    \end{corollary}

	\begin{remark}[Relation to Necessary Conditions for Identification]
		\cite{bonhomme2011} has \\ proved that (\ref{Cor02s}) is a necessary condition for positive information for $\psi(\lambda_0)$ when $\theta_{0}$ is finite-dimensional and known. Hence, (\ref{Cor02s}) is also a necessary condition for positive information for $\psi(\lambda_0)$ when $\theta_{0}$ is infinite-dimensional and unknown. Our characterization of orthogonality and the DML inferences that we propose allow for partial identification of $\psi(\lambda_0)$, but for simplicity of exposition, our three-step construction below and the applications to the empirical examples focus on the case where $\psi(\lambda_0)$ has positive information. In fact, we also provide sufficient conditions for regular identification of $\psi(\lambda_0)$ in the applications (i.e., identification with positive information) and LR estimators.  
	\end{remark}
	
	\begin{remark}[Orthogonality as Sufficient for Identification]
		If $\mathbb{E}\left[ m_0(Z,\lambda_{0})\right]$ is identified, \\(\ref{Cor02s}) provides a sufficient condition for identification of $\psi(\lambda_0)$, namely $\psi(\lambda_0)=\mathbb{E}\left[ m_0(Z,\lambda_{0})\right]$. Note that (\ref{Cor02s}) identifies at least one solution $m_0$ (e.g., the minimum variance solution, which is unique) for each fixed $\theta_0$. For any solution $m_0$, the next step in this identification argument is to show that $\mathbb{E}\left[ m_0(Z,\theta_{0})\right]$ is identified, which may happen even when $\theta_{0}$ is not identified. For example, this is the case in our example application to the high-dimensional random coefficient panel model, where, for example, $\mathbb{E}\left[ C'_2\alpha\right]=\mathbb{E}\left[ C'_2H(Y-W\beta_{0})\right]$ is identified even when $\beta_0$ is not (since $W\beta_{0}$ is identified). 
	\end{remark}
\begin{remark}[Relation with Differentiable Functionals and Finite Efficiency Bound]
\label{remarkDiff}
In \\ an important paper, \cite{van1991differentiable}  introduced a key concept of differentiable
functional $\mathbb{P}_{\lambda}\rightarrow\psi(\lambda),$ as a necessary
condition for regular $\sqrt{n}-$estimability of $\psi(\lambda_{0}).$ Theorem 3.1 in \cite{van1991differentiable} shows that $\mathbb{P}_{\lambda}\rightarrow\psi(\lambda)$
is a differentiable functional iff $c^{\prime}r_{\psi}$ belongs to the
range of $S_{\lambda}^{\ast}$ for \emph{all} $c\in\mathbb{R}^{d_{\psi}}.$
In contrast, the existence of orthogonal moment requires this to happen only for some $c\in\mathbb{R}%
^{d_{\psi}},$ see Remark \ref{RemMultFunc}, so the existence of orthogonal moments is weaker than
the differentiability of the functional. He also showed that the differentiability of
the functional is equivalent to the existence of a $d_{\psi}$-dimensional
measurable function $s_{\psi}^{\ast}$ with components in $\overline{\mathcal{R}(S_{\lambda})}$ and solving
$S_{\lambda}^{\ast}s_{\psi}^{\ast}=r_{\psi}$ with $I_{\psi
}:=\mathbb{E}\left[  s_{\psi}^{\ast}(Z)\left(  s_{\psi}^{\ast}(Z)\right)
^{\prime}\right]  $ positive definite. The function $s_{\psi}^{\ast}$ is the
efficient score, and $I_{\psi}$ is the (efficient) Fisher information for
$\psi(\lambda)$ at $\psi(\lambda_{0}).$ Relevance is equivalent to $s_{\psi
}^{\ast}\neq0,$ $I_{\psi}\neq0$ or $rank(I_{\psi})>0,$ so it is also weaker
than differentiability which requires full rank of $I_{\psi}$. The full rank condition is required in the previous DML literature.
\end{remark}

        \begin{remark}[On Global Robustness]
        Note that equation (\ref{Cor02s}), and hence any solution $m_0$, depends on the distribution of UH only through its support. Under the additional support condition that equation (\ref{Cor02s}) holds for a known set that contains the support of the UH (a common assumption in the functional differencing literature), the solution $m_0$ (and hence the orthogonal moment) will be globally robust to the distribution of UH. For further discussion, including conditions under which global robustness is necessary, see Section \ref{BonhommeFD} in the Supplementary Appendix and the first version of this paper in \cite{arganaraz2023existence}.
        \end{remark}

        \begin{remark}
        One important advantage of the presentation in Corollary \ref{CorLRmomentUHresult} is that checking equation (\ref{Cor01s}) might be easier than verifying (\ref{01s}). This is the case when the solution to (\ref{Cor02s}) does not depend on all the nuisance parameters of the model. We have exploited this feature in our applications. 
        \end{remark}
    
	\begin{remark}
		The previous remarks motivate a three-step sequential procedure that we introduce in the next section, where we first find a solution to (\ref{Cor02s}) and then modify that solution to also verify (\ref{Cor01s}).
	\end{remark}
	
	\begin{remark}
		If the distribution of $Y$ conditional on $\alpha$ and $X$ is $L_2$-complete, then there is at most one unique (up to a vector multiplication) solution to (\ref{02s}); see \cite{d2011completeness} for definitions and conditions for $L_2$-completeness.  
	\end{remark}
	
	\begin{remark}
		For outcomes with a finite support, the set of functions $r$ that satisfy the necessary condition for orthogonality in (\ref{Cor02s}) is finite-dimensional (of dimension at most given by the cardinality of the support) for each $x$. This is consistent with the prevalence of partial identification with discrete outcomes in models with UH, see, e.g., \cite{honore2006bounds} and \cite{chernozhukov2013average}. Nevertheless, for empirically relevant marginal effects, analytical solutions of this equation have been obtained for dynamic logit models by \cite{aguirregabiria2024identificationaveragemarginaleffects}, \cite{davezies2021identification} and \cite{dobronyi2021identificationdynamicpanellogit}.   
	\end{remark}
	
	\begin{remark}
		\label{remarkthetapsi}
		If the functional of interest is $\psi(\lambda_{0})=\theta_0$ and $\theta_0$ is finite-dimensional, we have $r_{\theta}=I_p$ and $r(\alpha,X,\theta_0)=\theta_{0}$. Hence, for $\psi(\lambda_{0})=\theta_0$, $r_\eta=0$, and equations (\ref{02s}) and (\ref{01s}) boil down to 
		\begin{equation}
			\label{fde}
			\mathbb{E}\left[  \left.  g(Z,\lambda_{0})\right\vert \alpha,X\right]
			=0,\text{ a.s}%
		\end{equation}
        and		
        \begin{equation*}
			\mathbb{E}\left[  s_{\theta}(Z,\lambda_{0})g(Z,\lambda_{0})\right]
			=c,
		\end{equation*}
		A related equation to \eqref{fde} has been used in the functional differencing literature, see \cite{bonhomme2012functional} and Section \ref{BonhommeFD} in the Supplementary Appendix. For some dynamic logit models, \cite{honore2023momentconditionsdynamicpanel} have obtained closed-form solutions to \eqref{fde}. For other functionals the appropriate conditions for orthogonality are (\ref{01s}) and (\ref{02s}).    
	\end{remark}
	
	To summarize the main result of this section, Theorem \ref{LRmomentUHresult} characterizes all Neyman-orthogonal moments for general smooth functionals \eqref{functionalUH} in models with nonparametric UH.\footnote{Extensions to semiparametric UH are also possible, see the first version of this paper in \cite{arganaraz2023existence}.} The existing DML literature had only limited results for such functionals. 
	
	\section{Construction of Relevant LR\ moments}
	\label{Construction}
	Theorem \ref{LRmomentUHresult}  is constructive, as it provides us with a method to construct debiased moments. In this section, we describe a three-step procedure that directly computes an LR moment function without estimating UH, as we next show. For simplicity of exposition, in this section we focus on the regular scalar case where $\psi(\lambda_0)$ has positive information.\footnote{This gives a strong sense of relevance, where the corresponding test has non-trivial power in all directions.}  
    
    Suppose that we can find a $g_0 \in L_2$ such that 
	\begin{equation}
		\label{cond1}
		\mathbb{E}\left[  \left.  g_{0}(Z,\rho_{01})\right\vert \alpha,X\right]
		=r(\alpha,X,\theta_{0}),\text{ a.s.},
	\end{equation}
    where $\rho_{01}=\rho_{1}(\lambda_0)$ may contain $\theta_0$ and additional nuisance parameters. 
    
	Moreover, suppose that we can find a $g_1 \in L_2$ such that 
	\begin{equation}
		\label{cond2}
		\mathbb{E}\left[  \left.  g_{1}(Z,\rho_{01})\right\vert \alpha,X\right]
		=0,\text{ a.s.}
	\end{equation}
	Finally, let $\Gamma_0$ be such that, for all $h\in \Delta(\lambda_0)$ and $\rho_{\tau1}=\rho_{1}(\lambda_\tau)$, 
	\begin{equation}
		\label{cond3}
		\frac{d \mathbb{E}\left[g_{0}(Z,\rho_{\tau1})-\Gamma_{0}(X)g_{1}(Z,\rho_{\tau1})\right]}{d\tau} =0.  
	\end{equation}
        In all our examples, we have found that it suffices to take $\Gamma_{0}(X)$ as a constant $\Gamma_{0}$. Therefore, for simplicity of exposition, we henceforth consider this case. Following the proof of Corollary \ref{CorLRmomentUHresult}, condition (\ref{cond3}) is equivalent to 
	\begin{equation*}
		S^{*}_\theta g_0 -r_{\theta} =\Gamma_{0}S^{*}_\theta g_1.
        \end{equation*}
	For example, for a finite-dimensional $\rho_{01}=\theta_0$ equation (\ref{cond3}) reads
	\begin{equation*}
		\mathbb{E}\left[  \frac{\partial g_{0}(Z,\theta_{0})}{\partial\theta^{\prime}%
		}\right]=\Gamma_{0}\mathbb{E}\left[  \frac{\partial g_{1}(Z,\theta_{0}%
			)}{\partial\theta^{\prime}}\right].  
	\end{equation*}
	This equation has a unique solution $\Gamma_0$ when, for example, $g_1$ satisfies a full column rank condition (i.e., $\theta_0$ is locally identified). We only need the existence of $\Gamma_0$ (a minimum norm solution $\Gamma_0$ can be selected, for example).
    
    In the general case, we can obtain an orthogonal moment for $\psi(\lambda_{0})=\mathbb{E}\left[
	r(\alpha,X,\theta_{0})\right]$ as 
	\begin{equation}
		g(Z,\rho_{0},\psi_{0})=g_{0}(Z,\rho_{01})-\psi_{0}-\Gamma
		_{0}g_{1}(Z,\rho_{01}),\label{LRUHg}%
	\end{equation}
    where $\rho_0=(\rho_{01},\rho_{02})$ and $\rho_{02}=\Gamma_{0}$.
    
        To see that \eqref{LRUHg} leads to an orthogonal moment, we just need to verify that it satisfies  \eqref{02s}-\eqref{01s} in Theorem \ref{LRmomentUHresult}, or \eqref{Cor02s}-\eqref{Cor01s} in Corollary \ref{CorLRmomentUHresult} with $m_0(Z,\rho_{0})=g_{0}(Z,\rho_{01})-\Gamma
	_{0}g_{1}(Z,\rho_{01}).$  Indeed,
        \begin{align*}
		\mathbb{E}\left[  \left.  g(Z,\rho_{0},\psi_0)\right\vert \alpha,X\right]&= \mathbb{E}\left[  \left.  g_0(Z,\rho_{01})\right\vert \alpha,X\right]-\psi_0-\Gamma_0\mathbb{E}\left[  \left.  g_1(Z,\rho_{01})\right\vert \alpha,X\right] \\ & = r\left(\alpha,X,\theta_0\right)-\psi_0,
	\end{align*}
    and by the definition of $\Gamma_0$, 
	\begin{align*}    
		S^{*}_\theta g &=S^{*}_\theta g_0-\Gamma_0S^{*}_\theta g_1 \\ &=r_\theta,
	\end{align*}
	where we have used that $S^{*}_\theta \psi_0=0$ by the zero mean property of scores. Hence, \eqref{02s}-\eqref{01s} hold, and \eqref{LRUHg} is an orthogonal moment for $\psi(\lambda_0)$. 
    
    We summarize this in the following three-step algorithm: 
	
	\bigskip \noindent \textbf{Algorithm for obtaining an orthogonal function for} $\psi(\lambda_{0})=\mathbb{E}\left[  r(\alpha
	,X,\theta_{0})\right]$:
	
	\begin{description}
		\item[Step 1:] Solve for $g_{0}\in L_2$ satisfying the equation%
		\[
		\mathbb{E}\left[  \left.  g_{0}(Z,\rho_{01})\right\vert \alpha,X\right]
		=r(\alpha,X,\theta_{0}),\text{ a.s.}%
		\]

		\item[Step 2:] Solve for $g_{1}\in L_2$ satisfying the equation%
		\[
		\mathbb{E}\left[  \left.  g_{1}(Z,\rho_{01})\right\vert \alpha,X\right]
		=0,\text{ a.s.}%
		\]

		\item[Step 3:] Solve $\Gamma_{0}$ satisfying the equation $S^{*}_\theta g_0 -r_{\theta} =\Gamma_{0}S^{*}_\theta g_1$, or equivalently, for all $h\in \Delta(\lambda_0)$ and $\rho_{\tau1}=\rho_{1}(\lambda_\tau)$, 
	\begin{equation*}
		\frac{d \mathbb{E}\left[g_{0}(Z,\rho_{\tau1})-\Gamma_{0}g_{1}(Z,\rho_{\tau1})\right]}{d\tau} =0.  
	\end{equation*}
		
	\end{description}
A orthogonal moment function for $\psi(\lambda_{0})$ can be then constructed as
		\begin{equation*}
			g(Z,\rho_{0},\psi_{0})=g_{0}(Z,\rho_{01})-\psi_{0}-\Gamma
			_{0}g_{1}(Z,\rho_{01}).
		\end{equation*}

\begin{remark} [Global Robustness and Additional Support Conditions]
  Suppose that the \\ following stronger version of Step 1 holds: there exists a known set $\mathcal{A}$ that contains the support of $\eta_{0}\left(  \alpha|x\right)$ for all $x$, such that
  \begin{equation}
   \label{Global}
		{\textstyle\int}
		g_0(z,\rho_{01})f_{y|\alpha,X}\left(  y|\alpha
		,x;\theta_{0}\right)d\mu_Y(y)=0, \text{ for all } (\alpha,x)\in\mathcal{A}\times\mathcal{X};
 \end{equation}
  where $\mathcal{X}$ is the support of $X$. Then, this $g_0$ will satisfy Step 1 and, importantly, since (\ref{Global}) does not involve the distribution of the UH, $\rho_{01}$ can be chosen not to depend on it. This is a global robustness property. This additional support conditions have been employed in the functional differencing literature; see Section \ref{BonhommeFD} for further discussion. 
\end{remark}        
\begin{remark}[Interpretations of each step]
    Step 1 is a necessary condition for the regular identification of the functional of interest (cf., \cite{bonhomme2011}). If $g_0$ is a locally robust moment already, Steps 2 and 3 are not needed (we can take $g_1=0$ and $\Gamma_0=0$). Otherwise, Steps 2 and 3 require that the influence function of the functional $\mu(\theta_0)=\mathbb{E}\left[ g_{0}(Z,\theta_{0})\right]$ is in the span generated by $g_1$. This follows from the representation, for all $\delta\in B(\theta_0)$,
    \begin{equation*}
		\frac{d \mathbb{E}\left[g_{0}(Z,\theta_{\tau})\right]}{d\tau} =\mathbb{E}\left[\Gamma_{0}g_{1}(Z,\theta_{0})S_{\theta}\delta\right].  
	\end{equation*}
    We can replace $\Gamma_{0}g_{1}(Z,\theta_{0})$ in the construction of the orthogonal moment with any other influence function of the functional $\theta \rightarrow\mathbb{E}\left[g_{0}(Z,\theta)\right]$. 
\end{remark}    
\begin{remark}
    The advantage of this three-step approach is that the fewer nuisance parameters in $g_0$ and $g_1$, the easier it is to apply this procedure. Our examples nicely illustrate this point. Suppose these moment functions $g_0$ and $g_1$ depend only on a subset $\rho_{01}$ of $\theta_0$, then the orthogonal moments will depend only on $\rho_{0}=(\rho_{01},\rho_{02})$, where $\rho_{02}=\Gamma_0$, and they will be globally robust to other nuisance parameters.         
\end{remark}
    
	\begin{remark}
		We have introduced an extra nuisance parameter $\rho_0$, making the dependence of the function $g$ on it explicit. This supposes no disagreement with our previous discussion as we let $g(Z,\rho_0,\psi_{0}) \equiv g(Z,\lambda_0)$, as $\rho_0 = \rho(\lambda_0)$ and $\psi_0 = \psi(\lambda_0)$. Furthermore, observe that, with $\rho_2 = \Gamma$, $\mathbb{E}\left[g(Z,\rho_{01},\rho_{2},\psi_{0}) \right] = 0$, for any $\rho_{2} \neq \rho_{02}$, implying that the moment is globally orthogonal to this additional nuisance parameter, which is key for our asymptotic theory; see Section \ref{asymptotictheory} of the Supplementary Appendix.   
	\end{remark}
	
	\begin{remark}
		Step 1 entails an integral equation of the first kind \cite[e.g., Section 3 in][]{CARRASCO20075633}. In Section \ref{illposedsection} of the Supplementary Appendix, we briefly discuss some important theoretical and practical points to solving ill-posed problems that are useful in implementing these methods in practice. For illustration, see the application to functionals of teacher value-added.  
	\end{remark}
	
	The moments resulting from Step 1, however, will not be LR\ for
	$\psi(\lambda_{0})$ in
	general, as they will be affected by the estimation of $\theta_0$, which motivates our three-step procedure. For finite-dimensional $\theta_0$, \cite{honore2024moment} have obtained analytical solutions for Step 2 in dynamic discrete choice panel data models, and \cite{bonhomme2012functional} has proposed numerical approximations for solutions in Step 2,  and estimators for $\theta_{0}$ when this parameter is identified. Step 3 involves Jacobians of the moments, whose estimation is discussed in \cite{bonhomme2012functional}, p. 1366. For high-dimensional $\theta_0$, we show how these steps are solved in our applications to the high-dimensional random coefficient model, Kotlarski model, and the teacher value-added example.

	
	
	
	\begin{remark}
		An important implication of regularity of an orthogonal moment is the differentiability of the functional $\mathbb{P}_{\lambda}\rightarrow\phi(\lambda)=\mathbb{E}\left[  g(Z,\lambda)\right]$. From $\mathbb{E}_{\tau}\left[  g(Z,\lambda_{\tau})\right]=0$, it follows that, for all paths with scores $h\in\Delta(\lambda_0)$, 
		\begin{align*}
			d\phi(\lambda_{\tau})/d\tau = & -\mathbb{E}\left[  g(Z,\lambda_{0})s(Z)\right] \\
			= & \langle S_{\lambda_{0}}^{\ast}(-g),h\rangle_{\mathbf{H}}.
		\end{align*}
		
		\noindent Hence, the functional $\phi(\lambda)$ is smooth with a Riesz representer in the range of $S_{\lambda_{0}}^{\ast}$. The differentiability then follows from Theorem 3.1 in \cite{van1991differentiable}. Differentiability is necessary but not sufficient for $\sqrt{n}-$ estimability, which will require additional assumptions; see the asymptotic theory in the Appendix. These assumptions allow for the possibility that $\theta_0$ is not (locally) point-identified.
	\end{remark}

\section{DML Inferences}
	\label{sectiontest}
	In this section, we propose a score-type test based on orthogonal moments for the hypothesis $H_{0}:\psi(\lambda_{0})=\psi_{0}$ vs $H_{0}:\psi(\lambda_{0})\neq\psi_{0}$, where we some abuse of notation, in this section $\psi_{0}$ denotes a specified value under the null (e.g., $\psi_{0}=0$). The proposed test is in the spirit of the Anderson-Rubin test, see \cite{anderson1949estimation}. For ease of our exposition, we collect all the involved nuisance parameters, i.e., parameters different from $\psi_0$, in the vector $\rho_0 = \left(\rho_{01},\rho_{02}\right)$, for some suitable partition. The nuisance parameter $\rho_0$ may contain $\eta_0$ or be part of it.  We write the debiased moment as follows: 
	\begin{equation}
		\label{decomeq}
		g\left(Z,\rho_{0},\psi_0\right) = \phi_0\left(Z,\rho_{01},\psi_0\right) + \phi_1\left(Z,\rho_{0},\psi_0\right), \;\;\;\; \bar{g}\left(\rho_0,\psi_0\right) = \mathbb{E}\left[g\left(Z,\rho_{0},\psi_0\right)\right].
	\end{equation}
	such that $\phi_0\left(Z,\rho_{01},\psi_0\right), \phi_1\left(Z,\rho_{0},\psi_0\right) \in L^0_2$. Decomposition \eqref{decomeq} is useful to develop our asymptotic theory in the Appendix. The correction term could be $\phi_1\left(Z,\rho_{0},\psi_0\right)=-\Gamma_{0}g_{1}(Z,\rho_{01})$ when $\phi_0\left(Z,\rho_{01},\psi_0\right) = g_{0}(Z,\rho_{01}) - \psi_0$. In this case, $\rho_{02}=\Gamma_{0}$, which satisfies the key global robustness property used in the asymptotic theory that $\mathbb{E}\left[\phi_1\left(Z,\rho_{01},\rho_{2},\psi_0\right)\right]=0$ for all $\rho_{2}$.  Our setting here generalizes \cite{chernozhukov2022locally} to a partial identification framework.\footnote{Recently,   \cite{bennett2023source} has studied locally robust inference on linear functionals in nonparametric instrumental variables models.}

	As in the current DML literature, an important element in our construction of orthogonal moments is cross-fitting. It has been commonly used in the semiparametric
	literature, see, e.g., \cite{bickel1982adaptive}, \cite{klaassen1987consistent}, and \cite{schick1986asymptotically}. Cross-fitting allows us to deal with high-dimensional estimators and avoid an ``own observation'' bias; see the discussion by \cite{chernozhukov2018double}. It is implemented as follows. Given the data $\left\{  Z_{i}\right\}  _{i=1}%
	^{n}$, we randomly partition the observation indices $(i=1,...,n)$ into $L$ groups
	$I_{\ell},$ $(\ell=1,...,L)$ of equal size. Let $\hat{\rho}_{\ell}$ denote an estimator of
	$\rho_{0}$  using all
	observations \textit{not} in $I_{\ell}$; see below for the specific choices of
	$\hat{\rho}_{\ell}$ in our examples. Thus, we here assume that there exists an estimator of the components in $\rho_0$ to implement our DML inferences.\footnote{Recall that $\rho_0$ might not necessarily contain the entire vector of nuisance parameters $(\theta_0,\eta_0)$.} If $\rho_{0}$ is
	low-dimensional there is no need to use cross-fitting, and we can set $\hat{\rho}_{\ell}=\hat{\rho}$ based on all the observations. The estimated orthogonal moment is then%
	\[
	\hat{g}_{i\ell}(\psi_{0})\equiv g(Z_{i},\hat{\rho}_{\ell},\psi_{0}).
	\]
	For a
	given (estimated) $k-$dimensional LR\ moment $\hat{g},$ a debiased cross-fitted
	sample moment functions is%
	\begin{equation}
		\mathbb{E}_{n,c}\left[  \hat{g}_{i\ell}(\psi_{0})\right] : =\frac{1}{n}%
		\sum_{\ell=1}^{L}\sum_{i\in I_{\ell}}\hat{g}_{i\ell}(\psi_{0}).\label{dmom}%
	\end{equation}
	
	Based on our estimated LR moments, \eqref{dmom} we propose a test to conduct robust inference on $\psi(\lambda_0)$. Let the cross-fitted weighting matrix be
	\[
	\check{W}_{n}=\frac{1}{n}\sum_{\ell=1}^{L}\sum_{i\in I_{\ell}}\hat
	{g}_{i\ell}(\psi_{0})\hat{g}_{i\ell}^{\prime}(\psi_{0}),
	\]
	which under regularity conditions consistently estimates $W_{g}=\mathbb{E}%
	\left[  g(Z,\rho_{0},\psi_0)g(Z,\rho_{0},\psi_0)^{^{\prime}}\right] .$ 
	\begin{remark}
		\label{SimpWg}
		When $g(z,\rho_{0},\psi_0)=m_0(z,\rho_{0})-\psi_0$, we can write 
		\begin{equation}
			W_{g}=\mathbb{E}\left[  (m_0(Z,\rho_{0})-\mathbb{E}%
			\left[  m_0(Z,\rho_{0})\right])(m_0(Z,\rho_{0})-\mathbb{E}%
			\left[  m_0(Z,\rho_{0})\right])' \right],\label{Wgm}
		\end{equation}
		which does not depend on $\psi_0$. The implementation of our inference is substantially simplified if a (cross-fitted) sample version of this expression for $W_{g}$ is used. In this scenario, we construct an estimator of $W_g$ using $\hat{g}_{i\ell}(\hat{\psi}_\ell)\equiv g\left(Z_i,\hat{\rho}_\ell,\hat{\psi}_\ell\right)$, where $\hat{\psi}_\ell$ estimates $\mathbb{E}%
		\left[  m_0(Z,\rho_{0})\right]$ for each $\ell$. 
	\end{remark}

	We allow for the possibility that $W_{g}$ is singular, as when there is partial identification. Define $0<k_{g}:=rank(W_{g})\leq k$. To construct a
	consistent estimator for $W_{g}^{\dagger},$ we follow a common nonparametric strategy in the literature, see \cite{andrews1987asymptotic}, or more recently \cite{lee2024locallyregularefficienttests}.\footnote{When information of the rank $k_{g}$ is available, more efficient and simpler to implement estimators of $W_{g}^{\dagger}$ might be available. Here, we focus on the most difficult case where $k_{g}$ is unknown, but of course, our results can also be applied to the simpler case where $k_{g}$ is known, with obvious modifications. Similar remarks apply to the set of eigenvalues and eigenvectors of $W_g$.} Let $\check{W}_{n}=\check{U}_{n}\check{\Lambda}_{n}\check{U}%
	_{n}^{^{\prime}}$ be the eigendecomposition of $\check{W}_{n}$,\footnote{Using either $\hat{g}_{i\ell}\left(\psi_0\right)$ or $\hat{g}_{i\ell}\left(\hat{\psi}_\ell\right)$.} with $\check{\lambda}_{n,i}$
	being the $i-th$ element of the diagonal matrix $\check{\Lambda}_{n}\ $of
	eigenvalues, where we drop the dependence on $\psi_{0}$ for simplicity of
	notation, and define
	\begin{equation}
		\hat{W}_n:=\check{U}_{n}\check{\Lambda}_{n}(\nu_{n})\check{U}%
		_{n}^{^{\prime}},\label{eq:trunc_matrix}%
	\end{equation}
	where $\check{\Lambda}_{n}(\nu_{n})$ is a diagonal matrix with $i-$th
	element equal to $\check{\lambda}_{n,i}$ if $\check{\lambda}_{n,i}%
	\geq\nu_{n}$ and zero otherwise, and $\nu_{n}$ is non-negative user-specific
	sequence that $\nu_{n}\longrightarrow0$ at a suitable rate; see Section \ref{asymptotictheory} of the Appendix. We derive specific $\nu_n$ for some of our examples below.
	
	Our feasible cross-fitted debiased score test statistic is
	\begin{equation}
		\label{teststat}
		\hat{C}_{n}(\psi_{0})=n\mathbb{E}_{n,c}\left[  \hat{g}_{i\ell}(\psi
		_{0})\right]  ^{\prime}\hat{W}_n^{\dagger}\mathbb{E}_{n,c}\left[  \hat{g}_{i\ell
		}(\psi_{0})\right]  .
	\end{equation}
	The proposed test for $H_{0}:\psi(\lambda_{0})=\psi_{0}$ vs $H_{0}:\psi(\lambda_{0})\neq\psi_{0}$ rejects the null hypothesis at $\zeta$ nominal level if
	$\hat{C}_{n}(\psi_{0})>c_{\zeta,\hat{k}_{g}}$, where $\hat{k}_{g}=rank(\hat{W})$ and $c_{\zeta,d}$ is the $1-\zeta$ quantile of a
	chi-square distribution with $d$ degrees of freedom. To define confidence
	regions for $\psi(\lambda_{0})$, let the parameter space be $\Psi=\{\psi(\lambda
	):\lambda\in\Lambda\}$ and the asymptotic confidence region 
	\[
	CR_{\zeta}=\{\psi\in\Psi:\hat{C}_{n}(\psi)\leq c_{\zeta,\hat{k}_{g}(\psi
		)}\},
	\]
	where we emphasize the dependence of the rank $\hat{k}_{g}$ on the parameter
	$\psi.$ Note that when the conditions of Remark \ref{SimpWg} hold, $\hat{k}_{g}$ and $\hat{W}$ do not depend on $\psi$. In Section \ref{asymptotictheory} of the  Appendix, we give primitive conditions for the asymptotic validity of the proposed tests as well as a detailed local power analysis of
	it.

	\section{Application to the Examples}
	\label{sectionExamples}
	
	\subsection{High-Dimensional Random Coefficient Model}
	\label{sectionhighdimensionalpanel}
	In this section, we apply our theory to the High-Dimensional Random Coefficient Model. 
		\subsubsection{Common Parameters}
	\label{commonpar}
	We first construct novel LR moments for $\psi(\lambda_0)=C_{1}^{\prime}\beta_{0}$ for a known $p\times p_{1}$ matrix $C_{1}$ of full column rank. In an important paper, \cite{chamberlain1992efficiency} proposed a GMM estimator for $\beta_{0}$ that is locally robust for a fixed $p$. He assumed $\mathbb{P}_{0}\left(rank(V^{\prime}V)=q\right)  =1$, and introduced the generalized within-group and between operators, respectively,
	\begin{align*}
		Q &  =I_{T}-V(V^{\prime}V)^{-1}V^{\prime},\\
		H &  =(V^{\prime}V)^{-1}V^{\prime}.
	\end{align*}
	Then, by the exogeneity condition (\ref{exo}) and $QV=0$ a.s., it follows that
	\begin{equation}
		\mathbb{E}\left[  \left.  W^{\prime}Q(Y-W\beta_{0})\right\vert \alpha
		,X\right]  =0,\text{ a.s.,}%
		\label{LRbeta}
	\end{equation}
	which leads to the LR moment function $g(Z,\beta_{0})=W^{\prime}Q(Y-W\beta
	_{0})$ for $\psi(\lambda_{0})=\beta_{0}$ when $p$ is fixed. 
	
	In high-dimensional settings where $p$ is large, possibly much larger than the sample size, Chamberlain's GMM estimator requires regularization, resulting in the large biases common of high-dimensional estimators. Moreover, in many applied settings, researchers are not interested in the whole vector $\beta_{0}$ but rather in some linear combination (e.g., a particular coefficient) $\psi(\lambda_{0})=C_{1}^{\prime}\beta_{0}$. The moment derived in \cite{chamberlain1992efficiency} is not LR for $\psi(\lambda_{0})=C_{1}^{\prime}\beta_{0}$, which is particularly problematic for cases where $p>>nT$. To see this, define the matrices  $H_{C_{1}}:=C_{1}(C_{1}^{\prime}C_{1})^{-1}$ and $Q_{C_{1}}:=I-C_{1}(C_{1}^{\prime}C_{1})^{-1}C^{\prime}_{1}$, and note the identity $\beta_0=H_{C_{1}}C_{1}^{\prime}\beta_0+Q_{C_{1}}\beta_0$. In Chamberlain's moment, now $Q_{C_{1}}\beta_0$ becomes a high-dimensional nuisance parameter, and the moment  $g(Z,\beta_{0})$ in (\ref{LRbeta}) is sensitive to it. 
	
	The form of the LR moment for $\psi(\lambda_{0})=C_{1}^{\prime}\beta_{0}$ and the assumptions needed for it follow from our general necessary and sufficient conditions for the existence of a relevant orthogonal moment. Here, we omit the technical details and give the LR moment directly. A more general definition of the generalized within-group operator is $Q=I_{T}-VV^{\dagger}$. Define also the $p\times p$ matrix $M:=\mathbb{E}\left[  W^{\prime}QW\right]$ and the $p_{1}\times p$ matrix $\rho_{02} := C^{\prime}_1M^{\dagger}$. We need the following assumption. 
	\begin{assumption}
		\label{RankHDpanelbeta}
		i) $\mathbb{E}\left[  \|W^{\prime}QY\|^2\right]<\infty$, $\mathbb{E}\left[  \|W^{\prime}QW\|^2\right]<\infty$, and $\mathbb{E}\left[\left|\left|W^{\prime}Q\left|\left|^2 \right|\right|\varepsilon\right|\right|^2\right] < \infty$ ; ii)  $\left|\left|\beta\right|\right| < C$, $\left|\left|\rho_2\right|\right| < C$, for $\left|\left|\beta - \beta_0 \right|\right|$  and $\left|\left|\rho_2 - \rho_{02} \right|\right|$ small enough; iii) the column space of $M$ contains that of $C_{1}$ and $rank(C_{1})=p_{1}$.  
	\end{assumption}
	Assumption \ref{RankHDpanelbeta} iii) requires that the variation in $W$ not collinear with $V$ (variation in  $QW$) is rich enough to identify the component $C'_1\beta_0$. This assumption implies the order conditions $rank(M)\geq p_{1}$ and $rank(V)<T$ a.s. Under Assumption \ref{RankHDpanelbeta} iii), $\rho_{02}$ is well defined. Before providing a formal result, it is worth noticing what the functions $g_0$, $g_1$, and $\Gamma_0$ look like in this situation. We have $r\left(\alpha,X,\beta_0\right) = C^{\prime}_1\beta_0 = \psi_0$. Next, if we let 
    $$
    g_0(Z,\rho_{0},\psi_0) = \rho_{02}W^{\prime}Q\left(Y-WH_{C_1}\psi_0 - WQ_{C_1}\beta_0\right) + \psi_0,
    $$
    where $\rho_0=(\beta_0,\rho_{02})$. By \eqref{LRbeta}, we obtain 
    $$
    \mathbb{E}\left[\left.g_0(Z,\rho_{0},\psi_0) \right|\alpha,X\right] = C^{\prime}_1\beta_0,
    $$
    as required by Step 1 of our algorithm above. Interestingly, direct calculation yields 
    $$
    \frac{d}{d\tau}\mathbb{E}\left[g_0(Z,\beta_\tau)\right] = 0.
    $$
Thus, $g_0$ is already orthogonal to $\beta_0$, which implies that the resulting moment is LR to, e.g., bias selection/bias regularization in $\beta_0$. Hence, we shall let $g_1 = 0$ and $\Gamma_0 = 0$, using our notation above. For these, Step 2 and Step 3 are therefore satisfied. We are now ready to give an identification result and a relevant LR moment for $\psi(\lambda_{0})=C_{1}^{\prime}\beta_{0}$.
	
	\begin{proposition}
		\label{propLRc1beta}
		Under Assumption \ref{RankHDpanelbeta}, $\psi_{0}=C_{1}^{\prime}\beta_{0}$ is identified as 
		\begin{equation}
			C_{1}^{\prime}\beta_{0}=\rho_{02}\mathbb{E}\left[  W^{\prime}QY\right],
		\end{equation} 
		and the following function is an LR moment function for $\psi_{0}$
		\begin{equation}
			\label{glrbeta}
g(Z,\rho_0,\psi_{0})=\rho_{02}W^{\prime}Q\left(Y-WH_{C_1}\psi_0 - WQ_{C_1}\beta_0\right).   
		\end{equation} 
	\end{proposition}
	
	The following algorithm describes the steps to estimate the LR moment. Define $Q_i:=I_{T}-V_iV_i^{\dagger}$.
	
	\bigskip \noindent \textbf{Algorithm for obtaining an LR moment and inferences for $\psi(\lambda_{0})=C_{1}^{\prime}\beta_{0}$}\textbf{:}
	
	\begin{description}
		\item[Step 1:] Compute $Q_i$ for $i=1,...,n.$ 
		
		\item[Step 2:] Compute for $\ell=1,...,L.$
		\[
		\check{M}_{\ell}=\frac{1}{n-n_{\ell}}%
		\sum_{l\neq\ell}^{L}\sum_{i\in I_{l}}W'_iQ_iW_i,
		\]
		where $n_{\ell}$ is the number of elements of the block $I_{\ell}$, and
		\[
		\hat{\rho}_{2\ell}=C_{1}^{\prime}\hat{M}^{\dagger}_{\ell},
		\] 
		for a suitable estimator $\hat{M}^{\dagger}_{\ell}$, e.g., following the procedure in Section \ref{moorepenroseest}. 
		
		\item[Step 3:] Estimate $\beta_{0}$ by the following penalized least squares problem
		
		\[
		\hat{\beta}_{\ell}=\underset{\beta}{\operatorname{argmin}} \frac{1}{\left(n-n_{\ell}\right)T}%
		\sum_{l\neq\ell}^{L}\sum_{i\in I_{l}}(Y_i-W_i\beta)'Q_i(Y_i-W_i\beta)+2c_{n}\sum^p_{j=1}\hat{\phi}_{j,\ell}\left|\beta_j\right|,
		\]
		
		where $c_{n}$ is a penalization parameter, i.e., a sequence of non-negative numbers converging to zero at a suitable rate and $\left\{\hat{\phi}_j\right\}^p_{j=1}$ are penalty loadings; see Section \ref{implementationlasso} for how we specify the tuning parameter and the penalty loadings in practice. In the same section, we provide an implementation procedure to solve for $\hat{\beta}_\ell$. 
		
		\end{description}
	
	Our penalized least squares estimator $\hat{\beta}_\ell$ extends the Lasso estimator of \cite{belloni2016inference} of high-dimensional (linear) panel models to account for multidimensional, possibly non-additive, UH. 
	
	\noindent Inference is then implemented following Section 5. This LR for $\psi(\lambda_{0})=C_{1}^{\prime}\beta_{0}$ is LR with respect to all nuisance parameters, including $\rho_{02}$ and $Q_{C_{1}}\beta_{0}$. When $p_{1}=1$, under Assumption \ref{RankHDpanelbeta} the asymptotic variance $W_g$ is positive. We recommend implementing the following cross-fitted variance estimator
	\[
	\hat{W}_{n}=\frac{1}{n}\sum_{\ell=1}^{L}\sum_{i\in I_{\ell}}
	({g}_{i\ell}(\hat{\psi}_{\ell})-\bar{g}_{\ell}(\hat{\psi}_{\ell}))^2,
	\]
	where  $\hat{\psi}_{\ell}=\hat{\rho}_{2\ell}\hat{R}_{\beta,\ell}$, with   
	\[
	\hat{R}_{\beta,\ell}=\frac{1}{n-n_{\ell}}%
	\sum_{l\neq\ell}^{L}\sum_{i\in I_{l}}W'_iQ_iY_i,\hspace{1cm}
	\]
	\[
	\bar{g}_{\ell}(\hat{\psi}_{\ell})=\frac{1}{n}\sum_{\ell=1}^{L}\sum_{i\in I_{\ell}}\hat
	{g}_{i\ell}(\hat{\psi}_{\ell}).
	\]

	\begin{proposition}
		\label{validinferpanelbeta}
		Suppose that  Assumption \ref{RankHDpanelbeta} i) and iii) above, and Assumption \ref{regcondpanelbeta} in Section \ref{secasymptotichighpaneldata} of the Supplementary Appendix hold. A test based on $\hat{C}_n(\psi_0)$ in (\ref{teststat}) for $\psi_0 = C_1^{\prime} \beta_0$, using the estimator of a LR moment function \eqref{glrbeta} and the variance-covariance matrix estimator 
		$$
		\check{W}_n = \frac{1}{n}\sum^L_{\ell=1}\sum_{i \in I_\ell}\left(\hat{g}_{i\ell}\left(\hat{\psi}_\ell\right) -\bar{g}_\ell\left(\hat{\psi_\ell}\right)\right)\left(\hat{g}_{i\ell}\left(\hat{\psi}_\ell\right) -\bar{g}_\ell\left(\hat{\psi_\ell}\right)\right)^{\prime},
		$$
		where $\hat{\psi}_\ell = \hat{\rho}_{2\ell} \hat{R}_{\beta,\ell}$, with $\hat{R}_{\beta,\ell} = (n - n_\ell)^{-1}\sum^L_{l \neq \ell} \sum_{i \in I_\ell} W_i^{\prime}Q_iY_i$, has size equal to $\zeta \in (0,1)$ asymptotically. The probability that $\psi_0 \in CR_{\zeta}$ converges to $(1-\zeta)$ as $n \rightarrow \infty$. Finally, the test presents a non-trivial local asymptotic power function.
	\end{proposition}
	A debiased estimator for $\psi_0$ is 
	\[
	\hat{\psi}=\frac{1}{n}\sum_{\ell=1}^{L}\hat{\rho}_{2\ell}\hat{R}_{\beta,\ell}.
	\]
	This estimator is asymptotically normal under the conditions of Proposition \ref{validinferpanelbeta}, as it follows from its proof in Section \ref{Proofs} of the Supplementary Appendix.

	\begin{remark}
		\label{rcausalinterrandom}
		Our high-dimensional random coefficient model might be interpreted through the lens of dynamic event-study designs. In ``staggered rollout" designs in which being treated is an absorbing state, we argue that our theory allows for more flexibility in modeling parallel trends assumptions; see Section \ref{sectioneventstudy} of the Supplementary Appendix.  
	\end{remark}

	\subsubsection{Average Marginal Effects}
	\label{averagemarginal}
	\cite{chamberlain1992efficiency} also derived the following moment function $g(Z,\beta_{0},\psi
	_{0})=H(Y-W\beta_{0})-\psi_{0}$ for the average marginal effects $\psi_{0}=\mathbb{E}\left[  \alpha\right]$ when $p$ is fixed. This moment is motivated from the exogeneity
	condition, since $HV=I_{q}$ a.s., and hence
	\[
	\mathbb{E}\left[  \left.  H(Y-W\beta_{0})\right\vert \alpha,X\right]
	=\alpha,\text{ a.s.}%
	\]
	Taking expectations both sides, we conclude $\mathbb{E}\left[  H(Y-W\beta
	_{0})-\psi_{0}\right]  =0.$ However, this moment is not LR for $\psi
	_{0}=\mathbb{E}\left[  \alpha\right]$, and it is in general, quite sensitive to the estimation of $\beta_{0}$ when
	$W$ and $V$ are correlated (which is often the case in applications). This
	sensitivity is particularly problematic for inference on $\psi_{0}$ when $p$
	is large, which will be the case in many applications. 
	
	We now derive novel LR moments for  
	$\psi(\lambda_{0}%
	)=\mathbb{E}\left[  C_{2}^{\prime}\alpha\right],$ for a known $q\times q_{2}$ matrix $C_{2}$ of full column rank, when possibly $p>>nT$. Define the generalized between operator $H:=V^{\dagger}$, and also define the $q\times p$ matrix $S_1:=\mathbb{E}\left[HW\right]$, and the $p_{2}\times p$ matrix $\Gamma_{0}:=-C_{2}^{\prime}S_1M^{\dagger}$.
	
	\begin{assumption}
		\label{RankHDpanelalpha}
		i) $\mathbb{E}\left[  \|HY\|^2\right]<\infty$ and $\mathbb{E}\left[  \|HW\|^2\right]<\infty$; ii)  $\left|\left|\beta\right|\right| < C$, $\left|\left|\Gamma\right|\right| < C$, for $\left|\left|\beta - \beta_0 \right|\right|$  and $\left|\left|\Gamma - \Gamma_0 \right|\right|$ small enough; and iii) the column space of $M$ contains that of $S_1^{\prime}C_{2}$ and $rank(C_{2})=q_2$. Moreover, iv) $C^{\prime}_2HV = C^{\prime}_2$.
	\end{assumption}

	This assumption implies the order conditions $rank(M)\geq q_{2}$ and $rank(V) > q_2$. Assumption \ref{RankHDpanelalpha} iii) requires that the variation in $W$ not colinear with $V$ (variation in  $QW$) is rich enough to identify the component $C'_2S_1\beta_0$ (as when $C'_1=C'_2S_1$ in the previous section. This assumption is needed for Steps 2 and 3 of the three-step algorithm.  Assumption \ref{RankHDpanelalpha} iv) is equivalent to the column space of $C_2$ beging contained in that of $V'$ (which is weaker than $rank(V)=q$ a.s.). This is needed for the identification of $\psi(\lambda_0)=\mathbb{E}\left[  C_{2}^{\prime}\alpha\right]$. Note that identification of $\beta_{0}$ or $\mathbb{E}\left[ \alpha\right]$ is not required for $\mathbb{E}\left[  C_{2}^{\prime}\alpha\right]$ to be identified. 
	
	\begin{proposition}
		\label{propLRc1alpha}
		Under Assumption \ref{RankHDpanelbeta} i) and Assumption \ref{RankHDpanelalpha} ii) and iv), $\psi_{0}=\mathbb{E}\left[  C_{2}^{\prime}\alpha\right]$ is identified as 
		\begin{equation}
			\psi_{0}=C_{2}^{\prime}\mathbb{E}\left[  H(Y-W\beta_{0})\right].
		\end{equation} 
		Additionally, if Assumption \ref{RankHDpanelalpha} holds, then the following function is an LR moment function for $\psi_{0}$
		\begin{equation}
			\label{galphalr}
			g(Z,\rho_0,\psi_{0})=(C_{2}^{\prime}H-\Gamma_{0}W'Q)(Y-W\beta_{0})-\psi_{0},   
		\end{equation}
		where $\rho_{01} = \beta_0$, $\rho_{02}=\Gamma_{0}=-C_{2}^{\prime}S_1M^{\dagger}$.
	\end{proposition}
	
	We describe an algorithm for the derivation of an LR moment for $\psi(\lambda_{0})=\mathbb{E}\left[  C_{2}^{\prime}\alpha\right]$ as well as for LR inferences based on it. Define $H_i:=V_i^{\dagger}$.
	
	\bigskip \noindent \textbf{Algorithm for obtaining an LR moment and inferences for $\psi(\lambda_{0}%
		)=\mathbb{E}\left[  C_{2}^{\prime}\alpha\right]$}\textbf{:}
	
	\begin{description}
		\item[Step 1:] Compute $Q_i$ and $H_i$ for $i=1,...,n.$ 
		
		\item[Step 2:] Compute for $\ell=1,...,L,$
		\[
		\check{M}_{\ell}=\frac{1}{n-n_{\ell}}%
		\sum_{l\neq\ell}^{L}\sum_{i\in I_{l}}W'_iQ_iW_i, \hspace{1cm}
		\hat{S}_{1\ell}=\frac{1}{n-n_{\ell}}%
		\sum_{l\neq\ell}^{L}\sum_{i\in I_{l}}H_iW_i,
		\]
		where $n_{\ell}$ is the number of elements of $I_{\ell}$, and
		\[
		\hat{\Gamma}_{\ell}=-C_{2}^{\prime}\hat{S}_{1\ell}\hat{M}^{\dagger}_{\ell},
		\] 
		for a suitable estimator $\hat{M}^{\dagger}_{\ell}$ (e.g., truncated singular value or nodewise lasso).
		
		\item[Step 3:] Estimate $\beta_{0}$ as in Step 3 of the previous algorithm.
		
		\item[Step 4:] An LR moment function for $\psi(\lambda_{0}%
		)=\mathbb{E}\left[  C_{2}^{\prime}\alpha\right]$ can be estimated by
		\begin{equation}
			\hat{g}_{i\ell}(\psi_{0})=(C_{2}^{\prime}H_i-\hat{\Gamma}_{\ell}W'_iQ_i)(Y_i-W_i\hat{\beta}_{\ell})-\psi_{0}. \label{LRHDpanelalpha}%
		\end{equation}
		
	\end{description}
	
	\noindent Inference is then implemented following Section 5. This moment for $\psi(\lambda_{0})=\mathbb{E}\left[  C_{2}^{\prime}\alpha\right]$ is LR with respect to all nuisance parameters, including $\Gamma_{0}$ and $\beta_{0}$. We recommend implementing the following cross-fitted variance estimator
	\[
	\hat{W}_{n}=\frac{1}{n}\sum_{\ell=1}^{L}\sum_{i\in I_{\ell}}
	({g}_{i\ell}(\hat{\psi}_{\ell})-\bar{g}_{\ell}(\hat{\psi}_{\ell}))^2,
	\]
	where  $\hat{\psi}_{\ell}=\hat{R}_{\alpha,\ell}$ for $\psi(\lambda_{0}%
	)=\mathbb{E}\left[  C_{2}^{\prime}\alpha\right]$, with   
	\[
	\hat{R}_{\alpha,\ell}=\frac{1}{n-n_{\ell}}%
	\sum_{l\neq\ell}^{L}\sum_{i\in I_{l}}
	(C_{2}^{\prime}H_i-\hat{\Gamma}_{\ell}W'_iQ_i)(Y_i-W_i\hat{\beta}_{\ell}),
	\]
	\[
	\bar{g}_{\ell}(\hat{\psi}_{\ell})=\frac{1}{n}\sum_{\ell=1}^{L}\sum_{i\in I_{\ell}}\hat
	{g}_{i\ell}(\hat{\psi}_{\ell}).
	\]

	\begin{proposition}
		\label{validinferpanelalpha}
		Suppose that Assumptions \ref{RankHDpanelbeta} i) and \ref{RankHDpanelalpha} i) and iii) above, and Assumption \ref{regcondpanelalpha} in Section \ref{secasymptotichighpaneldata} of the Supplementary Appendix hold. A test based on $\hat{C}_n(\psi_0)$ in (\ref{teststat}) for $\psi_0 = \mathbb{E}\left[C_2^{\prime} \alpha\right]$, using the estimator of a LR moment function \eqref{LRHDpanelalpha} and the variance-covariance matrix estimator 
		$$
		\check{W}_n = \frac{1}{n}\sum^L_{\ell=1}\sum_{i \in I_\ell}\left(\hat{g}_{i\ell}\left(\hat{\psi}_\ell\right) -\bar{g}_\ell\left(\hat{\psi_\ell}\right)\right)\left(\hat{g}_{i\ell}\left(\hat{\psi}_\ell\right) -\bar{g}_\ell\left(\hat{\psi_\ell}\right)\right)^{\prime},
		$$
		has size equal to $\zeta \in (0,1)$ asymptotically. The probability that $\psi_0 \in CR_{\zeta}$ converges to $(1-\zeta)$ as $n \rightarrow \infty$. Finally, the test presents a non-trivial local asymptotic power function.
	\end{proposition}
	An LR estimator for $\psi_0 = \mathbb{E}\left[C_2^{\prime} \alpha\right]$ can be  constructed as $\hat{\psi}=\frac{1}{L}\sum^L_{\ell=1}\hat{R}_{\alpha,\ell}$. This estimator is asymptotically normal, as it follows from the proof of Proposition \ref{validinferpanelalpha}.

	\subsubsection{Second Moments}
	\label{secondmoments}
	We aim to construct LR moments for quadratic functionals of the UH $\psi_0 = \mathbb{E}\left[\alpha^{\prime}\Omega \alpha\right]$, where $\Omega$ is a $q \times q$ known matrix. We consider the setting of \cite{arellano2012identifying}. These authors derive a system GMM for common parameters, variances of idiosyncratic errors, and variances of UH by employing the following additional generalized within-group and between operators, respectively,
	\begin{align*}
		\mathbb{Q} &  =I_{T^2}-[I_{T}-Q]\otimes[I_{T}-Q],\\
		\mathbb{H} &  =H\otimes H,
	\end{align*}
	where $\otimes$ denotes the Kronecker product. \cite{arellano2012identifying} consider the following covariance restrictions on the conditional variance $\Sigma_i:=\mathbb{V}\left[  \left.  \varepsilon_i\right\vert X_i\right]$, 
	\begin{equation}
		\label{Var}
		vec(\Sigma_i)=S_2\omega_{0i},
	\end{equation}
	where $vec$ is the vectorization operation, $S_2$ is a known selection matrix and $\omega_{0i}$ is an $m$ dimensional vector of parameters, possibly depending on $X_i$. 
	
	Define the parameter $\zeta_0:=(\beta_{0},\omega_0)$ and the generalized errors
	\begin{equation*}
		U\equiv U(X,\zeta_0):= (Y-W\beta_{0})\otimes(Y-W\beta_{0})-S_2\omega_0.
	\end{equation*}
	Also, to simplify the notation, denote $u(\beta)=Y-W\beta$. We apply the general algorithm for constructing an LR for $\psi_0$ as follows. First, we show that the moment function
    \begin{equation}
        \label{g0var}
	g_0(Z,\beta_0,\omega_0)=vec(\Omega)'\mathbb{H}\{(Y-W\beta_{0})\otimes(Y-W\beta_{0})-S_2\omega_0\}
    \end{equation}
    satisfies the following conditional moment restriction
	\[
	\mathbb{E}\left[  \left.  g_0(Z,\beta_0,\omega_0)\right\vert \alpha,X\right]
	=\alpha^{\prime}\Omega \alpha,\text{ a.s.}%
	\]
	
	Thus, taking expectations both sides, we conclude $\mathbb{E}\left[  g_0(Z,\beta_0,\omega_0)-\psi_{0}\right]  =0.$ However, this moment is not LR for $\psi_{0}$ because it is in general quite sensitive to the estimation of $\beta_{0}$ and the covariance parameters $\omega_0$. To orthogonalize this moment, we require the following assumptions. Recall the definition $M=\mathbb{E}\left[  W^{\prime}QW\right]$, let $\dot{\omega}_0$ be the Jacobian of $\omega_0$ and define the matrices $A:=vec(\Omega)'\mathbb{E}\left[ \mathbb{H}S_2\dot{\omega}_0\right]$, $B:=\mathbb{E}\left[ \mathbb{Q}S_2\right]$. Define also $\Gamma_{0\omega}:=AB^{\dagger}$, and $L:=-\mathbb{E}\left[  (vec(\Omega)'\mathbb{H}-\Gamma_{0\omega}\mathbb{Q})\{(W\otimes u(\beta_0))+(u(\beta_0)\otimes W)\}\right]$. 
	
	\begin{assumption}
		\label{RankHDpanelvar}
		(i) $\mathbb{E}\left[ \|\mathbb{Q}U\|^2\right]<\infty$ and $\mathbb{E}\left[  \|\mathbb{H}U\|^2\right]<\infty$; (ii) the column space of $M$ contains that of $L$; (iii) the column space of $B$ contains that of $A$. Moreover, $HV = I_q$.
	\end{assumption}
	
	\begin{proposition}
		\label{propLRvar}
		Under Assumption \ref{RankHDpanelbeta} i) and Assumption \ref{RankHDpanelvar} i) above, $\psi_0 = \mathbb{E}\left[\alpha^{\prime}\Omega \alpha\right]$ is identified as 
		\begin{equation}
			\psi_{0}=\mathbb{E}\left[  g_0(Z,\beta_0,\omega_0)\right],
		\end{equation} 
		where $g_0$ is defined in (\ref{g0var}). Additionally, if Assumption \ref{RankHDpanelvar} holds, then the following function is an LR moment function for $\psi_{0}$
		\begin{equation}
			g(Z,\rho_0,\psi_{0})=(vec(\Omega)'\mathbb{H}-\Gamma_{0\omega}\mathbb{Q})U-\Gamma_{0\beta}W'Qu(\beta_{0})-\psi_{0},   
		\end{equation}
		where  $\Gamma_{0\omega}=AB^{\dagger}$, $\Gamma_{0\beta}=LM^{\dagger}$, and $\Gamma_{0}=(\Gamma_{0\omega},\Gamma_{0\beta})$, given some suitable partition in $\rho_0$.
	\end{proposition}
	
	To implement this inference procedure, construct estimators: 
	$$
	\hat{A}_\ell = \frac{1}{n - n_\ell} \sum^L_{l \neq \ell} \sum_{i \in I_\ell} vec\left(\Omega\right)^{\prime}\mathbb{H}_iS_2\dot{\omega}_i,\;\;\;\; \hat{B}_\ell = \frac{1}{n - n_\ell} \sum^L_{l \neq \ell} \sum_{i \in I_\ell} S^{\prime}_2\mathbb{Q}_i,\;\;\; \hat{\Gamma}_{\omega \ell} = \hat{B}^{\dagger}_\ell \hat{A}_\ell,
	$$
	where  $\hat{B}^{\dagger}_\ell$ is a suitable estimator of the Moore-Penrose inverse of $B$ (e.g., spectral hard thresholding, as in Section \ref{moorepenroseest}).\footnote{In our empirical emplication below, we set the threshold to be $(\log(T^2)/n)^{1/2}$, which might be a conservative choice. Undertaking an optimal choice for the estimator of $B^{\dagger}$ is beyond the scope of this paper; see \cite{wainwright2019}.} Moreover, let\footnote{Possibly, results can be improved by using \textit{double} cross-fitted estimators $\hat{\Gamma}_{\omega \ell, \ell^{\prime}}$.} 
	$$
	\hat{L}_\ell = - \frac{1}{n - n_\ell} \sum^L_{l \neq \ell} \sum_{i \in I_\ell}  (vec(\Omega)'\mathbb{H}_i-\hat{\Gamma}_{\omega \ell}\mathbb{Q}_i)\{(W_i\otimes u(\hat{\beta}_\ell))+(u(\hat{\beta}_\ell)\otimes W_i)\},\;\;\; \hat{\Gamma}_{\beta \ell} = \hat{L}_\ell\hat{M}^{\dagger}_\ell.
	$$
	In addition, let 
	\begin{equation*}
		\hat{U}_{i\ell} =  (Y_i-W_i\beta_{\ell})\otimes(Y_i-W_i\hat{\beta}_{\ell})-S_2\hat{\omega}_\ell,
	\end{equation*}
	where $\hat{\omega}_\ell$ is an estimator of $\omega_0$, e.g., as in \cite{arellano2012identifying}, using observations not in $I_\ell$. A feasible LR moment is then
	\begin{equation}
		\label{gfvar}
		\hat{g}_{i\ell}(\psi_{0}) = (vec(\Omega)'\mathbb{H}_i-\hat{\Gamma}_{\omega \ell}\mathbb{Q}_i)\hat{U}_{i\ell}-\hat{\Gamma}_{\beta \ell}W_i'Q_iu(\hat{\beta}_\ell)-\psi_{0}
	\end{equation}
	An algorithm to implement LR inferences for $\psi_0 = \mathbb{E}\left[\alpha^{\prime}\Omega \alpha\right]$ follows in an analogous way as for average marginal effects. It is also possible to show the validity of an inference procedure based on \eqref{gfvar}. We omit the details for the sake of space.

	\subsection{Kotlarski model with factor loading}
	
	\label{Kotlarski}
	We provide in this section a recursive construction of LR moments for $\psi
	_{0k}:=\mathbb{E}\left[  \alpha^{k}\right]  ,$ $k\in\mathbb{N}$, in the Kotlarski model with a factor loading given by
	\begin{align*}
		Y_{1} &  =\alpha+\varepsilon_{1},\\
		Y_{2} &  =\beta_{0}\alpha+\varepsilon_{2},
	\end{align*}
	where $\alpha$ is a common unobserved factor, and $\varepsilon = (\varepsilon_1, \varepsilon_2)$ has independent nonobservable components with zero mean and densities $f_{\varepsilon_1}$ and  $f_{\varepsilon_2}$, respectively. The variables $\alpha$ and $\varepsilon$ are independent.  Define for $k\in\mathbb{N},$
	$\overrightarrow{\psi}_{0k}=(\psi_{01},...,\psi_{0k})^{\prime}\ $and the
	moments
	\[
	a_{0k}(Z,\beta_{0},\overrightarrow{\psi}_{0k})=\beta_{0}\psi_{0k}-Y_{2} %
	Y_{1}^{k-1}+\sum_{j=1}^{k-1}\left( \binom{k-1}{j-1} \beta_{0}Y_{1}^{k-j}-\binom{k-1}{j}Y_{2}Y_{1}^{k-j-1}\right)  \psi_{0j}.%
	\]
	The following lemma provides orthogonal moments for the multivariate functional $\overrightarrow{\psi
	}_{0k}$ if $\beta_{0}$ is known. Define 
	\[
	c_{k,j}:=\left\{
	\begin{array}
		[c]{cc}%
		(k-1)!\mathbb{E}\left[  \frac{Y_{1}^{k-j}\beta_{0}}{(k-j)!(j-1)!}-\frac
		{Y_{2}Y_{1}^{k-j-1}}{(k-j-1)!j!}\right]  & k>j\\
		0 & k<j\\
		\beta_{0} & k=j
	\end{array}
	\right.  .
	\]
	
	\begin{lemma}
		\label{lemmaKotlarski}
		In the Kotlarski model above, if the
		moments involved are finite, then
		\[
		\mathbb{E}\left[  a_{0k}(Z,\beta_{0},\overrightarrow{\psi}_{0k})\vert\alpha\right]
		=\sum_{j=1}^{k}c_{k,j}(\psi_{0j}-\alpha^j),\qquad k\in\mathbb{N}.
		\]
	\end{lemma}
	
	The moment functions $a_{0k}$ are not LR when $\beta_0$ is unknown and the parameter of interest is just the $k-$th moment. We provide LR
	moments in a recursive way as follows. Define, for $k\in\mathbb{N},$
	\begin{equation*}
		g_{0k}(Z,\beta_{0},\overrightarrow{\psi}_{0k},\Gamma_{0}) :=a_{0k}(Z,\beta_{0},\overrightarrow{\psi}_{0k}) +\sum_{j=1}^{k-1}\gamma_{k,j}a_{0j}(Z,\beta_{0},\overrightarrow{\psi}%
		_{0j}),
	\end{equation*}
	and
	\begin{equation*}
		g_{k}(Z,\beta_{0},\overrightarrow{\psi}_{0k},\Gamma_{0}) :=g_{0k}(Z,\beta_{0},\overrightarrow{\psi}_{0k})-\gamma_{k,0}(Y_{2}-\beta_{0}Y_{1})-\beta_{0}\psi_{0k},
	\end{equation*}
	
	where the sum from 1 to zero is zero, 
	\[
	b_{k}=\psi_{0k}+(k-1)!\sum_{j=1}^{k-1}\frac{\mathbb{E}\left[  Y_{1}^{k-j}\right]
	}{j!(k-1-j)!}\psi_{0j}.
	\]
	and the constants $\gamma_{k,j}$ are defined recursively as, for
	$j=k-1,k-2,...,1$%
	\[
	\gamma_{k,j}=-\frac{c_{k,j}+\sum_{h=j+1}^{k-1}\gamma_{k,h}c_{h,j}}{\beta_{0}},%
	\]
	and
	\[
	\gamma_{k,0}=\frac{b_{k}+\sum_{h=1}^{k-1}\gamma_{k,h}b_{h}}{\mathbb{E}\left[
		Y_{1}\right]  }.%
	\]
	For example,
	\begin{align*}
		\gamma_{k,k-1}  &  =-\frac{c_{k,k-1}}{\beta_{0}},\\
		\gamma_{k,k-2}  &  =-\frac{c_{k,k-2}+\gamma_{k,k-1}c_{k-1,k-2}}{\beta_{0}},\\
		\gamma_{k,k-3}  &  =-\frac{c_{k,k-3}+\gamma_{k,k-2}c_{k-2,k-3}+\gamma
			_{k,k-1}c_{k-1,k-3}}{\beta_{0}}.%
	\end{align*}
	Note that 
	\begin{align*}
		b_{k}  &  =\frac{\partial\mathbb{E}\left[  a_{0k}(Z,\beta_{0}%
			,\overrightarrow{\psi}_{0k})\right]  }{\partial\beta}\\
		c_{k,h}  &  =\frac{\partial\mathbb{E}\left[  a_{0k}(Z,\beta_{0}%
			,\overrightarrow{\psi}_{0k})\right]  }{\partial\psi_{h}}.
	\end{align*}
	Then, we have the following proposition. Define $\rho_{0k}:=(\beta_{0},\overrightarrow{\psi}_{0k-1},\Gamma_0)$ and write $g_{k}(Z,\beta_{0},\overrightarrow{\psi}_{0k},\Gamma_0)=g_{k}(Z,\rho_{0k},\psi_{0k})$. 
	
	\begin{proposition}
		\label{propositionkotlarski}
		In the Kotlarski model above, if the moments involved are finite,
		then $g_{k}(Z,\rho_{0k},\psi_{0k})$ is a LR\ moment for
		$\psi_{0k}=\mathbb{E}\left[  \alpha^{k}\right]$, as it satisfies
		\[
		\mathbb{E}\left[  g_{k}(Z,\rho_{0k},\psi_{0k})\vert\alpha\right]
		=-\beta_{0}\alpha^k,\qquad k\in\mathbb{N}.
		\]
		and
		\[
		\frac{\partial\mathbb{E}\left[  g_{k}(Z,\rho_{0k},\psi_{0k})\right]}{\partial\rho_k}
		=0,\qquad k\in\mathbb{N}.
		\]
	\end{proposition}
	
	The moment function $g_{k}$ satisfies the conditions of Theorem \ref{LRmomentUHresult} with $c=-\beta_{0}\neq0$, and hence, it is a relevant orthogonal moment. This construction avoids the nonparametric estimation of the distribution of latent variables, which are known to have very slow (logarithmic) rates of convergence, see, e.g., \cite{horowitz1996semiparametric}.
	
	\subsection{Teacher Value-Added}
	\label{TAV}
	
	In this application we follow \cite{gilraine2020new} and consider the model for estimated teacher fixed effects, given by $Y=\alpha+\varepsilon$, $\varepsilon=\theta_{0}u$, $u\sim N(0,1),$ $\alpha$
	independent of $u,$ with $\alpha\sim\eta_{0}$ for an unknown Lebesgue density $\eta
	_{0}$. This model with auxiliary moment conditions for $\theta_{0}$ is the model used in \cite{gilraine2020new}. In its most general version without auxiliary moments for $\theta_{0}$, this model has been extensively studied in the literature, see, e.g., \cite{carroll1988optimal}, \cite{fan1991optimal}, and \cite{ishwaran1999information}. In particular, it is well known that, without auxiliary moments and further assumptions, the model is not identified, see \cite{kiefer1956consistency}. In any case, we are not aware of any investigation on LR moments in this model, which is the goal of our paper. 
	
	To that end, a key condition for the existence of a relevant orthogonal moment for the functional $\psi _{0}=\mathbb{E}\left[r(\alpha ,\theta _{0})\right] $ is the existence of a $g_{0}\in
	L_{2},$ such that, a.s.
	
	\begin{equation}
		\mathbb{E}\left[ \left. g_{0}(Y,\theta _{0})\right\vert \alpha \right]
		=r(\alpha ,\theta _{0}),  \label{PM}
	\end{equation}%
	where $\theta _{0}$ is the unknown error's standard deviation. In its integral form, this equation is%
	\begin{equation*}
		\int_{}g_{0}(y,\theta _{0})\phi _{\theta _{0}}(y-\alpha)dy=r(\alpha ,\theta _{0}), \text{a.s.,}
	\end{equation*}%
	where, henceforth, we use the short notation $\phi _{\sigma }(x)=\phi \left(
	x/\sigma \right) /\sigma$ and recall $\phi\left(  \cdot\right)$ denotes the standard normal density. We consider two types of functionals in this application: (i) analytic functionals, as defined below; and (ii) CDFs, quantiles and policy functionals.
	\subsubsection{Analytic Functionals}
	Analytic functionals correspond to moments $r(\alpha ,\theta _{0})$ that are an entire function of $\alpha$, that is, a real analytic function in $\alpha$ with a Maclaurin series expansion $r(\alpha )=\sum_{j=0}^{\infty }r_{j}\alpha^{j}$, where we drop the dependence on $\theta _{0}$ in the coefficients $r_j$ for simplicity of notation. This class includes moments and other smooth functionals of teacher value-added. To implement our approach, we first solve the equation for moments of $\alpha$. To that end, let us introduce the Hermite polynomials $\{H_{j}(\cdot)\}_{j=0}^{\infty }.$\footnote{%
		The $j$-th Hermite polynomial is
		defined by $H_{j}(x)=(-1)^{j}\exp
		(-x^{2}/2)\partial ^{j}\exp (-x^{2}/2)/\partial x^{j}.$} Then, the convolution identity for Hermite polynomials, see, e.g., \cite{davis2024general}, yields 
	\begin{equation*}
		\mathbb{E}\left[ \left. \theta_0^{k}H_{k}(Y/\theta _{0})\right\vert \alpha \right]
		=\alpha^{k},
	\end{equation*}%
	and thus, $a_{0k}(Y,\theta _{0})=\theta_0^{k}H_{k}(Y/\theta _{0})$ is a solution of Step 1 in the direct approach for the $k-$th moment. Exploiting the Maclauring series, we conjecture that
	\begin{equation*} 
		g_0(Y,\theta_0 )=\sum_{j=0}^{\infty }\theta_0^{j}r_{j}H_{j}(Y/\theta _{0})
	\end{equation*}
	is a solution to (\ref{PM}) in the general analytic case, provided this solution is in $L_2$. Our next result gives primitive conditions for this and gives an LR moment for the functional. Recall, that after solving for $g_0$, the LR moment will be given by 
	\[
	g(Y,\theta_{0},\psi_{0})=g_{0}(Y,\theta_{0})-\psi_{0}-\Gamma_{0}g_1(Y,\theta_{0}),
	\]
	where $\Gamma_{0}$ solves the equation
	\[
	\mathbb{E}\left[  \frac{\partial g_{0}(Y,\theta_{0},\psi_{0})}{\partial\theta%
	}\right]=\Gamma_{0}\mathbb{E}\left[  \frac{\partial g_{1}(Y,\theta_{0}%
		)}{\partial\theta}\right],
	\]
	and $g_{1}$ is the auxiliary estimating equation for $\theta_0$ with finite variance and $\mathbb{E}\left[  \frac{\partial g_{1}(Y,\theta_{0}%
		)}{\partial\theta}\right]\neq0$.\footnote{Note that we can let $\rho_{01} = \theta_0$ and $\rho_{02} = \Gamma_0$.} We also assume that the derivative $\mathbb{E}\left[  \frac{\partial g_{0}(Y,\theta_{0},\psi_{0})}{\partial\theta%
	}\right]$ is well defined (this derivative can be obtained from well-known expressions of derivatives of Hermite polynomials and the derivatives of the coefficients of $r$). The following assumption is also required 
	\begin{equation*}
		\label{keysum}
		\sum_{j=0}^{\infty }\frac{j!\theta_0^{2j}r_{j}^{2}}{R ^{j}}<\infty,
	\end{equation*}%
	for some $R$ such that $0<R<1$.
	
	\begin{proposition}
		\label{propositionposteriormeans}
		In the teacher value-added model, under the stated assumptions, the above moment $g$ is an LR moment with finite variance for the analytic functional, provided $\mathbb{E}\left[ \exp\left( cY^2\right) \right] <\infty $ for $c=\frac{2R}{\theta^2_0(1+R)}$. 
	\end{proposition}
	
	\subsubsection{CDFs, quantiles and policy functionals}
	
	The literature of teacher value-added has long been interested in policy parameters such as the marginal gain in test scores resulting from replacing the bottom $\varphi$\% (typically, 5\%) of teachers. More generally, CDFs and quantiles are common parameters employed in the literature of models with nonparametric UH. In this section, we show that, as an important application of the necessity of our conditions, relevant orthogonal moments do not exist for these important parameters in teacher value-added models. This raises the question of the suitability of these policy parameters if the goal is to derive robust policy implications. 
	
	The following proposition shows that if $r$ satisfies equation (\ref{PM}), then it has to be an analytic function of $\alpha$. An implication of this is that CDFs, quantiles, and policy parameters with non-analytic Riesz representers with respect to $\eta$ cannot satisfy equation (\ref{PM}), and thus do not have a relevant orthogonal moment. 
	\begin{proposition}
		\label{propAnal}
		If $g_0\in L_2$ and $r$ satisfies \eqref{PM}, then $r$ is an analytic function.     
	\end{proposition}
	This proposition also motivates the approach of the previous section.
	
	Let us consider some examples, and their corresponding Riesz representer $r$. This Riesz representer is defined in general as 
	\begin{equation*}
		\frac{d\psi(\eta_\tau)}{d\tau}=\mathbb{E}[r(\alpha)b(\alpha)].
	\end{equation*}
	
	For the CDF at $\bar{\alpha}$, the corresponding Riesz representer is $r(\alpha)=\bm{1}(\alpha \leq \bar{\alpha})$. For $\varphi$-quantile it is well-known that $r(\alpha)=\bm{1}(\alpha \leq F^{-1}_\alpha(\varphi))/\eta_{0}(F^{-1}_\alpha(\varphi))$. For the policy parameter $\psi(\lambda_0) = -\mathbb{E}\left[\alpha \bm{1}\left\{ \alpha < F^{-1}_\alpha(\varphi)\right\}\right]$, by the chain rule and the fundamental theorem of calculus, $r(\alpha)$ $=(F^{-1}_\alpha(\varphi)-\alpha)\bm{1}(\alpha \leq F^{-1}_\alpha(\varphi))$. In all these cases, the corresponding $r$ is not analytic. Therefore, by our results, there does not exist an LR moment for these functionals.

	\section{Monte Carlo}
	\label{MC}
	The aim of this section is to illustrate that constructing relevant orthogonal moments is important. In particular, we present a numerical exercise that shows that those moments might lead to substantial gains in terms of efficiency, although it might involve estimating UH. 
	
	Consider a simpler version of the Kotlarski model of Section \ref{Kotlarski}. Specifically,
	\begin{equation}
		\label{eq:chaberlain_example}%
		\begin{split}
			Y_{1}  &  = \alpha + \varepsilon_1,\\
			Y_{2}  &  = \beta_{0}\alpha + \varepsilon_2,
		\end{split}
	\end{equation}
	where $\alpha$ is independent of $\varepsilon= (\varepsilon_1, \varepsilon_2)$, which has independent components, and $\beta_0 $ is unknown (with a true value of one).  In this section, for a better comparison with existing procedures, we consider the case where the marginal densities $f_{\varepsilon_1}$ and $f_{\varepsilon_2}$ are known. Particularly, their distributions are standard normals. However, $\eta_0 = f_{\alpha}$ is unknown.  The UH $\alpha$ is a discrete approximation to a standard normal with unknown support points; see below for the specification of $\eta_0$. We are interested in conducting LR inference on $\psi(\lambda_0) = \beta_0$, which is a special functional, e.g., it does not depend on the UH. This setting, with continuous $\alpha$,  has been used by \cite{bonhomme2012functional} (cf. Section 6) to evaluate the numerical performance of functional differencing, which we will exploit to have a benchmark for comparison. Functional differencing constructs moments orthogonal to UH (under a support condition), but does not necessarily deliver relevant moments.\footnote{This is a point acknowledged by Bonhomme; see Section 4.3 in \cite{bonhomme2012functional}.} Our construction algorithms allow us to purposely compute informative moments. Indeed, in this exercise, we can construct the most efficient of them all, as we show below.
	
	Notice that in this case, 
	$$
	f_{\beta_0,\eta_0}(y_1,y_2) = \int f_{z|\alpha}(z|\alpha;\beta_0)\eta_0 (\alpha)d\nu(\alpha),
	$$
	with
	$$
	f_{z|\alpha}(z|\alpha;\beta_0) = \phi\left(y_1 - \alpha\right)\times \phi\left(y_2 -\beta_0\alpha\right),
	$$
	where $\phi$ is the standard normal pdf and $\nu$ is the counting measure. 
	
	Since our target functional is $\psi(\lambda_{0}) = \beta_0 :=\theta_0$ and finite-dimensional, we can apply Remark \ref{remarkthetapsi}.  We aim to find a function $g$ such that\footnote{Observe that our second equation, $\mathbb{E}\left[s_\theta(Z,\lambda_0)g\left(Z,\lambda_0\right)\right] = c$, with $c \in \mathbb{R}$, holds in this case, provided that $s_\theta$ is in $L_2$.} 
	\begin{equation}
	\label{eqgMC}
		\mathbb{E}\left[\left. g\left(Z,\lambda_0\right) \right| \alpha\right] = 0.
	\end{equation}
    Notice that we are writing $g$ as a function of $\lambda_0$, instead of $\theta_0$ only. Our algorithm, explained in Section \ref{Construction}, suggests that is not necessary. The reason why we want $g$ to depend on $\lambda_0$ will be come clear below. To solve for $g$ in \eqref{eqgMC}, we implement a projection approach, where $g=\Pi_{\mathcal{N}(S_{\eta_{0}}^{*})}m$,
	$$
	S_{\eta}^{*}g = \mathbb{E}\left[  \left.  g(Z, \lambda_0)\right\vert \alpha\right],
	$$
	and $m$ is a given moment function, $m \in L^0_2$. A convenient candidate is 
	$$
	m\left(Y_1,Y_2,\lambda_0\right) = \frac{\partial}{\partial \beta} \log f_{\beta_0,\eta_0}(Y_1,Y_2),
	$$
	i.e., the score of the model with respect to $\beta$. This is an appealing choice since the resulting relevant LR $g$ will correspond to the efficient (semiparametric) score of $\beta_0$, which could derive in efficiency gains when inference is based on our test statistic; see \cite{vandervaart98}, Section 25.4, and \cite{lee2024locallyregularefficienttests}.  In Section \ref{implementationMC} of the Supplementary Appendix, we provide details regarding how to approximate $g=\Pi_{\mathcal{N}(S_{\eta}^{*})}m$ in practice. 
	
	In our simulations, the distribution of $\alpha$ is specified as follows
	\begin{equation}
		\eta_{0}\left(  \alpha= \alpha_{j}\right)  =%
		\begin{cases}
			\Phi\left(  \frac{\alpha_{j} + \alpha_{j+1}}{2}\right)  & \text{for}\;\;
			\alpha_{j} = -3.0,\\
			\Phi\left(  \frac{\alpha_{j} + \alpha_{j+1}}{2}\right)  - \Phi\left(
			\frac{\alpha_{j} + \alpha_{j-1}}{2}\right)  & \text{for}\;\; \alpha_{j} =
			-2.9,-2.8,\dots,4.9,\\
			1- \Phi\left(  \frac{\alpha_{j} + \alpha_{j-1}}{2}\right)  & \text{for}\;\;
			\alpha_{j} = 5.0,
		\end{cases}
		\label{eq:dist_het}%
	\end{equation}
	where $\Phi(.)$ is the standard normal distribution. Notice that \eqref{eq:dist_het} indicates that the true density of individual effects is discrete but it closely behaves as a standard normal; Section \ref{implementationMC} of the Supplementary Appendix discusses regularized estimation of \eqref{eq:dist_het}. All results are based on $1,000$ Monte Carlo repetitions.
	
	We are interested in comparing the performance of our test statistic and the naive plug-in one, which is based on $m(Z)$ without correcting for the first stage bias. To have a precise notion of the impact of the bias for inference, both tests are based on the same estimator of the asymptotic variance $\hat{W}$.  Additionally, we consider a Wald-test statistic using the functional differencing approach of \cite{bonhomme2012functional}. Specifically, we start with the function $\tilde{h}(y) = \varphi_2\left(y - \varrho\right)$ where $\varphi_2$ is the standard bivariate normal p.d.f. and $\varrho$ belongs to one of three increasing sets containing 9, 25, and 49 points, respectively.\footnote{These sets are $\left\{(0,0), (0,1), (0,-1),\cdots,(-1,-1)\right\}$, $\left\{(0,0), (0,1), (0,-1), (0,2), (0,-2)\cdots,(-2,-2)\right\}$, $\left\{(0,0), (0,1), (0,-1),(0,2), (0,-2), (0,3), (0,-3)\cdots,(-3,-3)\right\}$. } We can then obtain 9, 25, and 49 different functions $g_{FD}(y,\beta_0)$ by functional differencing. In fact, these are the same functions that \cite{bonhomme2012functional} considers in his simulations.\footnote{\cite{bonhomme2012functional}'s code is publicly available, which we have closely followed to reproduce the results here.} Then, we estimate $\beta_0$ using GMM based on $g_{FD}(y,\beta_0)$, and compute Wald-type statistics $W_{9}$, $W_{25}$ and $W_{49}$, where $W_j = (\hat{\beta}-\beta_0)^2/\widehat{AsyVar}(\hat{\beta})$, $j=9,25,49$, respectively.

	Table \ref{table:MCsize} compares rejection rates under the null hypothesis $H_0: \beta_0 =1$ of all the test statistics considered and different sample sizes. The most important point from this table is the importance of controlling for regularizarion/first stage bias to conduct inference on parameters of interest. In particular, we observe that failing to construct LR moments results in test statistics with rejection rates sufficiently above the nominal values to be a serious concern in applied settings. In contrast, our construction of debiased moments yields a test statistic that presents controlled sizes figures, even though we based inference in regularized estimators $\hat{\eta}_\ell$, as our theory suggests.  Functional differencing presents good performance in terms of size, uniformly in the number of moments constructed by such an approach. This is unsurprising since the Wald statistics are not affected by $\hat{\eta}_\ell$. 
	
	\begin{table}[H]
		\centering 
		\caption{Monte Carlo Results - Size}
		\label{table:MCsize}
		\begin{threeparttable}
			\begin{tabular}{clllllllllll}
				\hline \hline
				$n$ & Plug-in  & Plug-in  &  LR & LR  & $W_9$  & $W_9$  & $W_{25}$ & $W_{25}$ & $W_{49}$ & $W_{49}$ \\ 
				& 0.05 & 10  &  0.05 & 0.10  & 0.05  & 0.10  & 0.05 & 0.10 & 0.05 & 0.10 \\  
				\hline
				250 & 0.22 & 0.30 & 0.05 & 0.09 & 0.06 & 0.12 & 0.06 & 0.13 & 0.07 & 0.12 \\ 
				500 & 0.19 & 0.28 & 0.05 & 0.10 & 0.05 & 0.10 & 0.04 & 0.10 & 0.06 & 0.11 \\ 
				750 & 0.18 & 0.25 & 0.03 & 0.08 & 0.04 & 0.09 & 0.05 & 0.11 & 0.06 & 0.11 \\ 
				1,000 &  0.20 & 0.29 & 0.05 & 0.10 & 0.04 & 0.09 & 0.05 & 0.10 & 0.05 & 0.11 \\  
				\hline \hline
			\end{tabular}
			\begin{tablenotes}
				\scriptsize
				\item NOTE: The table shows size figures at $5\%$ and $10\%$ levels of our test statistic based on a non-LR moment (Plug-in), an LR-moment (LR), and Wald test statistics by functional differencing using 9, 25, and 49 moments ($W_9$, $W_{25}$, $W_{49}$, respectively). Results are based on $1,000$ Monte Carlo repetitions.
			\end{tablenotes}   
		\end{threeparttable}
	\end{table}
	
	To evaluate power figures we compute (size-adjusted) rejection rates under several local alternatives. To be specific, we consider equally spaced points for true $\beta_0$'s such that $\beta_0 = \bar{\beta} + \delta$, with $\delta \in \left\{-0.5,-0.475,\cdots,0.425,0.450\right\}$ and $\bar{\beta}=1$.  In this case, we only show the performance of our LR test statistics and the Wald-type ones. All them present larger rejection rates as the departure from the null becomes more important, although when a small number of moments is considered and for $n=250$, this behavior is not necessarily monotonous. More importantly, our test statistic is deemed more efficient as it presents power figures larger than the Wald-statistics.  This is the gain that our relevant LR moment delivers in this case. It is not surprising since our test, in this case, is based on an estimator of the efficient score for $\beta_0$, which is the optimal LR moment. Overall, our numerical experimentation leads us to conclude that the finite sample performance of our test, both in terms of size and power, is in line with our theoretical results, and shows the advantage of correcting for the first-stage bias and using relevant LR moments.

	\begin{figure}[H]%
		\caption{Monte Carlo Results - Power}
		\label{fig:MCpower}
		\begin{threeparttable}
			\centering
			\subfigure[$n = 250$]{\includegraphics[width=0.41\textwidth]{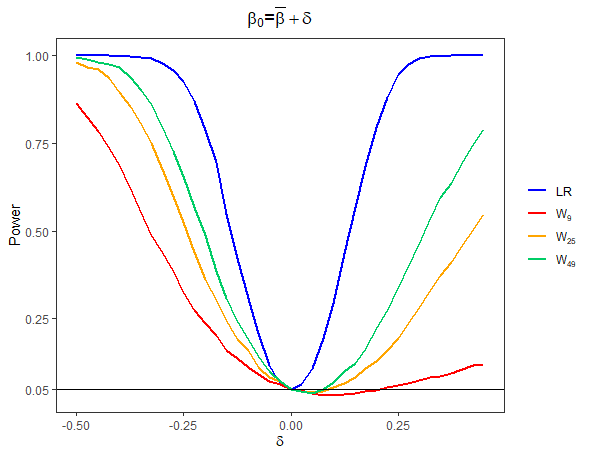}}\qquad
			\subfigure[$n = 1,000$]{\includegraphics[width=0.41\textwidth]{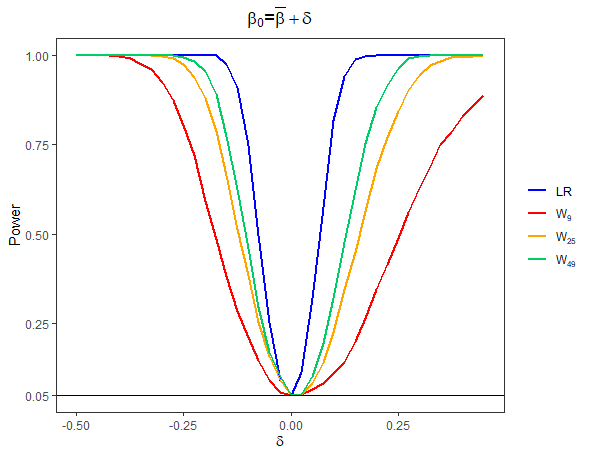}}
			\begin{tablenotes}
				\scriptsize
				\item NOTE: The figure shows size-normalized rejection rates for our test statistic based on an LR moment (LR) and Wald test statistics by functional differencing using 9, 25, and 49 moments ($W_9$, $W_{25}$, $W_{49}$, respectively). Results are based on $1,000$ Monte Carlo repetitions. 
			\end{tablenotes} 
		\end{threeparttable}
	\end{figure}
	
		\section{Application: The Effect of Smoking on Birth Weight}
	\label{empiricalapplication}
	We consider the following model: 
	\begin{equation}
		\label{empmodel1}
		y_{it} = \alpha_{1i} + \alpha_{2i}s_{it} + w_{it}^{\prime}\beta_0 + \varepsilon_{it},\;\;\; i=1,\cdots,n,\;\;\; t=1,\cdots,T, 
	\end{equation}
	where $i$ and $t$ index mothers and children, respectively, $y_{it}$ measures the child $t$'s birth weight of mother $i$, $s_{it}$ is equal to 1 if mother $i$ was smoking when she was pregnant with child $t$ and 0 otherwise, and $w_{it}$ is a vector collecting other determinants of weight at birth, including the sex of the child and the age of the mother. Notice that we can write \eqref{empmodel1} in our linear random coefficient model notation as 
	$$
	Y_i = W_{i}\beta_0 + V_{i}\alpha_i + \varepsilon_i, \;\;\; i=1,\cdots,n,
	$$
	where $Y_i = \left(y_{i1},\cdots,y_{iT}\right)^{\prime}$, $W_{i} = \left(w_{i1}, \cdots,w_{iT}\right)^{\prime}$, $V_i = \left(\imath_{T},\left(s_{i1},\cdots,s_{iT}\right)^{\prime}\right)$, where $\imath_T$ is a vector of ones with dimension $T=3$, and $\alpha_i = \left(\alpha_{i1}, \alpha_{i2}\right)^{\prime}$ varies across mothers ($q=2$). The intercept $\alpha_{i1}$ measures mother-specific unobserved characteristics (which could be partly genetic) that affect the weight of the child at birth whereas $\alpha_{i2}$ is the marginal effect of smoking. 
	
	\cite{abrevaya2006estimating}, assuming homogeneous $\alpha_{2i}$, finds large and significant negative effects of smoking on birth weight. This coincides with the results in \cite{arellano2012identifying} (AB, hereafter), who allow for UH across mothers. In both cases, the authors work with a restrictive specification for the additional controls in \eqref{empmodel1}. Differently, taking advantage of our theory for high-dimensional $W_i$, we will allow for a richer specification in covariates, addressing potential bias due to functional form assumptions on how controls enter in the model. Nevertheless, our results below generally confirm the findings obtained by the aforementioned works, suggesting a positive result: the existing empirical conclusions on the treatment effect of smoking on child birth weight are robust to the inclusion of a flexible model in covariates.
	
	Some comments regarding the data are in order. We use the same sample of mothers from \cite{abrevaya2006estimating} employed by AB. This data comes from the Natality Data Sets for the U.S. for the years 1990 and 1998. Abrebaya implements a method to match mothers to children as there is no unique identifiers in the original data. We use the ``matched panel $\#$3", which is the subsample less likely to suffer from large matching error. The restriction $rank(V_i) = q < T$ makes us focus on mothers who had at least three children during 1989-1998. In our data, all mothers have had three children during the period of analysis. In addition, we need the smoking status $s_{it}$ to be time-varying. Hence, we keep only mothers for whom $s_{it}$ changes at least once. Our final sample cointains 1,445 mothers.
	
	In the original specification of AB, $w_{it}$ includes the following variables: the sex of the child, the age of the mother at time of birth, dummy variables indicating number of conducted prenatal visits, and the value of an index---the ``Kessner" index, which measures the quality of prenatal care \citep[see][p. 496]{abrevaya2006estimating}. In this empirical exercise, we want to evaluate the robustness of this specification while still allowing for heterogeneous treatment. Specifically, we control for the aforementioned covariates along with a (cubic) B-spline basis expansion of five terms in age.\footnote{Our approach allows us to go beyond this specification. We, for instance, also considered a larger number of terms in the basis construction and interaction terms with the rest of the covariates in $w_{it}$; yet, some point estimates became reasonably implausible.} Our algorithm above will select the relevant terms in $w_{it}$.    
	
	We first conduct inference on some common parameters that appear in $\beta_0$, following the algorithm outlined in Section \ref{commonpar}.  The results are shown in the first panel of Table \ref{empresultstable}. Columns ``Est. (AB)" and ``CI$_{95}$ (AB)" display point estimates and $95\%$ confidence intervals by AB, respectively. Columns ``Est. (AE)" and ``CI$_{95}$ (AE)" show our own results. Our strategy, despite allowing for more flexibility, generally confirms the main conclusion obtained by AB: the sex of the child seems to be the only important control. In addition, both slope estimates for this variable are virtually equal: Boys tend to have, on average, 130 grams larger birth weights than girls. Although the rest of the regressors show the expected signs, they do not seem to be statistically significant. 
	
	Turning to heterogeneous treatment effects, we have estimated the average UH in the population of mothers with three children. Our results can be seen in the rows ``Mean $\alpha_{1i}$" and  ``Mean $\alpha_{2i}$", indicating the average of the random intercepts and slopes, respectively. For this, we have followed the algorithm in Section \ref{averagemarginal}. While our results virtually coincide with those in AB for the average treatment effect of smoking, our intercept coefficient differs. In particular, we estimate an average effect of -161.65 grams of smoking during pregnancy. This effect is statistically different from zero, and our $95\%$ confidence intervals are almost the same as the one implied by the results obtained by AB. Our intercept coefficient is $22\%$ larger and equal to 3,404 grams (and statistically different from zero). In addition, our confidence intervals do not fully overlap with those in AB and suggest larger estimates for the random intercept.
	
	Finally, we have estimated second moments for both UH parameters, following the guidelines in Section \ref{secondmoments}. From these calculations and our previous estimates, we have computed implied variance estimates, which appear in the rows ``Var $\alpha_{1i}$" and ``Var $\alpha_{2i}$".  We first assume iid homoskedastic errors.\footnote{In this case, $S_2 = vec(I_3)$ in \eqref{Var} and $\omega_{0i} = \sigma^2_0$, where $\sigma^2_{0}$ is the variance of the idiosyncratic errors.} Then, we allow for a more flexible variance-covariance matrix for the errors, following the same specification that Arellano and Bonhomme considered. In this case, the variance of errors for the $jth$ child is $a+b(j-1)$, where $a$ and $b$ are scalars.\footnote{Notice that the selection matrix $S_2$ must also change.} We refer to this model as the non-stationary errors case. In general, our results suggest significant dispersion,\footnote{Except for Var $\alpha_{1i}$ when iid errors are considered.} which is typically larger than that determined by AB. The most noticeable difference is observed for the variance (or standard deviation) of $\alpha_{2i}$ in the non-stationary errors case---our point estimate of its standard deviation is 18\% larger. Interestingly, while the estimates by AB indicate a reduction in both variances, our method implies an increase in the dispersion of the random intercept and slope.\footnote{This responds to the fact that the two approaches are based on different moments to estimate the variances.}  All in all, similarly to AB, our results suggest the presence of UH in the treatment effects of smoking on birth weights for this data; however, we find evidence of slightly more variability across mothers than previously determined.

	\section{Concluding Remarks}
	
	\label{Conclusions}
	
	In the recent decade, there has been an increased interest in orthogonal moments and DML inferences, due to their convenient properties in contexts with high-dimensional first-step estimators. Yet, their extensions to models with UH have remained limited. In this context, several important parameters are functionals of common components and UH. Typically, the distribution of UH is not identified, and if regularized plug-in approaches are used, this might result in slow rates for the target functionals and large biases that invalidate inferences. All of this justifies the importance of applying Neyman-orthogonality to these settings, which is the main goal of this paper.  
	
	We have shed light on DML for models with UH and general parameters of interest by providing necessary and sufficient conditions for the existence and relevance of Neyman-orthogonal moments. Crucially, our conditions are constructive for identification, estimation, and inference. We have obtained novel identification results and DML inferences for high-dimensional random coefficient models, the Kotlarski model with a factor loading, and teacher value-added models. We have introduced an algorithm to construct relevant LR moments for smooth functionals, which might not necessarily require estimating UH, as illustrated with our examples. We hope these ideas will be applied in different contexts. Our paper opens the door for the application of valid asymptotic inferences in models with UH and other parameters estimated by machine learning methods---including high-dimensional regressors---and efficient inferences, as illustrated by our Monte Carlo and empirical exercises. 
	
	There are important avenues for future research on this subject. On an applied note, we strongly believe our results are useful to a whole host of applications, which were not covered by us in this paper. These might include applications to other random coefficient models, production functions with heterogeneous technologies, duration models, individual-specific income profile models, and heterogeneous treatment effects. It would be interesting to apply our theory to these contexts to obtain reliable conclusions in relevant empirical settings, allowing for flexible specifications and UH. Furthermore, on a theoretical side, we have shown that some important functionals, such as CDFs, counterfactual measures, and quantiles of UH, fail to have relevant orthogonal moments, as defined in this paper. In the future, we plan to explore the extensions of our results to such parameters, further extending the applicability of our results.    
	
	\begin{sidewaystable}[htbp]
		\small
		\caption{Results--Effect of Smoking on Birth Weight}
		\label{empresultstable}
		\centering
		\begin{threeparttable}
			\begin{tabular}{P{2.7cm}P{1.2cm}P{5cm}P{2cm}P{5.1cm}}
				\hline \hline
				& \multicolumn{1}{>{\centering\arraybackslash}m{1.2cm}}{Est.}& \multicolumn{1}{>{\centering\arraybackslash}m{5cm}}{CI$_{95}$} & \multicolumn{1}{>{\centering\arraybackslash}m{2cm}}{Est.} & \multicolumn{1}{>{\centering\arraybackslash}m{5.1cm}}{CI$_{95}$} \\ 
				& (AB) & (AB) & (AE) & (AE) \\
				\hline \hline 
				\multicolumn{5}{c}{Common Parameters}  \\ \hline \\
Male & 129.81 & [85.21, 174.4] & 130.35 & [85.02, 175.68] \\ 
  Age & 39.03 & [-23.7, 101.77] & 2.95 & [-61.54, 67.45] \\ 
  Age-sq & -0.64 & [-1.77, 0.49] & -0.04 & [-1.15, 1.07] \\ 
  Kessner = 2 & -81.98 & [-185.22, 21.25] & -12.65 & [-46.03, 20.73] \\ 
  Kessner= 3 & -158.85 & [-319.31, 1.62] & -42.55 & [-121.25, 36.15] \\ 
  No visit & -18.01 & [-260.85, 224.83] & -1.21 & [-14.79, 12.36] \\ 
  Visit = 2 & 83.23 & [-22.46, 188.93] & 10.6 & [-27.59, 48.79] \\ 
  Visit = 3 & 135.81 & [-58.63, 330.25] & 3.65 & [-43.14, 50.44] \\ 
				\multicolumn{5}{c}{Heterogeneous Treatment: Mean}  \\ \hline \\
				Mean $\alpha_{1i}$ & 2782.38 & [1929.37, 3635.39] & 3403.55 & [2569.78, 4237.31] \\ 
				Mean $\alpha_{2i}$ & -161.26 & [-194.57, -127.95] & -161.65 & [-194.81, -128.49] \\ 
				\multicolumn{5}{c}{Heterogeneous Treatment: Variance (iid errors)}  \\ \hline \\
				Var $\alpha_{1i}$ & 127647.77 & [97932.05, 157363.48] & 115098.31 & [20540.43, 209656.19] \\ 
				Var $\alpha_{2i}$ & 98239.22 & [55756.76, 140721.68] & 107811.66 & [66180.42, 149442.91] \\ 
				\multicolumn{5}{c}{Heterogeneous Treatment: Variance (non-stationary errors)}  \\ \hline \\
				Var $\alpha_{1i}$ & 120422.97 & [73077.64, 167768.3] & 121239.72 & [34298.78, 208180.67] \\ 
				Var $\alpha_{2i}$ & 85673.17 & [17954, 153392.33] & 119105.57 & [64620.85, 173590.28] \\ 
				\hline \hline
			\end{tabular}
			\begin{tablenotes}
				\scriptsize
				\item NOTE: The table shows point estimates and $95\%$ confidence intervals for some common parameters and mean and variances of random intercepts and slopes in model \eqref{empmodel1}. These have been obtained following \cite{arellano2012identifying} (AB) and our own algorithms of Section \ref{sectionhighdimensionalpanel} (AE). The sample size is $n=1,445$.
			\end{tablenotes}   
		\end{threeparttable}
	\end{sidewaystable}

\clearpage
\phantomsection
\bibliographystyle{ectabib}
\bibliography{references}

\begin{thebibliography}{137}
\newcommand{\enquote}[1]{``#1''}
\expandafter\ifx\csname natexlab\endcsname\relax\def\natexlab#1{#1}\fi

\bibitem[\protect\citeauthoryear{Aaronson, Barrow, and Sander}{Aaronson
  et~al.}{2007}]{aaronson2007teachers}
\textsc{Aaronson, Daniel, Lisa Barrow, and William Sander} (2007):
  \enquote{\titlecap{Teachers and student achievement in the Chicago public
  high schools},} \emph{Journal of labor Economics}, 25 (1), 95--135.

\bibitem[\protect\citeauthoryear{Abowd, Kramarz, and Margolis}{Abowd
  et~al.}{1999}]{abowd1999high}
\textsc{Abowd, John~M, Francis Kramarz, and David~N Margolis} (1999):
  \enquote{\titlecap{High wage workers and high wage firms},}
  \emph{Econometrica}, 67 (2), 251--333.

\bibitem[\protect\citeauthoryear{Abrevaya}{Abrevaya}{2006}]{abrevaya2006estimating}
\textsc{Abrevaya, Jason} (2006): \enquote{\titlecap{Estimating the effect of
  smoking on birth outcomes using a matched panel data approach},}
  \emph{Journal of Applied Econometrics}, 21 (4), 489--519.

\bibitem[\protect\citeauthoryear{Aguirregabiria and Carro}{Aguirregabiria and
  Carro}{2024{\natexlab{a}}}]{aguirregabiria2024identification}
\textsc{Aguirregabiria, Victor and Jes{\'u}s~M Carro} (2024{\natexlab{a}}):
  \enquote{\titlecap{Identification of average marginal effects in fixed
  effects dynamic discrete choice models},} \emph{Review of Economics and
  Statistics}, 1--46.

\bibitem[\protect\citeauthoryear{Aguirregabiria and Carro}{Aguirregabiria and
  Carro}{2024{\natexlab{b}}}]{aguirregabiria2024identificationaveragemarginaleffects}
\textsc{Aguirregabiria, Victor and Jesus~M. Carro} (2024{\natexlab{b}}):
  \enquote{\titlecap{Identification of Average Marginal Effects in Fixed
  Effects Dynamic Discrete Choice Models},} .

\bibitem[\protect\citeauthoryear{Alvarez, Borovi{\v{c}}kov{\'a}, and
  Shimer}{Alvarez et~al.}{2016}]{alvarez2016decomposing}
\textsc{Alvarez, Fernando, Katar{\'\i}na Borovi{\v{c}}kov{\'a}, and Robert
  Shimer} (2016): \enquote{\titlecap{Decomposing duration dependence in a
  stopping time model},} Tech. rep., National Bureau of Economic Research.

\bibitem[\protect\citeauthoryear{Andersen}{Andersen}{1970}]{andersen1970asymptotic}
\textsc{Andersen, Erling~Bernhard} (1970): \enquote{Asymptotic properties of
  conditional maximum-likelihood estimators,} \emph{Journal of the Royal
  Statistical Society Series B: Statistical Methodology}, 32 (2), 283--301.

\bibitem[\protect\citeauthoryear{Anderson and Rubin}{Anderson and
  Rubin}{1949}]{anderson1949estimation}
\textsc{Anderson, Theodore~W and Herman Rubin} (1949): \enquote{Estimation of
  the parameters of a single equation in a complete system of stochastic
  equations,} \emph{The Annals of mathematical statistics}, 20 (1), 46--63.

\bibitem[\protect\citeauthoryear{Andrews}{Andrews}{1987}]{andrews1987asymptotic}
\textsc{Andrews, Donald~WK} (1987): \enquote{Asymptotic results for generalized
  Wald tests,} \emph{Econometric Theory}, 3 (3), 348--358.

\bibitem[\protect\citeauthoryear{Angrist, Hull, Pathak, and Walters}{Angrist
  et~al.}{2016}]{angrist2016interpreting}
\textsc{Angrist, Joshua, Peter Hull, Parag Pathak, and Christopher Walters}
  (2016): \enquote{\titlecap{Interpreting tests of school VAM validity},}
  \emph{American Economic Review}, 106 (5), 388--392.

\bibitem[\protect\citeauthoryear{Angrist, Hull, Pathak, and Walters}{Angrist
  et~al.}{2017}]{angrist2017leveraging}
\textsc{Angrist, Joshua~D, Peter~D Hull, Parag~A Pathak, and Christopher~R
  Walters} (2017): \enquote{\titlecap{Leveraging lotteries for school
  value-added: Testing and estimation},} \emph{The Quarterly Journal of
  Economics}, 132 (2), 871--919.

\bibitem[\protect\citeauthoryear{Arellano, Blundell, and Bonhomme}{Arellano
  et~al.}{2017}]{arellano2017earnings}
\textsc{Arellano, Manuel, Richard Blundell, and St{\'e}phane Bonhomme} (2017):
  \enquote{Earnings and consumption dynamics: a nonlinear panel data
  framework,} \emph{Econometrica}, 85 (3), 693--734.

\bibitem[\protect\citeauthoryear{Arellano and Bonhomme}{Arellano and
  Bonhomme}{2012}]{arellano2012identifying}
\textsc{Arellano, Manuel and St{\'e}phane Bonhomme} (2012):
  \enquote{\titlecap{Identifying distributional characteristics in random
  coefficients panel data models},} \emph{The Review of Economic Studies}, 79
  (3), 987--1020.

\bibitem[\protect\citeauthoryear{Arellano and Hahn}{Arellano and
  Hahn}{2007}]{arellano2007understanding}
\textsc{Arellano, Manuel and Jinyong Hahn} (2007): \enquote{Understanding bias
  in nonlinear panel models: Some recent developments,} \emph{Econometric
  Society Monographs}, 43, 381.

\bibitem[\protect\citeauthoryear{Arga{\~n}araz and Escanciano}{Arga{\~n}araz
  and Escanciano}{2023}]{arganaraz2023existence}
\textsc{Arga{\~n}araz, Facundo and Juan~Carlos Escanciano} (2023):
  \enquote{\titlecap{On the Existence and Information of Orthogonal Moments For
  Inference},} \emph{arXiv preprint arXiv:2303.11418}.

\bibitem[\protect\citeauthoryear{Argañaraz}{Argañaraz}{2024}]{arganarazconstructioncmr}
\textsc{Argañaraz, Facundo} (2024): \enquote{Automatic Debiased Machine
  Learning of Structural Parameters with General Conditional Moments,}
  \emph{Working paper}.

\bibitem[\protect\citeauthoryear{Begun, Hall, Huang, and Wellner}{Begun
  et~al.}{1983}]{begun1983information}
\textsc{Begun, Janet~M, W~Jackson Hall, Wei-Min Huang, and Jon~A Wellner}
  (1983): \enquote{\titlecap{Information and asymptotic efficiency in
  parametric-nonparametric models},} \emph{The Annals of Statistics}, 11 (2),
  432--452.

\bibitem[\protect\citeauthoryear{Belloni, Chen, Chernozhukov, and
  Hansen}{Belloni et~al.}{2012}]{belloni2012sparse}
\textsc{Belloni, Alexandre, Daniel Chen, Victor Chernozhukov, and Christian
  Hansen} (2012): \enquote{\titlecap{Sparse models and methods for optimal
  instruments with an application to eminent domain},} \emph{Econometrica}, 80
  (6), 2369--2429.

\bibitem[\protect\citeauthoryear{Belloni, Chernozhukov, Hansen, and
  Kozbur}{Belloni et~al.}{2016}]{belloni2016inference}
\textsc{Belloni, Alexandre, Victor Chernozhukov, Christian Hansen, and Damian
  Kozbur} (2016): \enquote{\titlecap{Inference in high-dimensional panel models
  with an application to gun control},} \emph{Journal of Business \& Economic
  Statistics}, 34 (4), 590--605.

\bibitem[\protect\citeauthoryear{Bennett, Kallus, Mao, Newey, Syrgkanis, and
  Uehara}{Bennett et~al.}{2023}]{bennett2023source}
\textsc{Bennett, Andrew, Nathan Kallus, Xiaojie Mao, Whitney Newey, Vasilis
  Syrgkanis, and Masatoshi Uehara} (2023): \enquote{Source condition double
  robust inference on functionals of inverse problems,} \emph{arXiv preprint
  arXiv:2307.13793}.

\bibitem[\protect\citeauthoryear{Bhatia}{Bhatia}{1997}]{bhatia1997}
\textsc{Bhatia, Rajendra} (1997): \emph{\titlecap{Matrix Analysis}}, Springer
  New York, NY.

\bibitem[\protect\citeauthoryear{Bickel}{Bickel}{1982}]{bickel1982adaptive}
\textsc{Bickel, Peter~J} (1982): \enquote{\titlecap{On adaptive estimation},}
  \emph{The Annals of Statistics}, 647--671.

\bibitem[\protect\citeauthoryear{Blandhol, Bonney, Mogstad, and
  Torgovitsky}{Blandhol et~al.}{2022}]{blandhollate}
\textsc{Blandhol, Christine, John Bonney, Magne Mogstad, and Alexander
  Torgovitsky} (2022): \enquote{When is TSLS Actually LATE?} Working Paper
  29709, National Bureau of Economic Research.

\bibitem[\protect\citeauthoryear{Bonhomme}{Bonhomme}{2011}]{bonhomme2011}
\textsc{Bonhomme, S.} (2011): \enquote{\titlecap{Panel Data, Inverse Problems,
  and the Estimation of Policy Parameters},} \emph{Unpublished manuscript}.

\bibitem[\protect\citeauthoryear{Bonhomme}{Bonhomme}{2012}]{bonhomme2012functional}
\textsc{Bonhomme, St{\'e}phane} (2012): \enquote{Functional Differencing,}
  \emph{Econometrica}, 80 (4), 1337--1385.

\bibitem[\protect\citeauthoryear{Bonhomme and Dano}{Bonhomme and
  Dano}{2024}]{bonhommeDano2024functional}
\textsc{Bonhomme, St{\'e}phane and Kevin Dano} (2024): \enquote{Functional
  Differencing in Networks,} \emph{Revue {\'e}conomique}, 75 (1), 147--175.

\bibitem[\protect\citeauthoryear{Bonhomme and Denis}{Bonhomme and
  Denis}{2024}]{bonhomme2024estimating}
\textsc{Bonhomme, St{\'e}phane and Angela Denis} (2024):
  \enquote{\titlecap{Estimating heterogeneous effects: applications to labor
  economics},} \emph{arXiv preprint arXiv:2404.01495}.

\bibitem[\protect\citeauthoryear{Bonhomme, Jochmans, and Weidner}{Bonhomme
  et~al.}{2024}]{bonhomme2024neyman}
\textsc{Bonhomme, St{\'e}phane, Koen Jochmans, and Martin Weidner} (2024):
  \enquote{\titlecap{A neyman-orthogonalization approach to the incidental
  parameter problem},} \emph{arXiv preprint arXiv:2412.10304}.

\bibitem[\protect\citeauthoryear{Bonhomme and Robin}{Bonhomme and
  Robin}{2010}]{bonhomme2010generalized}
\textsc{Bonhomme, St{\'e}phane and Jean-Marc Robin} (2010):
  \enquote{Generalized non-parametric deconvolution with an application to
  earnings dynamics,} \emph{The Review of Economic Studies}, 77 (2), 491--533.

\bibitem[\protect\citeauthoryear{Borusyak, Jaravel, and Spiess}{Borusyak
  et~al.}{2024}]{borusyak2024revisiting}
\textsc{Borusyak, Kirill, Xavier Jaravel, and Jann Spiess} (2024):
  \enquote{\titlecap{Revisiting event-study designs: robust and efficient
  estimation},} \emph{Review of Economic Studies}, 91 (6), 3253--3285.

\bibitem[\protect\citeauthoryear{Browning and Carro}{Browning and
  Carro}{2007}]{Browning_Carro_2007}
\textsc{Browning, Martin and Jesus Carro} (2007): \emph{Heterogeneity and
  Microeconometrics Modeling}, Cambridge University Press, 47–74, Econometric
  Society Monographs.

\bibitem[\protect\citeauthoryear{Bunting, Diegert, and Maurel}{Bunting
  et~al.}{2024}]{bunting2024heterogeneity}
\textsc{Bunting, Jackson, Paul Diegert, and Arnaud Maurel} (2024):
  \enquote{Heterogeneity, Uncertainty and Learning: Semiparametric
  Identification and Estimation,} Tech. rep., National Bureau of Economic
  Research.

\bibitem[\protect\citeauthoryear{Carrasco, Florens, and Renault}{Carrasco
  et~al.}{2007}]{CARRASCO20075633}
\textsc{Carrasco, Marine, Jean-Pierre Florens, and Eric Renault} (2007):
  \emph{\titlecap{Chapter 77 Linear Inverse Problems in Structural Econometrics
  Estimation Based on Spectral Decomposition and Regularization}}, vol.~6 of
  \emph{Handbook of Econometrics}, Elsevier.

\bibitem[\protect\citeauthoryear{Carroll and Hall}{Carroll and
  Hall}{1988}]{carroll1988optimal}
\textsc{Carroll, Raymond~J and Peter Hall} (1988): \enquote{Optimal rates of
  convergence for deconvolving a density,} \emph{Journal of the American
  Statistical Association}, 83 (404), 1184--1186.

\bibitem[\protect\citeauthoryear{Chamberlain}{Chamberlain}{1984}]{chamberlain1984panel}
\textsc{Chamberlain, Gary} (1984): \enquote{Panel Data,} \emph{Handbook of
  econometrics}, 2, 1247--1318.

\bibitem[\protect\citeauthoryear{Chamberlain}{Chamberlain}{1992}]{chamberlain1992efficiency}
---\hspace{-.1pt}---\hspace{-.1pt}--- (1992): \enquote{\titlecap{Efficiency
  bounds for semiparametric regression},} \emph{Econometrica: Journal of the
  Econometric Society}, 567--596.

\bibitem[\protect\citeauthoryear{Chen and Santos}{Chen and
  Santos}{2018}]{chen2018overidentification}
\textsc{Chen, Xiaohong and Andres Santos} (2018):
  \enquote{\titlecap{Overidentification in regular models},}
  \emph{Econometrica}, 86 (5), 1771--1817.

\bibitem[\protect\citeauthoryear{Chernozhukov, Chetverikov, Demirer, Duflo,
  Hansen, Newey, and Robins}{Chernozhukov
  et~al.}{2018}]{chernozhukov2018double}
\textsc{Chernozhukov, Victor, Denis Chetverikov, Mert Demirer, Esther Duflo,
  Christian Hansen, Whitney Newey, and James Robins} (2018):
  \enquote{\titlecap{Double/debiased machine learning for treatment and
  structural parameters},} \emph{The Econometrics Journal}, 21, C1--C68.

\bibitem[\protect\citeauthoryear{Chernozhukov, Escanciano, Ichimura, Newey, and
  Robins}{Chernozhukov et~al.}{2022}]{chernozhukov2022locally}
\textsc{Chernozhukov, Victor, Juan~Carlos Escanciano, Hidehiko Ichimura,
  Whitney~K Newey, and James~M Robins} (2022): \enquote{\titlecap{Locally
  robust semiparametric estimation},} \emph{Econometrica}, 90 (4), 1501--1535.

\bibitem[\protect\citeauthoryear{Chernozhukov, Fern{\'a}ndez-Val, Hahn, and
  Newey}{Chernozhukov et~al.}{2013}]{chernozhukov2013average}
\textsc{Chernozhukov, Victor, Iv{\'a}n Fern{\'a}ndez-Val, Jinyong Hahn, and
  Whitney Newey} (2013): \enquote{Average and quantile effects in nonseparable
  panel models,} \emph{Econometrica}, 81 (2), 535--580.

\bibitem[\protect\citeauthoryear{Chetty, Friedman, and Rockoff}{Chetty
  et~al.}{2014}]{chetty2014measuring}
\textsc{Chetty, Raj, John~N Friedman, and Jonah~E Rockoff} (2014):
  \enquote{\titlecap{Measuring the impacts of teachers II: Teacher value-added
  and student outcomes in adulthood},} \emph{American economic review}, 104
  (9), 2633--2679.

\bibitem[\protect\citeauthoryear{Chetty and Hendren}{Chetty and
  Hendren}{2018}]{chetty2018impacts}
\textsc{Chetty, Raj and Nathaniel Hendren} (2018): \enquote{\titlecap{The
  impacts of neighborhoods on intergenerational mobility II: County-level
  estimates},} \emph{The Quarterly Journal of Economics}, 133 (3), 1163--1228.

\bibitem[\protect\citeauthoryear{Clarke and Polselli}{Clarke and
  Polselli}{2025}]{clarke2025double}
\textsc{Clarke, Paul~S and Annalivia Polselli} (2025): \enquote{Double machine
  learning for static panel models with fixed effects,} \emph{The Econometrics
  Journal}, utaf011.

\bibitem[\protect\citeauthoryear{Cooper and Haltiwanger}{Cooper and
  Haltiwanger}{2006}]{cooper2006nature}
\textsc{Cooper, Russell~W and John~C Haltiwanger} (2006): \enquote{\titlecap{On
  the nature of capital adjustment costs},} \emph{The Review of Economic
  Studies}, 73 (3), 611--633.

\bibitem[\protect\citeauthoryear{Corman}{Corman}{1995}]{corman1995effects}
\textsc{Corman, Hope} (1995): \enquote{\titlecap{The effects of low birthweight
  and other medical risk factors on resource utilization in the pre-school
  years},} .

\bibitem[\protect\citeauthoryear{Corman and Chaikind}{Corman and
  Chaikind}{1998}]{corman1998effect}
\textsc{Corman, Hope and Stephen Chaikind} (1998): \enquote{\titlecap{The
  effect of low birthweight on the school performance and behavior of
  school-aged children},} \emph{Economics of Education Review}, 17 (3),
  307--316.

\bibitem[\protect\citeauthoryear{Cunha, Heckman, and Schennach}{Cunha
  et~al.}{2010}]{cunha2010estimating}
\textsc{Cunha, Flavio, James~J Heckman, and Susanne~M Schennach} (2010):
  \enquote{\titlecap{Estimating the technology of cognitive and noncognitive
  skill formation},} \emph{Econometrica}, 78 (3), 883--931.

\bibitem[\protect\citeauthoryear{Currie and Hyson}{Currie and
  Hyson}{1999}]{currie1999impact}
\textsc{Currie, Janet and Rosemary Hyson} (1999): \enquote{\titlecap{Is the
  impact of health shocks cushioned by socioeconomic status? The case of low
  birthweight},} \emph{American economic review}, 89 (2), 245--250.

\bibitem[\protect\citeauthoryear{Daubechies, Defrise, and De~Mol}{Daubechies
  et~al.}{2004}]{daubechies2004iterative}
\textsc{Daubechies, Ingrid, Michel Defrise, and Christine De~Mol} (2004):
  \enquote{\titlecap{An iterative thresholding algorithm for linear inverse
  problems with a sparsity constraint},} \emph{Communications on Pure and
  Applied Mathematics: A Journal Issued by the Courant Institute of
  Mathematical Sciences}, 57 (11), 1413--1457.

\bibitem[\protect\citeauthoryear{Davezies, D'Haultfoeuille, and Laage}{Davezies
  et~al.}{2021}]{davezies2021identification}
\textsc{Davezies, Laurent, Xavier D'Haultfoeuille, and Louise Laage} (2021):
  \enquote{Identification and estimation of average marginal effects in fixed
  effects logit models,} \emph{arXiv preprint arXiv:2105.00879}.

\bibitem[\protect\citeauthoryear{Davis}{Davis}{2024}]{davis2024general}
\textsc{Davis, Tom~P} (2024): \enquote{A general expression for Hermite
  expansions with applications,} \emph{The Mathematics Enthusiast}, 21 (1),
  71--87.

\bibitem[\protect\citeauthoryear{Dobbelaere and Mairesse}{Dobbelaere and
  Mairesse}{2013}]{dobbelaere2013panel}
\textsc{Dobbelaere, Sabien and Jacques Mairesse} (2013):
  \enquote{\titlecap{Panel data estimates of the production function and
  product and labor market imperfections},} \emph{Journal of Applied
  Econometrics}, 28 (1), 1--46.

\bibitem[\protect\citeauthoryear{Dobronyi, Gu, and il~Kim}{Dobronyi
  et~al.}{2021}]{dobronyi2021identificationdynamicpanellogit}
\textsc{Dobronyi, Christopher, Jiaying Gu, and Kyoo il~Kim} (2021):
  \enquote{\titlecap{Identification of Dynamic Panel Logit Models with Fixed
  Effects},} .

\bibitem[\protect\citeauthoryear{Dobronyi, Gu, il~Kim, and Russell}{Dobronyi
  et~al.}{2024}]{dobronyi2024identificationdynamicpanellogit}
\textsc{Dobronyi, Christopher, Jiaying Gu, Kyoo il~Kim, and Thomas~M. Russell}
  (2024): \enquote{Identification of Dynamic Panel Logit Models with Fixed
  Effects,} .

\bibitem[\protect\citeauthoryear{Doyle, Graves, and Gruber}{Doyle
  et~al.}{2019}]{doyle2019evaluating}
\textsc{Doyle, Joseph, John Graves, and Jonathan Gruber} (2019):
  \enquote{\titlecap{Evaluating measures of hospital quality: Evidence from
  ambulance referral patterns},} \emph{Review of Economics and Statistics}, 101
  (5), 841--852.

\bibitem[\protect\citeauthoryear{Duffy, Papageorgiou, and
  Perez-Sebastian}{Duffy et~al.}{2004}]{duffy2004capital}
\textsc{Duffy, John, Chris Papageorgiou, and Fidel Perez-Sebastian} (2004):
  \enquote{\titlecap{Capital-skill complementarity? Evidence from a panel of
  countries},} \emph{Review of Economics and Statistics}, 86 (1), 327--344.

\bibitem[\protect\citeauthoryear{D’Haultfoeuille}{D’Haultfoeuille}{2011}]{d2011completeness}
\textsc{D’Haultfoeuille, Xavier} (2011): \enquote{\titlecap{On the
  completeness condition in nonparametric instrumental problems},}
  \emph{Econometric Theory}, 27 (3), 460--471.

\bibitem[\protect\citeauthoryear{Efron}{Efron}{2016}]{efron2016empirical}
\textsc{Efron, Bradley} (2016): \enquote{\titlecap{Empirical Bayes
  deconvolution estimates},} \emph{Biometrika}, 103 (1), 1--20.

\bibitem[\protect\citeauthoryear{Engl, Hanke, and Neubauer}{Engl
  et~al.}{1996}]{engl1996regularization}
\textsc{Engl, Heinz~Werner, Martin Hanke, and Andreas Neubauer} (1996):
  \emph{Regularization of Inverse Problems}, vol. 375 of \emph{Mathematics and
  Its Applications}, Dordrecht: Kluwer Academic Publishers.

\bibitem[\protect\citeauthoryear{Escanciano}{Escanciano}{2022}]{escanciano2022semiparametric}
\textsc{Escanciano, Juan~Carlos} (2022): \enquote{\titlecap{Semiparametric
  identification and fisher information},} \emph{Econometric Theory}, 38 (2),
  301--338.

\bibitem[\protect\citeauthoryear{Escanciano and Li}{Escanciano and
  Li}{2021}]{escanciano2021optimal}
\textsc{Escanciano, Juan~Carlos and Wei Li} (2021): \enquote{Optimal linear
  instrumental variables approximations,} \emph{Journal of Econometrics}, 221
  (1), 223--246.

\bibitem[\protect\citeauthoryear{Evdokimov}{Evdokimov}{2010}]{evdokimov2010identification}
\textsc{Evdokimov, Kirill} (2010): \enquote{Identification and estimation of a
  nonparametric panel data model with unobserved heterogeneity,}
  \emph{Department of Economics, Princeton University}, 1 (5), 23.

\bibitem[\protect\citeauthoryear{Evdokimov and White}{Evdokimov and
  White}{2012}]{evdokimov2012some}
\textsc{Evdokimov, Kirill and Halbert White} (2012): \enquote{\titlecap{Some
  extensions of a lemma of Kotlarski},} \emph{Econometric Theory}, 28 (4),
  925--932.

\bibitem[\protect\citeauthoryear{Fan}{Fan}{1991}]{fan1991optimal}
\textsc{Fan, Jianqing} (1991): \enquote{\titlecap{On the optimal rates of
  convergence for nonparametric deconvolution problems},} \emph{The Annals of
  Statistics}, 1257--1272.

\bibitem[\protect\citeauthoryear{Ferreira and Menegatto}{Ferreira and
  Menegatto}{2013}]{ferreira2013positive}
\textsc{Ferreira, JC and Valdir~Ant{\^o}nio Menegatto} (2013):
  \enquote{POSITIVE DEFINITENESS, REPRODUCING KERNEL HILBERT SPACES AND
  BEYOND.} \emph{Annals of Functional Analysis}, 4 (1).

\bibitem[\protect\citeauthoryear{Finkelstein, Gentzkow, Hull, and
  Williams}{Finkelstein et~al.}{2017}]{finkelstein2017adjusting}
\textsc{Finkelstein, Amy, Matthew Gentzkow, Peter Hull, and Heidi Williams}
  (2017): \enquote{\titlecap{Adjusting risk adjustment--accounting for
  variation in diagnostic intensity},} \emph{The New England journal of
  medicine}, 376 (7), 608.

\bibitem[\protect\citeauthoryear{Fletcher, Horwitz, and Bradley}{Fletcher
  et~al.}{2014}]{fletcher2014estimating}
\textsc{Fletcher, Jason~M, Leora~I Horwitz, and Elizabeth Bradley} (2014):
  \enquote{\titlecap{Estimating the value added of attending physicians on
  patient outcomes},} Tech. rep., National Bureau of Economic Research.

\bibitem[\protect\citeauthoryear{Fox, il~Kim, and Yang}{Fox
  et~al.}{2016}]{fox2016simple}
\textsc{Fox, Jeremy~T, Kyoo il~Kim, and Chenyu Yang} (2016):
  \enquote{\titlecap{A simple nonparametric approach to estimating the
  distribution of random coefficients in structural models},} \emph{Journal of
  Econometrics}, 195 (2), 236--254.

\bibitem[\protect\citeauthoryear{Freyberger}{Freyberger}{2018}]{freyberger2018non}
\textsc{Freyberger, Joachim} (2018): \enquote{Non-parametric panel data models
  with interactive fixed effects,} \emph{The Review of Economic Studies}, 85
  (3), 1824--1851.

\bibitem[\protect\citeauthoryear{Friedman, Hastie, H{\"o}fling, and
  Tibshirani}{Friedman et~al.}{2007}]{friedman2007pathwise}
\textsc{Friedman, Jerome, Trevor Hastie, Holger H{\"o}fling, and Robert
  Tibshirani} (2007): \enquote{\titlecap{Pathwise coordinate optimization},}
  \emph{The annals of applied statistics}, 1 (2), 302--332.

\bibitem[\protect\citeauthoryear{Friedman, Hastie, and Tibshirani}{Friedman
  et~al.}{2010}]{friedman2010regularization}
\textsc{Friedman, Jerome, Trevor Hastie, and Rob Tibshirani} (2010):
  \enquote{\titlecap{Regularization paths for generalized linear models via
  coordinate descent},} \emph{Journal of statistical software}, 33 (1), 1.

\bibitem[\protect\citeauthoryear{Fruehwirth, Navarro, and Takahashi}{Fruehwirth
  et~al.}{2016}]{fruehwirth2016timing}
\textsc{Fruehwirth, Jane~Cooley, Salvador Navarro, and Yuya Takahashi} (2016):
  \enquote{\titlecap{How the timing of grade retention affects outcomes:
  Identification and estimation of time-varying treatment effects},}
  \emph{Journal of Labor Economics}, 34 (4), 979--1021.

\bibitem[\protect\citeauthoryear{Fu}{Fu}{1998}]{fu1998penalized}
\textsc{Fu, Wenjiang~J} (1998): \enquote{\titlecap{Penalized regressions: the
  bridge versus the lasso},} \emph{Journal of computational and graphical
  statistics}, 7 (3), 397--416.

\bibitem[\protect\citeauthoryear{Gilraine, Gu, and McMillan}{Gilraine
  et~al.}{2020}]{gilraine2020new}
\textsc{Gilraine, Michael, Jiaying Gu, and Robert McMillan} (2020):
  \enquote{\titlecap{A new method for estimating teacher value-added},} Tech.
  rep., National Bureau of Economic Research.

\bibitem[\protect\citeauthoryear{Goodman-Bacon}{Goodman-Bacon}{2021}]{goodman2021difference}
\textsc{Goodman-Bacon, Andrew} (2021):
  \enquote{\titlecap{Difference-in-differences with variation in treatment
  timing},} \emph{Journal of Econometrics}, 225 (2), 254--277.

\bibitem[\protect\citeauthoryear{Graham and Powell}{Graham and
  Powell}{2012}]{graham2012identification}
\textsc{Graham, Bryan~S and James~L Powell} (2012): \enquote{Identification and
  estimation of average partial effects in “irregular” correlated random
  coefficient panel data models,} \emph{Econometrica}, 80 (5), 2105--2152.

\bibitem[\protect\citeauthoryear{Guvenen}{Guvenen}{2009}]{guvenen2009empirical}
\textsc{Guvenen, Fatih} (2009): \enquote{\titlecap{An empirical investigation
  of labor income processes},} \emph{Review of Economic dynamics}, 12 (1),
  58--79.

\bibitem[\protect\citeauthoryear{Hack, Klein, and Taylor}{Hack
  et~al.}{1995}]{hack1995long}
\textsc{Hack, Maureen, Nancy~K Klein, and H~Gerry Taylor} (1995):
  \enquote{\titlecap{Long-term developmental outcomes of low birth weight
  infants},} \emph{The future of children}, 176--196.

\bibitem[\protect\citeauthoryear{Haider and Solon}{Haider and
  Solon}{2006}]{haider2006life}
\textsc{Haider, Steven and Gary Solon} (2006): \enquote{\titlecap{Life-cycle
  variation in the association between current and lifetime earnings},}
  \emph{American Economic Review}, 96 (4), 1308--1320.

\bibitem[\protect\citeauthoryear{Hanusheck}{Hanusheck}{2009}]{hanusheck09}
\textsc{Hanusheck, Eric~A.} (2009): \emph{Teacher Deselection}, Urban Institute
  Press, 47–74.

\bibitem[\protect\citeauthoryear{Hanushek}{Hanushek}{2011}]{hanushek2011economic}
\textsc{Hanushek, Eric~A} (2011): \enquote{\titlecap{The economic value of
  higher teacher quality},} \emph{Economics of Education review}, 30 (3),
  466--479.

\bibitem[\protect\citeauthoryear{Heckman and Singer}{Heckman and
  Singer}{1984{\natexlab{a}}}]{heckman1984method}
\textsc{Heckman, James and Burton Singer} (1984{\natexlab{a}}):
  \enquote{\titlecap{A method for minimizing the impact of distributional
  assumptions in econometric models for duration data},} \emph{Econometrica:
  Journal of the Econometric Society}, 271--320.

\bibitem[\protect\citeauthoryear{Heckman and Singer}{Heckman and
  Singer}{1984{\natexlab{b}}}]{heckman1984identifiability}
---\hspace{-.1pt}---\hspace{-.1pt}--- (1984{\natexlab{b}}):
  \enquote{\titlecap{The identifiability of the proportional hazard model},}
  \emph{The Review of Economic Studies}, 51 (2), 231--241.

\bibitem[\protect\citeauthoryear{Heckman}{Heckman}{2001}]{heckman2001micro}
\textsc{Heckman, James~J} (2001): \enquote{\titlecap{Micro data, heterogeneity,
  and the evaluation of public policy: Nobel lecture},} \emph{Journal of
  political Economy}, 109 (4), 673--748.

\bibitem[\protect\citeauthoryear{Honor{\'e}}{Honor{\'e}}{1992}]{honore1992trimmed}
\textsc{Honor{\'e}, Bo~E} (1992): \enquote{\titlecap{Trimmed LAD and least
  squares estimation of truncated and censored regression models with fixed
  effects},} \emph{Econometrica: Journal of the Econometric Society}, 533--565.

\bibitem[\protect\citeauthoryear{Honor{\'e} and Tamer}{Honor{\'e} and
  Tamer}{2006}]{honore2006bounds}
\textsc{Honor{\'e}, Bo~E and Elie Tamer} (2006): \enquote{Bounds on parameters
  in panel dynamic discrete choice models,} \emph{Econometrica}, 74 (3),
  611--629.

\bibitem[\protect\citeauthoryear{Honor{\'e} and Weidner}{Honor{\'e} and
  Weidner}{2024}]{honore2024moment}
\textsc{Honor{\'e}, Bo~E and Martin Weidner} (2024): \enquote{\titlecap{Moment
  conditions for dynamic panel logit models with fixed effects},} \emph{Review
  of Economic Studies}.

\bibitem[\protect\citeauthoryear{Honoré and Weidner}{Honoré and
  Weidner}{2023}]{honore2023momentconditionsdynamicpanel}
\textsc{Honoré, Bo~E. and Martin Weidner} (2023): \enquote{\titlecap{Moment
  Conditions for Dynamic Panel Logit Models with Fixed Effects},} .

\bibitem[\protect\citeauthoryear{Horowitz}{Horowitz}{1999}]{horowitz1999semiparametric}
\textsc{Horowitz, Joel~L} (1999): \enquote{Semiparametric estimation of a
  proportional hazard model with unobserved heterogeneity,}
  \emph{Econometrica}, 67 (5), 1001--1028.

\bibitem[\protect\citeauthoryear{Horowitz and Markatou}{Horowitz and
  Markatou}{1996}]{horowitz1996semiparametric}
\textsc{Horowitz, Joel~L and Marianthi Markatou} (1996):
  \enquote{\titlecap{Semiparametric estimation of regression models for panel
  data},} \emph{The Review of Economic Studies}, 63 (1), 145--168.

\bibitem[\protect\citeauthoryear{Hu and Schennach}{Hu and
  Schennach}{2008}]{hu2008instrumental}
\textsc{Hu, Yingyao and Susanne~M Schennach} (2008): \enquote{Instrumental
  variable treatment of nonclassical measurement error models,}
  \emph{Econometrica}, 76 (1), 195--216.

\bibitem[\protect\citeauthoryear{Ibragimov and Khasminskii}{Ibragimov and
  Khasminskii}{1981}]{ibragimovkhasminskii1981}
\textsc{Ibragimov, I.A. and R.Z. Khasminskii} (1981):
  \emph{\titlecap{Statistical Estimation: Asymptotic Theory}}, Springer-Verlag,
  New York.

\bibitem[\protect\citeauthoryear{Ishwaran}{Ishwaran}{1999}]{ishwaran1999information}
\textsc{Ishwaran, Hemant} (1999): \enquote{Information in semiparametric
  mixtures of exponential families,} \emph{The Annals of Statistics}, 27 (1),
  159--177.

\bibitem[\protect\citeauthoryear{Kane and Staiger}{Kane and
  Staiger}{2008}]{kane2008estimating}
\textsc{Kane, Thomas~J and Douglas~O Staiger} (2008):
  \enquote{\titlecap{Estimating teacher impacts on student achievement: An
  experimental evaluation},} Tech. rep., National Bureau of Economic Research.

\bibitem[\protect\citeauthoryear{Kato, Sasaki, and Ura}{Kato
  et~al.}{2021}]{kato2021robust}
\textsc{Kato, Kengo, Yuya Sasaki, and Takuya Ura} (2021): \enquote{Robust
  inference in deconvolution,} \emph{Quantitative Economics}, 12 (1), 109--142.

\bibitem[\protect\citeauthoryear{Kiefer and Wolfowitz}{Kiefer and
  Wolfowitz}{1956}]{kiefer1956consistency}
\textsc{Kiefer, Jack and Jacob Wolfowitz} (1956): \enquote{Consistency of the
  maximum likelihood estimator in the presence of infinitely many incidental
  parameters,} \emph{The Annals of Mathematical Statistics}, 887--906.

\bibitem[\protect\citeauthoryear{Klaassen}{Klaassen}{1987}]{klaassen1987consistent}
\textsc{Klaassen, Chris~AJ} (1987): \enquote{\titlecap{Consistent estimation of
  the influence function of locally asymptotically linear estimators},}
  \emph{The Annals of Statistics}, 15 (4), 1548--1562.

\bibitem[\protect\citeauthoryear{Kline, Rose, and Walters}{Kline
  et~al.}{2022}]{kline2022systemic}
\textsc{Kline, Patrick, Evan~K Rose, and Christopher~R Walters} (2022):
  \enquote{\titlecap{Systemic discrimination among large US employers},}
  \emph{The Quarterly Journal of Economics}, 137 (4), 1963--2036.

\bibitem[\protect\citeauthoryear{Klosin and Vilgalys}{Klosin and
  Vilgalys}{2022}]{klosin2022estimating}
\textsc{Klosin, Sylvia and Max Vilgalys} (2022): \enquote{\titlecap{Estimating
  continuous treatment effects in panel data using machine learning with an
  agricultural application},} \emph{arXiv preprint arXiv:2207.08789}.

\bibitem[\protect\citeauthoryear{Koedel and Rockoff}{Koedel and
  Rockoff}{2015}]{koedel2015value}
\textsc{Koedel, Cory and Jonah~E Rockoff} (2015):
  \enquote{\titlecap{Value-added modeling: A review},} \emph{Economics of
  Education Review}, 47, 180--195.

\bibitem[\protect\citeauthoryear{Koenker and Mizera}{Koenker and
  Mizera}{2014}]{koenker2014convex}
\textsc{Koenker, Roger and Ivan Mizera} (2014): \enquote{Convex optimization,
  shape constraints, compound decisions, and empirical Bayes rules,}
  \emph{Journal of the American Statistical Association}, 109 (506), 674--685.

\bibitem[\protect\citeauthoryear{Krantz and Parks}{Krantz and
  Parks}{2002}]{krantzparks2002}
\textsc{Krantz, Steven~G. and Harold~R. Parks} (2002): \emph{\titlecap{A Primer
  of Real Analytic Functions}}, Springer Science \& Business Media, New York.

\bibitem[\protect\citeauthoryear{Krasnokutskaya}{Krasnokutskaya}{2011}]{krasnokutskaya2011identification}
\textsc{Krasnokutskaya, Elena} (2011): \enquote{Identification and estimation
  of auction models with unobserved heterogeneity,} \emph{The Review of
  Economic Studies}, 78 (1), 293--327.

\bibitem[\protect\citeauthoryear{Kress}{Kress}{1989}]{kress1989}
\textsc{Kress, Rainer} (1989): \emph{\textit{\titlecap{Linear Integral
  Equations}}}, Springer Berlin, Heidelberg.

\bibitem[\protect\citeauthoryear{Lee}{Lee}{2024}]{lee2024locallyregularefficienttests}
\textsc{Lee, Adam} (2024): \enquote{Locally Regular and Efficient Tests in
  Non-Regular Semiparametric Models,} .

\bibitem[\protect\citeauthoryear{Lee and Mesters}{Lee and
  Mesters}{2024}]{lee2024locally}
\textsc{Lee, Adam and Geert Mesters} (2024): \enquote{\titlecap{Locally robust
  inference for non-Gaussian linear simultaneous equations models},}
  \emph{Journal of Econometrics}, 240 (1), 105647.

\bibitem[\protect\citeauthoryear{Lehmann and Romano}{Lehmann and
  Romano}{2005}]{lehmann2005testing}
\textsc{Lehmann, Erich~L. and Joseph~P. Romano} (2005): \emph{Testing
  Statistical Hypotheses}, Springer Texts in Statistics, Springer, 3rd ed.

\bibitem[\protect\citeauthoryear{Lewbel}{Lewbel}{2022}]{lewbel2022kotlarski}
\textsc{Lewbel, Arthur} (2022): \enquote{\titlecap{Kotlarski with a factor
  loading},} \emph{Journal of Econometrics}, 229 (1), 176--179.

\bibitem[\protect\citeauthoryear{Lewbel and Pendakur}{Lewbel and
  Pendakur}{2017}]{lewbel2017unobserved}
\textsc{Lewbel, Arthur and Krishna Pendakur} (2017): \enquote{Unobserved
  preference heterogeneity in demand using generalized random coefficients,}
  \emph{Journal of Political Economy}, 125 (4), 1100--1148.

\bibitem[\protect\citeauthoryear{Lewbel, Schennach, and Zhang}{Lewbel
  et~al.}{2024}]{lewbel2024identification}
\textsc{Lewbel, Arthur, Susanne~M Schennach, and Linqi Zhang} (2024):
  \enquote{Identification of a triangular two equation system without
  instruments,} \emph{Journal of Business \& Economic Statistics}, 42 (1),
  14--25.

\bibitem[\protect\citeauthoryear{Li and Vuong}{Li and
  Vuong}{1998}]{li1998nonparametric}
\textsc{Li, Tong and Quang Vuong} (1998): \enquote{\titlecap{Nonparametric
  estimation of the measurement error model using multiple indicators},}
  \emph{Journal of Multivariate Analysis}, 65 (2), 139--165.

\bibitem[\protect\citeauthoryear{Lillard and Weiss}{Lillard and
  Weiss}{1979}]{lillard1979components}
\textsc{Lillard, Lee~A and Yoram Weiss} (1979): \enquote{\titlecap{Components
  of variation in panel earnings data: American scientists 1960-70},}
  \emph{Econometrica: Journal of the Econometric Society}, 437--454.

\bibitem[\protect\citeauthoryear{Luenberger}{Luenberger}{1997}]{luenberger1997optimization}
\textsc{Luenberger, David~G} (1997): \emph{\titlecap{\textit{Optimization by
  vector space methods}}}, John Wiley \& Sons.

\bibitem[\protect\citeauthoryear{Magnus and Neudecker}{Magnus and
  Neudecker}{2019}]{magnus2019}
\textsc{Magnus, J.~R. and H.~Neudecker} (2019): \emph{\titlecap{Matrix
  Differential Calculus with Applications in Statistics and Econometrics}},
  John Wiley \& Sons.

\bibitem[\protect\citeauthoryear{Mairesse and Griliches}{Mairesse and
  Griliches}{1990}]{MairesseGriliches1990}
\textsc{Mairesse, Jacques and Zvi Griliches} (1990): \enquote{Heterogeneity in
  Panel Data: Are There Stable Production Functions?} in \emph{Essays in Honor
  of Edmond Malinvaud, Vol. 3}, ed. by Paul Champsaur, Michel Deleau,
  Jean-Michel Grandmont, Guy Laroque, Roger Guesnerie, Claude Henry,
  Jean-Jacques Laffont, Jacques Mairesse, Alain Monfort, and Yves Younes,
  Cambridge, MA: MIT Press.

\bibitem[\protect\citeauthoryear{Masten}{Masten}{2018}]{masten2018random}
\textsc{Masten, Matthew~A} (2018): \enquote{Random coefficients on endogenous
  variables in simultaneous equations models,} \emph{The Review of Economic
  Studies}, 85 (2), 1193--1250.

\bibitem[\protect\citeauthoryear{Mundlak}{Mundlak}{1978}]{mundlak1978pooling}
\textsc{Mundlak, Yair} (1978): \enquote{On the Pooling of Time Series and Cross
  Section Data,} \emph{Econometrica: journal of the Econometric Society},
  69--85.

\bibitem[\protect\citeauthoryear{Narasimhan and Efron}{Narasimhan and
  Efron}{2020}]{GmodelinR}
\textsc{Narasimhan, Balasubramanian and Bradley Efron} (2020):
  \enquote{deconvolveR: A G-Modeling Program for Deconvolution and Empirical
  Bayes Estimation,} \emph{Journal of Statistical Software}, 94 (11), 1–20.

\bibitem[\protect\citeauthoryear{Navarro and Zhou}{Navarro and
  Zhou}{2017}]{navarro2017identifying}
\textsc{Navarro, Salvador and Jin Zhou} (2017): \enquote{\titlecap{Identifying
  agent's information sets: An application to a lifecycle model of schooling,
  consumption and labor supply},} \emph{Review of Economic Dynamics}, 25,
  58--92.

\bibitem[\protect\citeauthoryear{Navjeevan, Pinto, and Santos}{Navjeevan
  et~al.}{2023}]{navjeevan2023identification}
\textsc{Navjeevan, Manu, Rodrigo Pinto, and Andres Santos} (2023):
  \enquote{Identification and estimation in a class of potential outcomes
  models,} \emph{arXiv preprint arXiv:2310.05311}.

\bibitem[\protect\citeauthoryear{Newey}{Newey}{1990}]{newey1990semiparametric}
\textsc{Newey, Whitney~K} (1990): \enquote{\titlecap{Semiparametric efficiency
  bounds},} \emph{Journal of applied econometrics}, 5 (2), 99--135.

\bibitem[\protect\citeauthoryear{Neyman}{Neyman}{1959}]{neyman1959}
\textsc{Neyman, J.} (1959): \enquote{\titlecap{Optimal Asymptotic Tests of
  Composite Statistical Hypothesis},} \emph{Probability and Statistics: The
  Harald Cramer Volume}, 213--234.

\bibitem[\protect\citeauthoryear{Nybom and Stuhler}{Nybom and
  Stuhler}{2016}]{nybom2016heterogeneous}
\textsc{Nybom, Martin and Jan Stuhler} (2016): \enquote{\titlecap{Heterogeneous
  income profiles and lifecycle bias in intergenerational mobility
  estimation},} \emph{Journal of Human Resources}, 51 (1), 239--268.

\bibitem[\protect\citeauthoryear{Rafi}{Rafi}{2024}]{rafi2024nonparametric}
\textsc{Rafi, Ahnaf} (2024): \enquote{Nonparametric inference for a class of
  functionals in the random coefficients logit model,} \emph{Working paper}.

\bibitem[\protect\citeauthoryear{Rao}{Rao}{1983}]{rao1983nonparametric}
\textsc{Rao, BLS~Prakasa} (1983): \emph{\titlecap{Nonparametric functional
  estimation}}, Academic Press.

\bibitem[\protect\citeauthoryear{Rao and Mitra}{Rao and
  Mitra}{1972}]{rao1972generalized}
\textsc{Rao, C~Radhakrishna and Sujit~Kumar Mitra} (1972):
  \enquote{\titlecap{Generalized inverse of a matrix and its applications},} in
  \emph{Proceedings of the Berkeley Symposium on Mathematical Statistics and
  Probability}, University of California Press, Berkeley, vol.~1, 601--620.

\bibitem[\protect\citeauthoryear{Santos}{Santos}{2011}]{santos2011instrumental}
\textsc{Santos, Andres} (2011): \enquote{Instrumental variable methods for
  recovering continuous linear functionals,} \emph{Journal of Econometrics},
  161 (2), 129--146.

\bibitem[\protect\citeauthoryear{Schick}{Schick}{1986}]{schick1986asymptotically}
\textsc{Schick, Anton} (1986): \enquote{\titlecap{On asymptotically efficient
  estimation in semiparametric models},} \emph{The Annals of Statistics},
  1139--1151.

\bibitem[\protect\citeauthoryear{Semenova, Goldman, Chernozhukov, and
  Taddy}{Semenova et~al.}{2023}]{semenova2023inference}
\textsc{Semenova, Vira, Matt Goldman, Victor Chernozhukov, and Matt Taddy}
  (2023): \enquote{\titlecap{Inference on heterogeneous treatment effects in
  high-dimensional dynamic panels under weak dependence},} \emph{Quantitative
  Economics}, 14 (2), 471--510.

\bibitem[\protect\citeauthoryear{Sz{\'e}kely and Rao}{Sz{\'e}kely and
  Rao}{2000}]{szekely2000identifiability}
\textsc{Sz{\'e}kely, GJ and CR~Rao} (2000): \enquote{Identifiability of
  distributions of independent random variables by linear combinations and
  moments,} \emph{Sankhy{\=a}: The Indian Journal of Statistics, Series A},
  193--202.

\bibitem[\protect\citeauthoryear{Tseng}{Tseng}{2001}]{tseng2001convergence}
\textsc{Tseng, Paul} (2001): \enquote{\titlecap{Convergence of a block
  coordinate descent method for nondifferentiable minimization},} \emph{Journal
  of optimization theory and applications}, 109, 475--494.

\bibitem[\protect\citeauthoryear{Van~den Berg}{Van~den
  Berg}{2001}]{van2001duration}
\textsc{Van~den Berg, Gerard~J} (2001): \enquote{\titlecap{Duration models:
  specification, identification and multiple durations},} in \emph{Handbook of
  Econometrics}, Elsevier, vol.~5, 3381--3460.

\bibitem[\protect\citeauthoryear{Van~der Vaart}{Van~der
  Vaart}{1988}]{vandervaartthesis}
\textsc{Van~der Vaart, AW} (1988): \enquote{\titlecap{Statistical estimation in
  large parameter spaces},} \emph{Thesis. CWI tract 44, Centrum voor Wiskunde
  en Informatics, Amsterdam}.

\bibitem[\protect\citeauthoryear{Van~der Vaart}{Van~der
  Vaart}{1991}]{van1991differentiable}
\textsc{Van~der Vaart, Aad} (1991): \enquote{On differentiable functionals,}
  \emph{The Annals of Statistics}, 178--204.

\bibitem[\protect\citeauthoryear{Van~der Vaart}{Van~der
  Vaart}{1998}]{vandervaart98}
\textsc{Van~der Vaart, A.~W.} (1998): \emph{\titlecap{\textit{Asymptotic
  Statistics}}}, Cambridge University Press, New York.

\bibitem[\protect\citeauthoryear{Wainwright}{Wainwright}{2019}]{wainwright2019}
\textsc{Wainwright, Martin~J.} (2019): \emph{\titlecap{High-Dimensional
  Statistics: A Non-Asymptotic Viewpoint}}, Cambridge University Press.

\bibitem[\protect\citeauthoryear{Yakusheva, Lindrooth, and Weiss}{Yakusheva
  et~al.}{2014}]{yakusheva2014nurse}
\textsc{Yakusheva, Olga, Richard Lindrooth, and Marianne Weiss} (2014):
  \enquote{\titlecap{Nurse value-added and patient outcomes in acute care},}
  \emph{Health services research}, 49 (6), 1767--1786.

\end{thebibliography}

    \newpage
	\begin{center}
		{\Huge{ \textsc{Supplementary Appendix}}}
	\end{center}

    \begin{appendix}
		
		\section{Proofs of Main Results}
		\label{Proofs}
		
		\noindent \textbf{Proof of Theorem \ref{LRmomentUHresult}:} From the regularity of model and the moment, for any regular moment function $g$ and parametric submodel we have, for $\epsilon>0,$
		\begin{equation}
			\mathbb{E}_{\tau}\left[  g(Z,\lambda_{\tau})\right]  =0,\text{ all }\tau
			\in\lbrack0,\epsilon). \label{null}%
		\end{equation}
		Differentiating this equation, we obtain by the chain rule and (\ref{GI})
		\begin{align}
			0  &  =\frac{d}{d\tau}\mathbb{E}_{\tau}\left[  g(Z,\lambda_{0})\right]
			+\frac{d}{d\tau}\mathbb{E}\left[  g(Z,\lambda_{\tau})\right] \nonumber\\
			&  =\mathbb{E}\left[  g(Z,\lambda_{0})s_{0}(Z)\right]  +\frac{d}{d\tau
			}\mathbb{E}\left[  g(Z,\lambda_{\tau})\right], \label{Chain}%
		\end{align}
		where $s_{0}=S_{\lambda_{0}}h=S_{\theta}\delta+S_{\eta}b,\text{ }h=(\delta,b)\in
		\Delta(\lambda_{0})\subset\mathbf{H}=\mathcal{H}_{\theta}\times
		\mathcal{H}_{\eta}$ such that $\psi(\lambda_\tau) = \psi(\lambda_0)$. 
		For this restricted path, it must hold that 
		\begin{equation}
			\frac{d\psi(\lambda_{\tau})}{d\tau}=\langle r_{\theta},\delta\rangle_{\mathcal{H_\theta}}+\langle r,b\rangle_{\mathcal{H_\eta}}=0.
			\label{zeroder}   
		\end{equation}
		
		If conditions (\ref{02s}) and (\ref{01s}) hold, then
		
		\begin{align*}
			\mathbb{E}\left[  g(Z,\lambda_{0})s_{0}(Z)\right] & = \mathbb{E}\left[  g(Z,\lambda_{0})S_{\lambda_{0}}h(Z)\right]\\ 
			& = \mathbb{E}\left[  g(Z,\lambda_{0})S_{\theta_{0}}\delta(Z)\right]+\mathbb{E}\left[  g(Z,\lambda_{0})S_{\eta_{0}}b(Z)\right]\\ 
			& =\langle S^{*}_{\theta_{0}}g,\delta\rangle_{\mathcal{H_\theta}}+\langle S^{*}_{\eta_{0}}g,b\rangle_{\mathcal{H_\eta}}\\
			& =c\langle r_{\theta},\delta\rangle_{\mathcal{H_\theta}}+c\langle r,b\rangle_{\mathcal{H_\eta}}\\
			& =0 \text{ by (\ref{zeroder})}.  
		\end{align*}
		Thus, this zero moment restriction, together with (\ref{Chain}), implies that the moment is orthogonal.
		
		Suppose now the moment is orthogonal, so again
		by (\ref{Chain}), the moment $\mathbb{E}\left[  g(Z,\lambda_{0})s_{0}(Z)\right] =0$. From the third equality in the last displayed equation, it follows that  
		\begin{equation*}
			\langle S^{*}_{\theta_{0}}g,\delta\rangle_{\mathcal{H_\theta}}+\langle S^{*}_{\eta_{0}}g,b\rangle_{\mathcal{H_\eta}}\\
			=0,  
		\end{equation*}
		for all deviations $h=(\delta,b)$ satisfying (\ref{zeroder}). This implies that there is a constant $c$, possibly zero, such that $S^{*}_{\theta_{0}}g=cr_\theta$ and $S^{*}_{\eta_{0}}g=cr$. 
        
        We now show that the orthogonal moments are relevant iff $c\neq0$. From the regularity of the moment and (\ref{Chain}), it follows that  for any $h \in \overline{\Delta(\lambda_0)}$ such that (\ref{zeroder}) does not hold,
		\begin{align}
			-\frac{d}{d\tau}\mathbb{E}\left[g\left(Z,\lambda_\tau\right)\right] & = \mathbb{E}\left[g\left(Z,\lambda_0\right)s_0\left(Z\right)\right] \nonumber \\ & = \mathbb{E}\left[g\left(Z,\lambda_0\right)S_{\lambda_0}h\left(Z\right)\right] \nonumber  \\ & = \langle S^{*}_{\theta_{0}}g,\delta\rangle_{\mathcal{H_\theta}}+\langle S^{*}_{\eta_{0}}g,b\rangle_{\mathcal{H_\eta}} \nonumber  \\ & = c\langle r_{\theta},\delta\rangle_{\mathcal{H_\theta}}+c\langle r,b\rangle_{\mathcal{H_\eta}} \nonumber  \\ & \neq 0, \label{relth}
		\end{align}
		iff $c\neq0$. $\blacksquare$ \bigskip

        \noindent\textbf{Proof of Corollary \ref{CorLRmomentUHresult}:} By the information equality and definition of $r_\theta$, the condition $S^{*}_\theta g=r_\theta$ is equivalent to 
        \begin{equation*}
			-\frac{d\mathbb{E}\left[  g(Z,\theta_{\tau},\eta_{0})\right]  }{d\tau}=\frac{d\psi(\theta_{\tau},\eta_{0}) }{d\tau}, \quad \textit{for all } \delta\in B(\theta_0).
		\end{equation*}
        By substituting $g$, this is equivalent to
      
        	\begin{equation*}
			\frac{d\mathbb{E}\left[  m_0(Z,\theta_{\tau},\eta_{0})\right]  }{d\tau}=0, \quad \textit{for all } \delta\in B(\theta_0),
		\end{equation*}
         $\blacksquare$ \bigskip

		\noindent \textbf{Proof of Proposition \ref{propLRc1beta}:} By $QV = 0$ a.s., mean independence, and Assumption \ref{RankHDpanelbeta}, $\Gamma_0\mathbb{E}\left[W^{\prime}QY\right] = \Gamma_0 \mathbb{E}\left[W^{\prime}QW\right]\beta_0=C_1'\beta_0$, which shows the identification result.
		
		We next prove the second part of the proposition. The boundness conditions imply that we can compute derivatives of the proposed LR moment function with respect to $\beta$ and $\Gamma$. Next, we verify that \eqref{glrbeta} is indeed an LR moment function. First, we note that, by the exogeneity condition (\ref{exo}) and $QV=0$ a.s., we obtain that 
		$$
		\mathbb{E}\left[g\left(Z,\beta_0, \Gamma_0,\psi_0\right)\right] = \Gamma_0\mathbb{E}\left[W^{\prime}Q\left(Y - W\beta_0\right)\right] =  0,
		$$
		as $\mathbb{E}\left[W^{\prime}Q\left(Y - W\beta_0\right)\right] =  0$. Second, note that the function above has well-defined second moments, by $\mathbb{E}\left[\left|\left|W^{\prime}Q\varepsilon\right|\right|^2\right]$ being finite. Third, notice that by our always-assumed smoothness conditions that allow us to interchange derivatives and integrals, 
		\begin{equation*}
			\begin{split}
				\frac{d}{d \tau}\mathbb{E}\left[g\left(Z,\beta_\tau, \Gamma_0,\psi_0\right)\right] & = \Gamma_0\mathbb{E}\left[W^{\prime}QW\right]\delta_\beta - \Gamma_0\mathbb{E}\left[W^{\prime}QW\right] C_1\left(C^{\prime}_1C_1\right)^{-1}C^{\prime}_1\delta_\beta \\ & = \Gamma_0M\delta_\beta - \Gamma_0MC_1\left(C^{\prime}_1C_1\right)^{-1}C^{\prime}_1\delta_\beta\\ &  = C^{\prime}_1 \delta_\beta - C^{\prime}_1 \delta_\beta \\ & = 0,
			\end{split}
		\end{equation*}
		where $\left|\left|\delta_\beta\right|\right| < \infty$ and $\beta_\tau = \beta_0 + \tau \delta_\beta$. Since the moment does not depend on UH, the previous result shows that it is LR with respect to $\eta_0$. Finally, invoking again the zero mean property of the function, it is not difficult to show that 
		\begin{equation*}
			\begin{split}
				\frac{d}{d \tau}\mathbb{E}\left[g\left(Z,\lambda_0, \Gamma_\tau,\psi_0\right)\right] & = \delta_\Gamma \mathbb{E}\left[W^{\prime}Q\left(Y - W\beta_0\right)\right] =  0,
			\end{split}
		\end{equation*}
		for $\Gamma_\tau = \Gamma_0 + \tau\delta_{\Gamma}$, $\left|\left|\delta_{\Gamma}\right|\right| < \infty$, and the proof is complete. $\blacksquare$ \bigskip
		
		\noindent \textbf{Proof of Proposition \ref{validinferpanelbeta}:} See Section \ref{secasymptotichighpaneldata}. $\blacksquare$ \bigskip
		
		\noindent \textbf{Proof of Proposition \ref{propLRc1alpha}:} By mean independence and $C^{\prime}_2HV = C^{\prime}_2$, we obtain $C^{\prime}_2\mathbb{E}\left[H\left(Y-W\beta_0\right)\right] = \mathbb{E}\left[C^{\prime}_2HV\alpha\right] = \mathbb{E}\left[C^{\prime}_2\alpha\right]$, which shows the identification result. 
		
		As above the boundness conditions in the proposition imply that derivatives of the moment of \eqref{galphalr} are well-defined. Notice that
		\begin{align}
			\mathbb{E}\left[g\left(Z,\lambda_0, \Gamma_0,\psi_0\right)\right] & = \mathbb{E}\left[C_2^{\prime}HV\alpha\right] - \psi_0 \nonumber \\ & = \psi_0 - \psi_0 \nonumber\\ & = 0, \nonumber
		\end{align}
		where the first equality uses $QV=0$ a.s. and (\ref{exo}), and the second equality is implied by the fact that $C^{\prime}_2HV = C^{\prime}_2$. Furthermore, by our maintained assumptions allowing us to interchange derivatives and integrals, we obtain that, for $\left|\left|\delta_\beta\right|\right| < \infty$,
		\begin{equation*}
			\begin{split}
				\frac{d}{d \tau}\mathbb{E}\left[g\left(Z,\lambda\tau, \Gamma_0,\psi_0\right)\right] & = -\mathbb{E}\left[\left(C^{\prime}_2H - \Gamma_0W^{\prime}Q\right)W\right]\delta_\beta \\ & = -C^{\prime}_2S\delta_\beta  + \Gamma_0 M\delta_\beta  \\ &= 0,
			\end{split}
		\end{equation*}
		where $\lambda_\tau = \left(\beta_\tau,\eta_\tau\right)$, with $\beta_\tau = \beta_0 + \tau \delta_\beta$. This result shows that the moment is locally insensitive to $\lambda_0$. Finally, using a similar argument as in the proof of Proposition \ref{propLRc1beta}, we can conclude that \eqref{galphalr} also yields a LR moment that is robust with respect to $\Gamma_0$ using the path $\Gamma_\tau = \Gamma_0 + \tau \delta_\Gamma$, $\left|\left|\delta_{\Gamma}\right|\right| < \infty$, which finishes the proof. $\blacksquare$
		
		\bigskip  \noindent \textbf{Proof of Proposition \ref{validinferpanelalpha}:} See Section \ref{secasymptotichighpaneldata}. $\blacksquare$
		
		\bigskip \noindent   \textbf{Proof of Proposition \ref{propLRvar}:} It follows that 
		$g_0(Z,\beta_0,\omega_0)=vec(\Omega)'\mathbb{H}\{(Y-W\beta_{0})\otimes(Y-W\beta_{0})-S_2\omega_i\}$ satisfies the following conditional moment restriction
		\[
		\mathbb{E}\left[  \left.  g_0(Z,\beta_0,\omega_0)\right\vert \alpha,X\right]
		=vec(\Omega)'(\alpha\otimes\alpha)=\alpha^{\prime}\Omega \alpha,\text{ a.s.}%
		\]
		Thus, taking expectations on both sides concludes the identification result. For local robustness with respect to $\omega_0$, take derivatives of the moment with respect to $\omega_0$ to obtain 
		$$
		\Gamma_{0\omega}B-A,
		$$ 
		which is zero by the definition of $\Gamma_{0\omega}$. Likewise, the derivative with respect to $\beta_0$ is also zero by the definition of $L=-\mathbb{E}\left[  (vec(\Omega)'\mathbb{H}-\Gamma_{0\omega}\mathbb{Q})\{(W\otimes u(\beta_0))+(u(\beta_0)\otimes W)\}\right]$ and $\Gamma_{0\beta}=LM^{\dagger}$. Finally, the LR to $\Gamma_{0\omega}$ and $\Gamma_{0\beta}$ follow from the zero-mean restriction of the errors $U$ and $u(\beta_0)$. $\blacksquare$

		\bigskip 
		\noindent \textbf{Proof of Lemma \ref{lemmaKotlarski}:} 
		Define the function $R(t):=\mathbb{E}\left[  iY_{2}\exp(itY_{1})\right]  ,$ and
		the characteristic functions $\phi_{\alpha}(t)$ and $\phi_{1}(t)$ of
		$\alpha$ and $Y_{1},$ respectively. Let $\phi_{\alpha}^{(1)}(t)$
		denote the derivative of $\phi_{\alpha}(t).$ By the independence of the
		components, and $\mathbb{E}\left[  \varepsilon_{2}\right]  =0,$ it follows that,
		for all $t\in\mathbb{R},$%
		\begin{align*}
			R(t)\phi_{\alpha}(t)  & =\beta_{0}\mathbb{E}\left[  i\alpha%
			\exp(it\alpha)\exp(it\varepsilon_{1})\right]  \phi_{\alpha}(t)\\
			& =\beta_{0}\phi_{\alpha}^{(1)}(t)\phi_1(t).
		\end{align*}
		From this equality, considering an expansion in terms of moments we obtain for
		all $t\in\mathbb{R},$%
		\[
		\left(
		{\displaystyle\sum\limits_{j=0}^{k-1}}
		i^{j+1}\frac{\mathbb{E}\left[  Y_{2}Y_{1}^{j}\right]  }{j!}t^{j}\right)
		\left(
		{\displaystyle\sum\limits_{j=0}^{k-1}}
		i^{j}\frac{\mathbb{E}\left[  \alpha^{j}\right]  }{j!}t^{j}\right)
		=\beta_{0}\left(
		{\displaystyle\sum\limits_{j=0}^{k-1}}
		i^{j+1}\frac{\mathbb{E}\left[  \alpha^{j+1}\right]  }{j!}t^{j}\right)
		\left(
		{\displaystyle\sum\limits_{j=0}^{k-1}}
		i^{j}\frac{\mathbb{E}\left[  Y_{1}^{j}\right]  }{j!}t^{j}\right)  .
		\]
		Multiplying the corresponding expansions and equating the coefficients, we
		arrive at%
		\[%
		{\displaystyle\sum\limits_{j=0}^{k-1}}
		\frac{\mathbb{E}\left[  Y_{2}Y_{1}^{k-1-j}\right]  }{\left(  k-1-j\right)
			!}\frac{\mathbb{E}\left[  \alpha^{j}\right]  }{j!}=\beta_{0}%
		\frac{\mathbb{E}\left[  \alpha^{k}\right]  }{(k-1)!}+%
		\beta_0{\displaystyle\sum\limits_{j=1}^{k-1}}
		\frac{\mathbb{E}\left[  Y_{1}^{k-j}\right]  }{\left(  k-j\right)  !}%
		\frac{\mathbb{E}\left[  \alpha^{j}\right]  }{(j-1)!}%
		\]
		and hence, to the expression for $a_{0k}(Z,\beta_{0},\overrightarrow{\psi}%
		_{0k}).$ From the zero unconditional mean of $a_{0k}$, and the definition
		\begin{equation*}
			c_{k,h}  =\frac{\partial\mathbb{E}\left[  a_{0k}(Z,\beta_{0}%
				,\overrightarrow{\psi}_{0k})\right]  }{\partial\psi_{h}}.
		\end{equation*}
		the expression for the conditional mean of $a_{0k}$ given $\alpha$ follows. 
		$\blacksquare$

		\bigskip 
		\noindent \textbf{Proof of Proposition \ref{propositionkotlarski}:} 
		Note that
		\begin{align*}
			\frac{\partial\mathbb{E}\left[  g_{k}(Z,\beta_{0},\overrightarrow{\psi}%
				_{0k},\Gamma_{0})\right]  }{\partial\beta} &  =b_{k}-\gamma_{k,0}\psi_{01}\\
			&  +\sum_{j=1}^{k-1}\gamma_{k,j}b_{j}\\
			&  =0,
		\end{align*}
		since $\gamma_{k,0}\psi_{01}=b_{k}+\sum_{h=1}^{k-1}\gamma_{k,h}b_{h}.$
		Clearly,
		\[
		\frac{\partial\mathbb{E}\left[  g_{k}(Z,\beta_{0},\overrightarrow{\psi}%
			_{0k},\Gamma_{0})\right]  }{\partial\gamma_{k,j}}=\left\{
		\begin{array}
			[c]{cc}%
			\mathbb{E}\left[  Y_{2}-\beta_{0}Y_{1}\right]  =0 & j=0\\
			\mathbb{E}\left[  a_{0j}(Z,\beta_{0},\overrightarrow{\psi}_{0j})\right]  =0 &
			j>0
		\end{array}
		\right.  .
		\]
		Also,
		\begin{align*}
			\frac{\partial\mathbb{E}\left[  g_{k}(Z,\beta_{0},\overrightarrow{\psi}%
				_{0k},\Gamma_{0})\right]  }{\partial\psi_{0h}}  & =\frac{\partial
				\mathbb{E}\left[  a_{0k}(Z,\beta_{0},\overrightarrow{\psi}_{0k},\Gamma
				_{0})\right]  }{\partial\psi_{0h}}+\sum_{j=1}^{k-1}\gamma_{k,j}\frac
			{\partial\mathbb{E}\left[  a_{0j}(Z,\beta_{0},\overrightarrow{\psi}%
				_{0j})\right]  }{\partial\psi_{0h}}\\
			& =c_{k,h}+\sum_{j=1}^{k-1}\gamma_{k,j}c_{j,h}\\
			& =c_{k,h}+\gamma_{k,h}\beta_{0}+\sum_{j=h+1}^{k-1}\gamma_{k,j}c_{j,h}\\
			& =0,
		\end{align*}
		since
		\[
		\gamma_{k,h}=-\frac{c_{k,j}+\sum_{j=h+1}^{k-1}\gamma_{k,j}c_{j,h}}{\beta_{0}}.
		\]
		These results imply that 
		\[
		\mathbb{E}\left[  g_{k}(Z,\beta_{0},\overrightarrow{\psi}_{0k})\vert\alpha\right]
		=\beta_{0}(\psi_{0k}-\alpha^k)-\beta_{0}\psi_{0k},\qquad k\in\mathbb{N}.
		\]
		
		$\blacksquare$
		
		\bigskip \noindent \textbf{Proof of Proposition \ref{propositionposteriormeans}:} That $g_{0}$ is the solution of (\ref{PM}) follows by substitution and the convolution identity. It remains to show that the function has finite variance. To that end, we embed the function in a bigger space. Denote by $\mathcal{H}_{R}$ the set of functions having the expansion
		\begin{equation*}
			g(y)=\sum_{j=0}^{\infty }g_{j}H_{j}(y/\theta_0),
		\end{equation*}%
		such that 
		\begin{equation*}
			\label{sum}
			\sum_{j=0}^{\infty }\frac{j!g_{j}^{2}}{R ^{j}}<\infty .
		\end{equation*}%
		Under the assumptions of the proposition, $g_0\in\mathcal{H}_{R}$. Furthermore, we also require that $\mathbb{E}\left[ g_{0}^{2}(Y)\right]
		<\infty $. It is known, that $\mathcal{H}_{R}$ is the Mehler's Reproducing Kernel Hilbert Space (RKHS), with kernel
		\begin{equation*}
			K_{R}(x,y)=\exp \left( \frac{2R xy-R^2(x^2+y^2)}{\theta_0^2(1-R)^2 }\right) .
		\end{equation*}%
		Also, from \cite{ferreira2013positive} we have that $\mathbb{E}\left[ K_{R}(Y,Y) \right] <\infty$ is sufficient for $\mathbb{E}\left[ g_{0}^{2}(Y)\right]
		<\infty $, which concludes the proof. $\blacksquare$
		
		\bigskip \noindent \textbf{Proof of Proposition \ref{propAnal}:} The proof follows from Theorem 2.7.1, pg. 49-50, in \cite{lehmann2005testing} after expanding the square in the Gaussian density. $\blacksquare$

	\section{General Asymptotic Theory}
		\label{asymptotictheory}
		\subsection{Preliminary results}
		In this section, we provide theoretical guarantees for the test statistic introduced in Section \ref{sectiontest}. A key result will follow by theoretical results developed by \cite{chernozhukov2022locally}, who study, under general conditions, asymptotic theory for debiased moments.  Let
		$$
		\hat{g}\left(\psi_0\right) = \frac{1}{n}\sum^L_{\ell=1} \sum_{i \in I_\ell} \left[\phi_0\left(Z_i,\hat{\rho}_{1\ell},\psi_0\right) + \phi_1\left(Z_i,\hat{\rho}_{1\ell},\hat{\rho}_{1\ell},\psi_0\right)\right].
		$$
		An important result in the development of our asymptotic theory is to show that
		
		\begin{equation}
			\label{eqopmoment}
			\sqrt{n}\hat{g}(\psi_0)=\frac{1}{\sqrt{n}}\sum_{i=1}^{n}g
			(W_{i},\rho_{01},\rho_{02},\psi_0)+o_{p}(1).
		\end{equation}
		To this end, let $F_0$ be the true distribution of the data and assume
		
		\begin{assumption}
			\label{aconvtozero}
			$\mathbb{E}\left[\left\Vert g(Z,\rho_{01},\rho_{02},\psi_0)\right\Vert ^{2}\right]<\infty$ and
			\begin{itemize}
				\item[i)] $\int\left\Vert \phi_0(z,\hat{\rho}_{1\ell},\psi_{0}%
				)-\phi_0(z,\rho_{01},\psi_{0})\right\Vert ^{2}F_{0}%
				(dz)\overset{p}{\longrightarrow}0;$
				\item[ii)] $\int\left\Vert \phi_1(z,\hat{\rho}_{1\ell},\rho_{02}%
				,\psi_{0})-\phi_1(z,\rho_{01},\rho_{02},\psi_{0})\right\Vert ^{2}%
				F_{0}(dz)\overset{p}{\longrightarrow}0;$
				\item[iii)] $\int\left\Vert \phi_1(z,\rho_{01},\hat{\rho}_{2\ell}%
				,\psi_0)-\phi_1(z,\rho_{01},\rho_{02},\psi_{0})\right\Vert
				^{2}F_{0}(dz)\overset{p}{\longrightarrow}0.$
			\end{itemize}
		\end{assumption}
		The conditions in Assumption \ref{aconvtozero} are general mean-square consistency conditions for $\hat{\rho}_{\ell}$. Let
		\begin{equation}
			\label{delta}
			\hat{\Delta}_{\ell}(z):=\phi_1(z,\hat{\rho}_{1\ell},\hat{\rho}_{2\ell}%
			,\psi_0)-\phi_1(z,\rho_{01},\hat{\rho}_{2\ell},\psi_0)-\phi_1(z,\hat{\rho}_{1\ell},\rho_{02},\psi_{0})+\phi_1(z,\rho
			_{01},\rho_{02},\psi_{0}).
		\end{equation}
		
		\begin{assumption}
			\label{adelta}
			For each $\ell = 1,\cdots,L$, one of the following conditions hold:
			\begin{itemize}
				\item[i)]
				$$
				\sqrt{n}\int\hat{\Delta}_{\ell}(z)F_{0}(dz)\overset{p}{\longrightarrow
				}0,\text{ }\int\left\Vert \hat{\Delta}_{\ell}(z)\right\Vert ^{2}%
				F_{0}(dz)\overset{p}{\longrightarrow}0,
				$$
				\item[ii)] $\sum_{i\in I_{\ell}}\left\Vert \hat{\Delta}_{\ell}%
				(Z_{i})\right\Vert /\sqrt{n}\overset{p}{\longrightarrow}0$,
				\item[iii)] $\sum_{i\in I_{\ell}}\hat{\Delta}_{\ell}(Z_{i})/\sqrt{n}%
				\overset{p}{\longrightarrow}0.$
			\end{itemize}
		\end{assumption}
		Any of the previous conditions impose a rate condition on
		$\hat{\Delta}_{\ell}(z)$ such that  its average goes to zero faster than $\sqrt{n}.$
		
		\begin{assumption}
			\label{adoublefrechet}
			For each $\ell=1,...,L$,
			\begin{itemize}
				\item[i)] $\int%
				\phi_1(z,\rho_{01},\hat{\rho}_{2\ell},\psi_0)F_{0}(dz)=0$, with probability approaching one;
				\item[ii)] $\left|\left|\hat{\rho}_{1\ell} - \rho_{01}\right|\right| = O_p(n^{-1/4})$ and there exists a $C>0$ such that $\left|\left|\bar{g}(\rho_{1},\rho_{01},\psi_0)\right|\right| \leq C \left|\left|\rho_{1} - \rho_{01}\right|\right|^2$ for all $\rho_1$ with $\left|\left|\rho_{1} - \rho_{01}\right|\right|^2$ small enough, or $\sqrt{n}\bar{g}\left(\hat{\rho}_{1\ell}, \rho_{02},\psi_0\right)\overset{p}{\longrightarrow}0.$
			\end{itemize}
		\end{assumption}
		Assumption \textit{i)} requires a zero mean condition for $\phi_1(\cdot,\rho_{01},\hat{\rho}_{2\ell},\psi_0)$. Sufficient conditions to achieve will depend on the particular setting. For instance, in \cite{arganarazconstructioncmr}, a sufficient condition is that $\hat{\rho}_{2\ell}$ belongs to $L_2$, with high probability. Assumption \textit{ii)} imposes that the nuisance parameters collected in $\rho_{01}$ are estimated at a rate faster than $n^{-1/4}$, a requirement that will typically be satisfied by several machine learners under general circumstances and for the mean-square norm. Since $\bar{g}$ is an orthogonal moment, a sufficient condition for the second part of \textit{ii)} is that $\bar{g}(\cdot,\rho_{02},\psi_0)$ is  twice Fréchet differentiable on a neighborhood of $\rho_{01}$ (Proposition 7.3.3 of \cite{luenberger1997optimization}). If this do not hold, we can assume \textit{iii)} instead.  We next state:

		\begin{lemma}
			\label{lopmoment}
			Let Assumptions \ref{aconvtozero}-\ref{adoublefrechet}. Then, Equation \eqref{eqopmoment} holds.
		\end{lemma}
		\bigskip \noindent \textbf{Proof of Lemma \ref{lopmoment}:} The result of Lemma \ref{lopmoment} is analogous to Lemma 8 of \cite{chernozhukov2022locally}. The analogy is as follows. The nuisance parameters $\rho_{01}$ and $\rho_{02}$ are $\gamma_0$ and $\alpha_0$ in \cite{chernozhukov2022locally}, respectively, and our LR function $g$ is $\psi$ in that paper. Accordingly, our functions $g_0$ and $\phi$ are $g$ and $\phi$ in \cite{chernozhukov2022locally}, respectively. The only difference is that we focus on $\psi_0$, which is $\theta_0$ in \cite{chernozhukov2022locally} and this is always fixed. In particular, $\phi(\lambda_0)$ is evaluated at $\psi_0$ instead of an estimator of it, as it would be the case in \cite{chernozhukov2022locally}, so we do not need convergence conditions for it. Considering this, Assumptions \ref{aconvtozero}-\ref{adoublefrechet} are weaker versions of the assumptions required by Lemma 8 of \cite{chernozhukov2022locally}. Hence, the result of our lemma follows. $\blacksquare$
		
		\bigskip Lemma \ref{lopmoment} is analogous to Lemma 8 of \cite{chernozhukov2022locally}, and hence the conditions that our lemma requires are analogous to the assumptions required by such a paper. We leverage their theory to provide general conditions under which our test statistic has good properties asymptotically.
		
		Let $\tilde{g}_i = g\left(Z_i,\rho_0,\psi_0\right) - \mathbb{E}\left[g\left(Z,\rho_0,\psi_0\right)\right]$ and $W_n = \frac{1}{n} \sum^n_{i=1} \tilde{g}_i \tilde{g}_i^{\prime}$, and for simplicity let $\hat{\tilde{g}}_{i\ell}$ be  $\hat{g}_{i\ell}(\psi_0) - \bar{g}_\ell\left(\psi_0\right)$ or $\hat{g}_{i\ell}(\hat{\psi}_\ell) - \bar{g}_\ell\left(\hat{\psi}_\ell\right)$ (as in Remark \ref{SimpWg}), where $\bar{g}_\ell\left(\psi_0\right) = n^{-1}\sum^L_{\ell = 1} \sum_{i \in I_\ell} \hat{g}_\ell(\psi_0)$, depending on how an estimator for $W_g$ is constructed. The following assumption, which accommodates either situation,  is key to show that $\hat{W}_n$ is $\nu_n-$consistent for $\tilde{W}_n$.
		
		\begin{assumption}
			\label{anormgconverg}
			$\left|\left|\hat{\tilde{g}}_{i\ell} - \tilde{g}_i\right|\right|^2 = O_p(\nu_n)$, where $\nu_n$ denotes a non-negative sequence such that $\nu_n \rightarrow 0$, for each $\ell = 1,\cdots,L$ and all $i$.
		\end{assumption}

		\begin{lemma}
			\label{lWconv}
			Let $\mathbb{E}\left[\left\Vert g(Z,\rho_{01},\rho_{02},\psi)- \mathbb{E}\left[g(Z,\rho_{01},\rho_{02},\psi)\right]\right \Vert ^{2}\right]<\infty$, for all $\psi$. In addition, let Assumption \ref{anormgconverg} hold. Then, $\left|\left|\check{W}_n - W_g\right|\right| = O_p(\nu_n)$.
		\end{lemma}

		\bigskip \noindent \textbf{Proof of Lemma \ref{lWconv}:}  By the triangle inequality and the Cauchy-Schwarz's inequality,
		\begin{equation*}
			\begin{split}
				\left|\left|\check{W}_n - W_n\right|\right| & \leq \sum^L_{\ell = 1} \frac{1}{n} \sum_{i \in I_\ell}\left(\left|\left|\hat{\tilde{g}}_{i\ell} - \tilde{g}_i\right|\right|^2 + 2\left|\left|\tilde{g}_i\right|\right|\left|\left|\hat{\tilde{g}}_{i\ell} - \tilde{g}_i\right|\right|\right) \\ & \leq \underbrace{\sum^L_{\ell = 1} \frac{1}{n} \sum_{i \in I_\ell}\left|\left|\hat{\tilde{g}}_{i\ell} - \tilde{g}_i\right|\right|^2}_{O_p(\nu_n)}  + \underbrace{2 \sum^L_{\ell = 1} \left(\frac{1}{n} \sum_{i \in I_\ell} \left|\left|\tilde{g}_i\right|\right|^2\right)^{1/2}}_{O_p(1)} \underbrace{\left(\frac{1}{n} \sum_{i \in I_\ell}\left|\left|\hat{\tilde{g}}_{i\ell} - \tilde{g}_i\right|\right|^2\right)^{1/2}}_{O_p(\nu_n)} \\  & = O_p(\nu_n),
			\end{split}
		\end{equation*}
		Furthermore, by the Central Limit Theorem, $\left|\left|W_n - W_g \right|\right| = O_p\left(n^{-1/2}\right)$. Then, the conclusion follows by the triangle inequality. $\blacksquare$
		
		\bigskip We next show that the estimator of the Moore-Penrose inverse, proposed by \eqref{eq:trunc_matrix}, is consistent for $W^{\dagger}_g$.
		
		\begin{lemma}
			\label{lrankWhat}
			Let the assumptions of Lemma \ref{lWconv} hold.  Then, $\hat{W}_n \overset{p}{\longrightarrow} W_g$ and $rank\left(\hat{W}_n\right) = rank\left(W_g\right)$ with probability approaching one, where $\hat{W}_n$ is defined in \eqref{eq:trunc_matrix}. As a result, $\left|\left|\hat{W}^{\dagger}_n - W^{\dagger}_g\right|\right| = o_p(1)$.
		\end{lemma}
		
		\bigskip \noindent \textbf{Proof of Lemma \ref{lrankWhat}:} The result follows by Proposition S1 in \cite{lee2024locally}. $\blacksquare$

		\subsection{Local Asymptotic Analysis of Size and Power}
		\label{LocalPower}
		\bigskip Now we can analyze the asymptotic properties, in terms of asymptotic size and  local power, of the test statistic of Section \ref{sectiontest}. We are interested in testing
		$$
		H_0: \psi(\lambda_0) = \psi_0\;\;\;\; vs. \;\;\; H_{1n}: \psi(\lambda_0) = \psi_{0,n} \equiv\psi_0+1/\sqrt{n}.
		$$
		Notice that we are restricting attention to local alternatives, and a maintained assumption throughout is that the corresponding scores satisfy the properties we discussed in the main text. In particular, under $H_0$, we consider scores $s = s_0 \in \overline{\mathcal{T}_0}$ while under the local alternative we have scores $s \in \overline{\mathcal{T}}$ such that $s = h_\psi + s_0$, where $0\neq h_\psi \in \tilde{S}_\psi$. Additionally, we consider measures $\mathbb{P} \in \mathcal{P}_0$ under $H_0$ while $\mathbb{P}_{\psi_{0,n}} \in \mathcal{P}$ under $H_{1n}$. The local asymptotic power function of the test statistic is defined as
		$$
		\pi_g\left(s\right) := \lim_{n\rightarrow \infty} \mathbb{P}_{\psi_{0,n}}\left(\hat{C}_n\left(\psi_0\right) > c_{\zeta,\hat{k}_g}\right),
		$$
		for any function $g \in L^0_2$, where recall that $c_{\zeta,d}$ is the $1 - \zeta$ quantile of a $\chi^2_d$, a chi-square distribution with $d$ degrees of freedom. Additionally, with some abuse of notation, let $\chi^2_d(\omega)$ denote a random variable that presents a chi-square distribution with $d$ degrees of freedom and non-centrality parameter $\omega$.
		
		\begin{proposition}
			\label{psizetest}
			Suppose  i) $k_g = rank\left(W_g\right) >0$; ii) \eqref{eqopmoment} holds; iii)  $\hat{W}_n \overset{p}{\longrightarrow} W_g$ and $rank\left(\hat{W}_n\right) = rank\left(W_g\right)$ with probability approaching one. Then, under $H_0$, $\pi_g(s_0) = \zeta$.
		\end{proposition}
		
		\bigskip \noindent \textbf{Proof of Proposition \ref{psizetest}:} For all $s_0 \in \overline{\mathcal{T}}_0$, by orthogonality, we have $\mathbb{E}\left[g\left(Z,\rho_0,\psi_0\right)s_0(Z)\right]=0$. Hence, by LeCam's third lemma, under $P_{\psi_{0,n}}$
		
		$$
		\sqrt{n} \mathbb{E}_n\left[g\left(Z_i,\rho_0,\psi_0\right)\right] \overset{d}{\longrightarrow} Z \sim N_k\left(0,W_g\right),
		$$
		where $\mathbb{E}_n$ denotes the sample mean operator. Since $\left|\left|\hat{W}^{\dagger}_n - W^{\dagger}_g\right|\right| = o_p(1)$, which is a consequence of \textit{iii)} (see Proposition S1 in \cite{lee2024locally}), by the continuous mapping theorem and Theorem 7.3 in \cite{rao1972generalized}, under $\mathbb{P}_{\psi_{0,n}}$,
		$$
		n\mathbb{E}_n\left[g\left(Z_i,\rho_0,\psi_0\right)\right] \hat{W}^{\dagger}_n \mathbb{E}_n\left[g\left(Z_i,\rho_0,\psi_0\right)\right] \overset{d}{\longrightarrow} ZW^{\dagger}_gZ \sim \chi^2_{k_g}.
		$$
		Notice that, by assumption, $\mathbb{P}\left(c_{\zeta,\hat{k}_g} = c_{\zeta,k_g}\right) = \mathbb{P}\left(\hat{k}_g = k_g\right) \rightarrow 1$. This implies that $c_{\zeta,\hat{k}_g} \overset{\mathbb{P}_{\psi_{0,n}}}{\longrightarrow} c_{\zeta,k_g}$, by contiguity and LeCam's first lemma. Thus, by the continuous mapping theorem, we obtain $\hat{C}_n\left(\psi_0\right) - c_{\zeta,\hat{k}_g} \overset{d}{\longrightarrow} ZW^{\dagger}_gZ - c_{\zeta,k_g}$. By the Portmanteau theorem, under  $H_{0},$
		$$
		\mathbb{P}_{\psi_{0,n}}\left(\hat{C}_n\left(\psi_{0}\right) - c_{\zeta,\hat{k}_g} > 0\right) \longrightarrow \mathbb{L}\left(ZW^{\dagger}_gZ - c_{\zeta,k_g} >0\right) = \zeta,
		$$
		where $\mathbb{L}$ is the law of $ZW^{\dagger}_gZ$. $\blacksquare$
		
		\bigskip \noindent Remark that \textit{ii)}  incorporates Lemma \ref{lopmoment} while \textit{iii)} is implied by Lemma \ref{lrankWhat}. The following proposition states the validity of confidence intervals $CR_\zeta$ constructed by inverting out rest statistic.
		
		\begin{proposition}
			\label{pcivalidity}
			Let the assumptions in Proposition \ref{psizetest} hold. Then,
			$$
			\lim_{n \rightarrow \infty} \mathbb{P}_{\psi_{0,n}}\left(\psi_0 \in CR_\zeta\right) = 1-\zeta.
			$$
		\end{proposition}
		
		\bigskip \noindent \textbf{Proof of Proposition \ref{pcivalidity}:}  Simply notice that for any $s_0 \in \overline{\mathcal{T}}_0$,
		$$
		\lim_{n \rightarrow \infty} \mathbb{P}_{\psi_{0,n}}\left(\psi_0 \in CR_\zeta\right) = 1-  \lim_{n\rightarrow \infty} \mathbb{P}_{\psi_{0,n}}\left(\hat{C}_n\left(\psi_0\right) > c_{\zeta,\hat{k}_g}\right).
		$$
		Then, apply Proposition \ref{psizetest} to obtain the desired result. $\blacksquare$
		
		\bigskip Next, we turn to local asymptotic power considerations. Define a drift parameter
		$$
		\mu \equiv \mu\left(g,h_\psi\right): = \mathbb{E}\left[g\left(Z,\rho_0,\psi_0\right)h_\psi(Z)\right],
		$$
		and let $\Delta = \mu^{\prime}W^{\dagger}_g\mu$. As we show below, $\mu$ can equivalently be written as the square of the $L_2-$norm of $\Pi_{\mathcal{V}}h_\psi$, where $\mathcal{V}$ is the linear span of $g$ in $L_2$, i.e., $\Delta = \left|\left|\Pi_\mathcal{V}h_{\psi}\right|\right|^2_2$.
		
		\begin{proposition}
			\label{ppowertest}
			Suppose  i) $k_g = rank\left(W_g\right) >0$; ii) \eqref{eqopmoment} holds; iii)  $\hat{W}_n \overset{p}{\longrightarrow} W_g$ and $rank\left(\hat{W}_n\right) = rank\left(W_g\right)$ with probability approaching one. Then, under $H_{1n}$, the local asymptotic power function $\pi_g(s)$ satisfies
			$$
			\pi_g(s) = \mathbb{L}\left(\chi^2_{k_g}(\Delta) > c_{\zeta,k_g}\right),
			$$
			where $\mathbb{L}$ is the law of $\chi^2_{k_g}(\Delta)$. Additionally,
			\begin{itemize}
				\item[i)] If $\Pi_\mathcal{V}h_{\psi}(Z) \neq 0$ a.s.,  $\pi_g(s) > \zeta$;
				\item[ii)] If $\Pi_\mathcal{V}h_{\psi}(Z) = 0$ a.s.,  $\pi_g(s) = \zeta$;
				\item[iii)] Let $\left\{\tilde{s}_{j\psi}\right\}^{d_\psi}_{j=1}$, $d_\psi < \infty$, be an orthonormal basis for $\tilde{S}_\psi$ such that $h_\psi = \tilde{s}_\psi^{\prime}\delta_s$, with $\delta_s \in \mathbb{R}^{d_\psi}$, for all $h_\psi \in \tilde{S}_\psi$.  Then, observe that $\pi_g$ only depends on $s$ through $\delta_s$, and
				\begin{equation}
					\label{upperboundpower}
					\inf_{\delta_s \in S(a)} \pi_g(\delta_s) \leq \inf_{\delta_s \in S(a)} \pi_{\tilde{s}_\psi}(\delta_s),
				\end{equation}
			\end{itemize}
			where $S(a) = \left\{\delta \in \mathbb{R}^{d_\psi}: \delta^{\prime} \delta \geq a\right\}$.
		\end{proposition}
		
		\bigskip \noindent \textbf{Proof of Proposition \ref{ppowertest}:} Similarly, as in the proof of Proposition \ref{psizetest}, by LeCam's third lemma, we have, under $\mathbb{P}_{\psi_{0,n}}$
		\begin{equation*}
			\sqrt{n} \mathbb{E}_n\left[g\left(Z_i,\rho_0,\bar{\psi}\right)\right] \overset{p}{\longrightarrow} Z \sim  N_k\left(\mathbb{E}\left[g\left(Z,\rho_0,\bar{\psi}\right)s(Z)\right],W_g\right), 	
		\end{equation*}
		and notice, by orthogonality,
		$$
		\mathbb{E}\left[g\left(Z,\rho_0,\bar{\psi}\right)h_\psi(Z)\right] = \mathbb{E}\left[g\left(Z,\rho_0,\bar{\psi}\right)h_\psi(Z)\right].
		$$
		Then, under the maintained assumptions, by the continuous mapping theorem and Theorem 7.3 in \cite{rao1972generalized}, under $\mathbb{P}_{\psi_{0,n}}$,
		$$
		n\mathbb{E}_n\left[g\left(Z_i,\rho_0,\bar{\psi}\right)\right] \hat{W}^{\dagger}_n \mathbb{E}_n\left[g\left(Z_i,\rho_0,\bar{\psi}\right)\right] \overset{d}{\longrightarrow} ZW^{\dagger}_gZ \sim \chi^2_{k_g}(\Delta).
		$$
		Observe that $\Delta = \beta^{\prime}W_g \beta$, where $\beta = W^{\dagger}_g \mu$ is the minimum norm solution to
		$$
		\mathbb{E}\left[g\left(Z,\rho_0,\bar{\psi}\right)g\left(Z,\rho_0,\bar{\psi}\right)^{\prime}\right]\beta = \mathbb{E}\left[g\left(Z,\rho_0,\bar{\psi}\right)h_\psi(Z)\right].
		$$
		Notice that $\Pi_{\mathcal{V}} h_\psi(Z) = \beta^{\prime}g\left(Z,\rho_0,\bar{\psi}\right)$. As a result, we obtain, $\Delta = \left|\left|\Pi_{\mathcal{V}} h_\psi\right|\right|^2_2$. Using the same arguments as in the proof of Proposition \ref{psizetest}, we conclude that
		\begin{equation}
			\label{lawrest}
			\mathbb{P}_{\psi_{0,n}}\left(\hat{C}_n\left(\bar{\psi}\right) - c_{\zeta,\hat{k}_g} > 0\right) \longrightarrow \mathbb{L}\left(ZW^{\dagger}_gZ - c_{\zeta,k_g} >0\right),
		\end{equation}
		where $\mathbb{L}$ is the law of $ZW^{\dagger}_gZ$. The result in \eqref{lawrest} implies the asymptotic limit of the local power function $\pi_g$ and conclusions \textit{i)} and \textit{ii)}.

		Statement \textit{iii)} is analogous to Lemma 3.2 (ii) in \cite{chen2018overidentification}. It is analogous since the asymptotic local power function $\pi_{\tilde{s}_\psi}$ is that of a test based on $\tilde{s}_{\psi}$, which directs its power specifically towards $\tilde{S}_\psi = \overline{\mathcal{T}_0}^{\perp} \cap \overline{\mathcal{T}}$, just as the statistic $\hat{\mathbb{G}}_n$ based on an orthonormal basis for $\overline{T}(P)^{\perp} \cap \overline{M}(P)^{\perp}$ in \cite{chen2018overidentification}. Note that Assumption 2.1 and 3.1 of \cite{chen2018overidentification}, required by Lemma 3.2 (ii), holds in our context. To be precise, Assumption 2.1 asks for iid that and linear tangent spaces, which we assume throughout the paper. Assumption 3.1 requires that $n^{-1/2}\mathbb{E}_n\left[\tilde{s}_{j\psi}(Z_i)\right]$ is bounded uniformly in $j$ and that it has a tight nondegenerate centered Gaussian measure asymptotically, which holds as each $\tilde{s}_{j\psi} \in L^0_2$. The result in \cite{chen2018overidentification} implies that we shall consider scores such that $s \in \overline{T}$ and $\left|\left|\Pi_{\overline{\mathcal{T}_0}^\perp}s\right|\right|^2_2 \geq a$,\footnote{where $\left|\left|f\right|\right|_2$ denotes mean square norm of a generic $f \in L_2$.} but in our case since $s = \tilde{s}_\psi^{\prime}\delta_s + s_0 $, for any $s \in \overline{\mathcal{T}}$ and some $\delta_s \in \mathbb{R}^{d_\psi}$, and $\tilde{s}_\psi$ is an orthonormal basis for $\tilde{S}_\psi$, we simply obtain $\left|\left|\Pi_{\overline{\mathcal{T}_0}^\perp}s\right|\right|^2_2 = \delta_s^{\prime}\delta_s$, which explains the restriction in $S(a)$. Hence, \textit{iii)} follows, which completes the proof. $\blacksquare$

		\section{Regularity Conditions for Models with UH}
		\label{regpaneldatagral}
		Regularity of our model is pinned down by three conditions: \textit{i)} Differentiability in Quadratic Mean; \textit{ii)} Linearity of the tangent space of the model; and \textit{iii)} Continuity of the score operator. We will be precise about these three conditions below and see that \textit{ii)} is implied by \textit{i)} in our setting. 
 		
		We define the measure $\mu$ on $\mathcal{Z} = \mathcal{Y} \times \mathcal{X}$, where $\mathcal{Y}$ is the support of $Y$ and $\mathcal{X}$ is the support of $X$. Let $\mathcal{A}_0$ be the support of $\alpha$. Define $\mu := \mu_Y \times \mu_X$, where $\mu_Y$ is a $\sigma$-finite measure, and $\mu_X$ is the probability distribution of $X$. The density of the data $Z = (Y,X)$, wrt the measure $\mu$ is given by 
		\begin{equation}
			f_{\lambda_{0}}\left(  z\right)  =\int f_{Y|\alpha,X}\left(  y|\alpha
			,x;\theta_{0}\right)  \eta_{0}(\alpha|x)d\alpha,\label{mixt2a}%
		\end{equation}
		where $\eta_0$ is the conditional density of $\alpha$, depending on $X = x$. For simplicity, we consider the Lebesgue measure, but other measures are allowed.
		
		 Let $\mathbf{H}=\mathcal{H}_{\theta}\times \mathcal{H}_{\eta} $ be a Hilbert space with inner product $\langle(\delta_{1},b_{1}%
		),(\delta_{2},b_{2})\rangle_{\mathbf{H}}:=\langle\delta_{1},\delta_{2}%
		\rangle_{\mathcal{H}_{\theta}}+\langle b_{1},b_{2}\rangle_{\mathcal{H}_{\eta}%
		}$, and where $\mathcal{H}_{\theta}$ and $\mathcal{H}_{\eta}\ $are Hilbert
		spaces endowed with the inner products $\langle\cdot,\cdot\rangle
		_{\mathcal{H}_{\theta}}$ and $\langle\cdot,\cdot\rangle_{\mathcal{H}_{\eta}},$
		respectively$.$ Since we are considering a fixed effects approach, let $\mathcal{H}_{\eta}=L_2(\eta_0\times \mu_X)$.  The inner product associated to $\mathcal{H}_{\eta}$ is then simply $\langle b_{1},b_{2}\rangle_{\mathcal{H}_{\eta}%
		}=\mathbb{E}\left[  b_{1}(\alpha,X)b_{2}(\alpha,X)\right]$. 
		
		 Let $B(\eta_0) \subseteq \mathcal{H}_\eta$ be  the set of measurable functions $b$ in $L_2(\eta_0 \times \mu_X)$, that are bounded with\footnote{Note that $B(\eta_0)$ is dense in $\mathcal{H}_\eta \cap L^0_2(\eta_0)$. }
		$$
		\int b\left(\alpha,x\right) \eta_0\left(\left. \alpha \right|x\right) d\alpha = 0, \;\;\; \mu_X-\text{a.s.}
		$$
		We define the paths through $\eta_0$  as
		$$
		\eta_\tau := \eta_0 (1+\tau b),
		$$
		where $\tau \in \left(0, \epsilon\right)$. Note that $\eta_\tau$ is a density for $\tau$ sufficiently small. Let $B(\theta_0)\subseteq \mathcal{H}_\theta$, and let $\Theta$ be an open set, we define 
		$$
		\theta_{\tau} : = \theta_0 + \tau \delta, \;\;\; \delta \in B(\theta_0),
		$$
		where the deviation $\delta$ is such that $\theta_{\tau} \in \Theta$ and $f_{Y|\alpha,X}$ is a bounded density for $\tau$ sufficiently small. Let $\mathcal{D}$ be the space of strictly positive functions in $L^{\infty}(\mu \times \eta_0)$ (with some abuse of notation).

		\begin{assumption}
			\label{aregconddqm}
			i) The mapping $\tau \in \left[0,\varepsilon\right) \mapsto \eta_\tau$ is differentiable in the sense that, as $\tau$ goes to zero (from non-negative values), for some $b \in B(\eta_0)$,
			$$
			\int \int \left[\frac{\eta^{1/2}_\tau\left(\alpha|x\right) - \eta^{1/2}_0\left(\alpha|x\right)}{\tau} - \frac{1}{2}b\left(\alpha,x\right)\eta^{1/2}_0(\alpha|x)\right]^2d\alpha  d\mu_X(x)\rightarrow 0, 
			$$
			ii) $f_{Y|\alpha,X}: \Theta \mapsto \mathcal{D}$ is Fréchet differentiable with Fréchet differential $\Delta \left(Z,\alpha,\delta\right)$, which is linear and continuous in $\delta$, tangentially to $B(\theta_0)$. In addition, let $\Delta(Z,\alpha,\delta)$ be measurable; (iii) There exists a dominating function $M$ such that 
			$$
			\frac{\Delta^2(Z,\alpha,\delta)}{f_{Y|\alpha,X}(Y|\alpha,X;\theta_0)} \leq M(Z,\alpha), \;\;\; a.s;
			$$
			where
			$$
			\int \int \left|M(z,\alpha) \right| \eta_0(\alpha|x) d\alpha d\mu(z) < \infty.
			$$
            iii) $f^{-1}_{Y|\alpha,X} \in L_2\left(\mu \times \eta_0\right)$.
		\end{assumption}
		\bigskip  Note that Assumption \ref{aregconddqm} \textit{ii)} implies (with some abuse of notation) that $f^{1/2}_{Y|\alpha,X}(y|\alpha,x;\theta_0): \mathcal{D} \mapsto L^2(\eta_0 \times \mu)$ is Fréchet differentiable with derivative $f_{Y|\alpha,X}^{-1/2}/2$. This implies that, as $\tau\rightarrow0$,
		\begin{equation}
			\label{eq:dqmcond}
			\begin{split}
				& \int \int \left[f^{1/2}_{Y|\alpha,X}(y|\alpha,x;\theta_0 + \tau \delta) - f^{1/2}_{Y|\alpha,X}(y|\alpha,x;\theta_0) - \frac{1}{2} [\ell_\theta\delta](z,\alpha) f^{1/2}_{Y|\alpha,X}(y|\alpha,x;\theta_0)\right]^2 \eta_0(\alpha|x) d\alpha d\mu(z) \\ & \rightarrow 0,
			\end{split}
		\end{equation}
		where 
		$$
		 [\ell_\theta\delta](Z,\alpha) := \frac{\Delta(Z,\alpha,\delta)}{f_{Y|\alpha,X}(Y|\alpha,X;\theta_0)} \equiv \ell_\theta(Y\vert\alpha,X,\theta_0	).
		$$
		
		\begin{proposition}
			\label{DQMpanelgral}
			Let Assumption \ref{aregconddqm} i)-ii) hold. Then, 
			\begin{equation}
				\label{eq:dqmcondbig}
				\int \left[\frac{f^{1/2}_{\lambda_{\tau}}(z) - f^{1/2}_{\lambda_{0}}(z)}{\tau} - \frac{1}{2}\left([S_\theta \delta] (z) + [S_\eta b](z)\right)f^{1/2}_{\lambda_{0}(z)}\right]^2 d\mu(z) \rightarrow 0, 
			\end{equation}
			where 
			\begin{equation*}
				\begin{split}
					S_\theta \delta (Z) &= \mathbb{E}\left[\left. \ell_\theta\delta(Z,\alpha)\right|Z\right], \\
					S_\eta b (Z) & = \mathbb{E}\left[\left.b(\alpha,X)\right|Z\right].
				\end{split}
			\end{equation*}
		\end{proposition}
		
		\bigskip \noindent \textbf{Proof of Proposition \ref{DQMpanelgral}:} By iterated expectations $S_\eta b \in L^0_2$. We next show that $S_\theta \delta \in L_2$. Note that, by the Cauchy-Schwartz's inequality,
		\begin{equation*}
			\begin{split}
				\int \left[S_\theta \delta\right]^2(z) f_{\lambda_0}(z) d\mu(z) \leq \int \int \frac{\Delta^2\left(z,\alpha,\delta\right)}{f_{Y|\alpha,X}\left(y|\alpha,x;\theta_0\right)} \eta_0\left(\alpha|x\right) d\alpha d\mu(z),
			\end{split}
		\end{equation*}
		which is finite by \textit{(ii)}. By Theorem 5.13 of \cite{vandervaartthesis}, $S_\theta \delta$ is indeed a score for model $\theta \mapsto f_{(\theta,\eta_0)}$. In fact, observe that, by \textit{(i)} and \eqref{eq:dqmcond},
		\begin{equation*}
			\begin{split}
				& \int \int \left[f^{1/2}_{Y|\alpha,X}\left(y|\alpha,x;\theta_0 + \tau \delta\right)\eta_\tau\left(\alpha|x\right) - f^{1/2}_{Y|\alpha,X}\left(y|\alpha,x;\theta_0\right)\eta_0\left(\alpha|x\right)  \right.\\ & \left. - \frac{1}{2}\left(\frac{\Delta\left(z,\alpha,\delta\right)}{f_{Y|\alpha,X}\left(y|\alpha,x;\theta_0\right)} + b\left(\alpha,x\right)\right)f^{1/2}_{Y|\alpha,X}\left(y|\alpha,x;\theta_0\right)\eta^{1/2}_0(\alpha|x)\right]^2 d\alpha d\mu(z) \\ & \leq 3 \int \int \left[\tau^{-1}\left(f^{1/2}_{Y|\alpha,X}\left(y|\alpha,x;\theta_0 + \tau \delta\right) - f^{1/2}_{Y|\alpha,X}\left(y|\alpha,x;\theta_0\right)\right) \right. \\ & \left. - \frac{1}{2} \frac{\Delta\left(z,\alpha,\delta\right)}{f_{Y|\alpha,X}\left(y|\alpha,x;\theta_0\right)}f^{1/2}_{Y|\alpha,X}\left(y|\alpha,x;\theta_0\right)\right]^2\eta_0(\alpha|x)d\alpha d\mu(z) \\ & + 3 \int \int \left[\tau^{-1}\left(\eta^{1/2}_\tau\left(\alpha|x\right) - \eta^{1/2}_0(\alpha|x)\right) - \frac{1}{2}b\left(\alpha,x\right)\eta^{1/2}_0\left(\alpha|x\right)\right]^2f_{Y|\alpha,X}\left(y|\alpha,x;\theta_0 + \tau \delta\right) d\alpha d\mu(z) \\ & + \frac{3}{4} \int \int \left[b\left(\alpha,x\right)\eta^{1/2}_0(\alpha|x)\right]^2\left[f^{1/2}_{Y|\alpha,X}\left(y|\alpha,x;\theta_0 \tau \delta\right) - f^{1/2}_{Y|\alpha,X}\left(y|\alpha,x;\theta_0\right)\right]^2 d\alpha d\mu(z) \\ & = o(1) + C\int \int \left[f^{1/2}_{Y|\alpha,X}\left(y|\alpha,x;\theta_0 \tau \delta\right) - f^{1/2}_{Y|\alpha,X}\left(y|\alpha,x;\theta_0\right)\right]^2 \eta_0(\alpha|x) d\alpha d\mu(z) \\ & \rightarrow 0,
			\end{split}
		\end{equation*}
		where the last display again uses \eqref{eq:dqmcond}. Therefore, \eqref{eq:dqmcondbig} follows By Lemma 5.18 of \cite{vandervaartthesis}.\footnote{Note that $S_\theta \delta$ satisfies the zero-mean restriction, which can be shown similarly as Theorem 7.2 of \cite{vandervaart98}. } $\blacksquare$
		
		\bigskip Observe that both $S_\theta$ and $S_\eta$ are linear operators from $B(\theta_0)$ and $B(\eta_0)$, respectively, to $L_2$ . In addition, an implication of Proposition \ref{DQMpanelgral} is that the score of the model corresponding to the path $\lambda_\tau$ satisfies the chain rule, specifically, 
		$$
		S_\lambda h(Z):= S_\theta \delta (Z)  + S_\eta b (Z),\;\;\; h=(\delta,b) \in \Delta(\lambda_{0}) \subset \mathbf{H},
		$$
		where $\Delta(\lambda_{0}) := B(\theta_0) \times B(\eta_0)$. Hence, the tangent space of the model, given by $\mathcal{T}=\{S_{\lambda}h:h\in \mathbf{H}\}$ is linear, an important regularity property of this model. 
		
		

        \bigskip \begin{proposition}
        \label{Adjointsgral}

        If Assumption \ref{aregconddqm} iii) hold, then $S_\theta: B(\theta_0) \mapsto L_2$ is a continuous operator. Moreover, the adjoint operators of $S_{\theta}$ and $S_{\eta}$ are well-defined and given by, respectively, $S^{*}_\theta = \ell^{*}_\theta$ and
        \begin{equation*}
            \begin{split}
                 S^{*}_\eta g\left(\alpha,X\right) & = \mathbb{E}\left[\left.g(Z)\right|\alpha,X\right],\\
            \end{split}
        \end{equation*}
        where $\ell^{*}_\theta$ is the adjoint of $\ell_\theta$ restricted to $L_2$.
        \end{proposition}

\bigskip \noindent \textbf{Proof of Proposition \ref{Adjointsgral}:} To show continuity of $S_\theta$, observe that 
\begin{equation*}
    \begin{split}
        \left|\left| S_\theta \delta\right|\right|^2 & \leq \mathbb{E}\left[\left[\ell_\theta \delta \right]^2(Z,\alpha)\right] \\ & = \mathbb{E}\left[\frac{\left|\Delta(Z,\alpha,\delta)\right|^2}{f^2_{Y|\alpha,X}(Y|\alpha,X;\theta_0)}\right] \\ &  \leq C \mathbb{E}\left[\frac{1}{f^2_{Y|\alpha,X}(Y|\alpha,X;\theta_0)}\right] \left|\left| \delta\right|\right|^2_{\mathcal{H}_\theta},
    \end{split}
\end{equation*}
where $\left|\left| \delta\right|\right|^2_{\mathcal{H}_\theta} = \left<\delta,\delta\right>_{\mathcal{H}_\theta}$, the first inequality follows by Jensen's inequality and the law of iterated expectations, the second equality is by definition, and the last inequality uses that $\Delta: B(\theta_0) \mapsto \mathcal{D}$ is bounded. Note that the previous displays also implies that $\ell_\theta: B(\theta_0) \mapsto L_2(\mu \times \eta_0)$ is continuous. Hence, its adjoint, $\ell^{*}_\theta: L_2(\mu \times \eta_0)  \mapsto B(\theta_0)$ exists and is well-defined \citep[see, e.g., Theorem 2.21 of][]{CARRASCO20075633}. Additionally, 
\begin{equation*}
\begin{split}
      \mathbb{E}\left[S_\theta \delta(Z) g(Z)\right] & = \mathbb{E}\left[\ell_\theta \delta \left(Z,\alpha\right)g\left(Z\right)\right] \\ & = \left<\delta,\ell^{*}_\theta g\right>_{\mathcal{H}_\theta},
\end{split}
\end{equation*}
  where the first equality follows by the law of iterated expectations and the second one uses the definition of the adjoint of $\ell_\theta$, defined on $L_2$.

Furthermore, as $b \in L_2(\eta_0)$, $S_\eta$ is bounded and by the law of iterated expectations,
        \begin{equation*}
            \begin{split}
                \mathbb{E}\left[\left[S_\eta\right] b(Z) g(Z)\right] &= \mathbb{E}\left[\mathbb{E}\left[\left. b\left(\alpha,X\right)\right|Z\right]g(Z)\right] \\ & = \mathbb{E}\left[\mathbb{E}\left[\left. g(Z)\right|\alpha,X\right] b\left(\alpha,X\right)\right]\\ & = \left<b,S^{*}_\eta g \right>_{\mathcal{H}_\eta}.
            \end{split}
        \end{equation*}
where the last equality follows by definition of $\left<\cdot,\cdot \right>_{\mathcal{H}_\eta}$. This completes the proof. $\blacksquare$ 

			\section{Asymptotic Results of High-Dimensional Random Coefficient Model}
	\label{secasymptotichighpaneldata}
	In this section, we provide theoretical asymptotic guarantees for the inference procedure implemented by the algorithms outlined in Section \ref{sectionhighdimensionalpanel}.

	\begin{assumption}
		\label{regcondpanelbeta}
		Suppose that \textit{i)}  $\left|\left|\beta\right|\right| < C$, $\left|\left|\check{M}^{\dagger}\right|\right| < C$, for $\left|\left|\beta - \beta_0 \right|\right|$  and $\left|\left|\check{M}^{\dagger} - M^{\dagger} \right|\right|$ small enough; \textit{ii)} $\|C_1\|$, $\mathbb{E}\left[\|WQ_{C_1}\|^2\right]$,  and $\mathbb{E}\left[\left|\left|W^{\prime}QWH_{C_1} \right|\right|^2\right]$ are all bounded;  \textit{iii)} $\left|\left|\hat{\beta}_\ell - \beta_0 \right|\right| = O_p\left(n^{-1/4}\right)$ and $\left|\left|\hat{M}^{\dagger}_\ell - M^{\dagger} \right|\right| = O_p\left(n^{-1/4}\right)$, for each $\ell = 1,\cdots,L$; \textit{iv)} $rank\left(W_g\right) >0$. 
	\end{assumption}
	
	\bigskip \noindent \textbf{Proof of Proposition \ref{validinferpanelbeta}:} To prove the results of the proposition, we need to verify the conditions of Lemmas \ref{lopmoment} and \ref{lrankWhat}. First, we verify Assumptions \ref{aconvtozero}-\ref{adoublefrechet}. Notice that, in this case, we have 
	\begin{equation*}
		\begin{split}
			g\left(Z,\rho_{1},\rho_{2},\psi\right) & = \rho_2 W^{\prime}Q\left(Y - WH_{C_1}\psi - WQ_{C_1}\beta\right) \\ 
			& = \phi_0\left(Z,\rho_{1},\psi_0\right) + \phi_1\left(Z,\rho_{1},\rho_{2},\psi\right), \\
		\end{split}
	\end{equation*}
	where
	\begin{equation*}
		\begin{split}
			\phi_0\left(Z,\rho_{1},\psi\right) & = Y - WH_{C_1}\psi - WQ_{C_1}\beta, \\
			\phi_1\left(Z,\rho_{1},\rho_{2},\psi\right) & = \left(\rho_2 W^{\prime}Q - I_T\right)\left(Y - WH_{C_1}\psi - WQ_{C_1}\beta\right),
		\end{split}
	\end{equation*}
	with $\rho_{1} = \beta$. Hence, by Assumption \ref{RankHDpanelbeta} \textit{i)}-\textit{ii)},
	\begin{equation*}
		\begin{split}
			\mathbb{E}\left[\left|\left|g\left(Z,\rho_{01},\rho_{02},\psi_0\right)\right|\right|^2\right]  & \leq \mathbb{E}\left[\left|\left|\rho_{02}W^{\prime}Q\right|\right|^2 \left|\left|\varepsilon\right|\right|^2\right] \\ 
			& < \infty.
		\end{split}
	\end{equation*}
	Moreover, by Assumption \ref{regcondpanelbeta} \textit{i)}-\textit{iii)}, 
	\begin{equation*}
		\begin{split}
			\int \left|\left| \phi_0\left(z,\hat{\rho}_{1\ell},\psi_0\right) - \phi_0\left(z,\rho_{01},\psi_0\right)\right|\right|^2 F_0(dz) & \leq \underbrace{\left|\left|\hat{\beta}_\ell - \beta_0\right|\right|^2}_{o_p(1)} \underbrace{ \mathbb{E}\left[\left|\left|WQ_{C_1}\right|\right|^2\right]}_{O(1)} \\ & = o_p(1).
		\end{split}
	\end{equation*}
	\begin{equation*}
		\begin{split}
			\int \left|\left| \phi_1\left(z,\hat{\rho}_{1\ell},\rho_{02},\psi_0\right) - \phi_1\left(z,\rho_{01},\rho_{02},\psi_0\right)\right|\right|^2 F_0(dz)  \leq & C\underbrace{\left|\left|\hat{\beta}_\ell - \beta_0\right|\right|^2}_{o_p(1)} \\ &\underbrace{ \left(\mathbb{E}\left[\left|\left|\rho_{02}W^{\prime}QWQ_{C_1}\right|\right|^2\right] + \mathbb{E}\left[\left|\left|WQ_{C_1}\right|\right|^2\right]\right)}_{O(1)} \\ & = o_p(1).
		\end{split}
	\end{equation*}
	Also, 
	\begin{equation*}
		\begin{split}
			\int \left|\left| \phi_1\left(z,\rho_{01},\hat{\rho}_{2\ell},\psi_0\right) - \phi_1\left(z,\rho_{01},\rho_{02},\psi_0\right)\right|\right|^2 F_0(dz)  & \leq C \underbrace{\left|\left|\hat{M}^{\dagger}_\ell - M^{\dagger}\right|\right|^2}_{o_p(1)} \underbrace{ \mathbb{E}\left[\left|\left|W^{\prime}Q\left(Y - W\beta_0\right)\right|\right|^2\right]}_{O(1)} .   
		\end{split}
	\end{equation*}
	Thus, Assumption \ref{aconvtozero} holds. Next, in this scenario, 
	$$
	\hat{\Delta}_\ell(Z) = -C^{\prime}_1\left(\hat{M}^{\dagger}_\ell - M^{\dagger}\right)W^{\prime}QWQ_{C_1}\left(\hat{\beta}_\ell - \beta_0\right).
	$$
	By our boundness conditions, for all $\ell$, simple algebra shows 
	\begin{equation*}
		\begin{split}
			\left|\left|\sqrt{n}\int\hat{\Delta}_\ell(z) F_0(dz)\right|\right| & \leq C \sqrt{n} \left|\left|\hat{M}^{\dagger}_\ell - M^{\dagger}\right|\right|\left|\left|\hat{\beta}_\ell - \beta_0\right|\right| \\ & = o_p(1),
		\end{split}
	\end{equation*}
	where the last equality follows by \textit{iii)}. Similarly, we can show that 
	\begin{equation*}
		\begin{split}
			\int \left|\left|\hat{\Delta}_\ell(z)\right|\right|^2 F_0(dz) & \leq C \underbrace{\left|\left|\hat{M}^{\dagger}_\ell - M^{\dagger}\right|\right|^2}_{o_p(1)}\underbrace{\left|\left|\hat{\beta}_\ell - \beta_0\right|\right|^2}_{o_p(1)}\\
			& = o_p(1).
		\end{split}
	\end{equation*}
	Thus, Assumption \ref{adelta} \textit{i)} holds. Furthermore, by mean independence and the law of iterated expectations, with probability approaching one, 
	\begin{equation*}
		\begin{split}
			\int \phi_1\left(z,\rho_{01},\hat{\rho}_{2\ell},\psi_0\right) F_0(dz) & =\hat{\rho}_{2\ell} \underbrace{\int W^{\prime}Q\left(Y - W\beta_0\right)F_0(dz)}_{= 0} \\ & = 0. 
		\end{split}
	\end{equation*}
	This result implies that Assumption \ref{adoublefrechet} \textit{i)} holds. That $\left|\left|\rho_{02}\right|\right|$ and $\mathbb{E}\left[\left|\left|W^{\prime}QWQ_{C_1}\right|\right|^2\right]$ are finite imply that $\bar{g}\left(\cdot,\rho_{02},\psi_0\right)$ is twice Fréchet differentiable and thus by Proposition 7.3.3. of \cite{luenberger1997optimization}, there exists a $C>0$ such that $\left|\left|\bar{g}(\rho_{1},\rho_{01},\psi_0)\right|\right| \leq C \left|\left|\rho_{1} - \rho_{01}\right|\right|^2$ for all $\rho_1$ with $\left|\left|\rho_{1} - \rho_{01}\right|\right|^2$ small enough. Then, \ref{adoublefrechet} \textit{ii)} is satisfied. Hence, we have verified Assumptions \ref{aconvtozero}-\ref{adoublefrechet} and Lemma \ref{lopmoment} applies. 
	
	Next, we would like to determine $\nu_n$ in Assumption \ref{anormgconverg}. First, notice that adding and subtracting $\rho_{02}W^{\prime}QWH_{C_1}(\psi_0 - \hat{\psi}_\ell)$ yields
	\begin{align}
		\left|\left|\hat{g}_{i\ell}\left(\hat{\psi}_\ell\right) - \hat{g}_{i\ell}\left(\psi_0\right)\right|\right|^2 \leq &  C\left\{\left|\left|\rho_{02}\right|\right|^2\left|\left|W^{\prime}QWH_{C_1}\right|\right|^2 \left|\hat{\psi}_\ell - \psi_0 \right|^{2} \right. \nonumber \\ 
		& + \left. \left|\left|C_1 \right|\right|^2\left|\left|\hat{M}^{\dagger}_\ell - M^{\dagger} \right|\right|^2 \left|\left|W^{\prime}QWH_{C_1} \right|\right|^2 \left| \hat{\psi}_\ell - \psi_0\right|^2\right\} \nonumber \\  = & O_p\left(\left|\left|\hat{\psi}_\ell - \psi_0\right|\right|^2\right), \label{Oppsi}
	\end{align}
	where the last equality follows by observing that, provided assumption Assumption \ref{regcondpanelbeta} \textit{ii)}, $\left|\left|W^{\prime}QWC_{1}\left(C^{\prime}_1C_1\right)^{-1}\right|\right|^2$ is $O_p(1)$ by the Markov's inequality. Then, by  the Central Limit Theorem, for all $\ell$,
	\begin{align}
		\left|\left|\hat{\psi}_\ell - \psi_0\right|\right|^2 & \leq \left|\left|C_1\right|\right|^2 \left|\left|\hat{M}^{\dagger}_\ell\right|\right|^2 \left|\left|\frac{1}{n-n_\ell}\sum^L_{l \neq \ell} \sum_{i \in I_\ell} W_iQ_i\varepsilon_i\right|\right|^2 \nonumber \\ & = O_p\left(n^{-1}\right) \label{eqbound1}.
	\end{align}
	Plugging \eqref{eqbound1} into \eqref{Oppsi} yields that $\left|\left|\hat{g}_{i\ell}\left(\hat{\psi}_\ell\right) - \hat{g}_{i\ell}\left(\psi_0\right)\right|\right|^2 =  O_p\left(n^{-1}\right)$. Furthermore, observe that 
	\begin{align}
		\left|\left|\hat{g}_{i\ell}\left(\psi_0\right) - g_{i}\left(\psi_0\right)\right|\right|^2 \leq & C \left|\left|\phi_0\left(Z,\hat{\rho}_{1\ell},\psi_0\right) - \phi_0\left(Z,\rho_{01},\psi_0\right) \right|\right|^2 \nonumber \\ 
		& + C\left|\left|\phi_1\left(Z,\hat{\rho}_{1\ell},\rho_{02},\psi_0\right) - \phi_1\left(Z,\rho_{01},\rho_{02},\psi_0\right) \right|\right|^2 \nonumber \\ &  + C\left|\left|\phi_1\left(Z,\rho_{01}, \hat{\rho}_{2\ell}, \psi_0\right) - \phi_1\left(Z,\rho_{01},\rho_{02},\psi_0\right) \right|\right|^2 + C \left|\left|\hat{\Delta}_\ell(Z) \right|\right|^2 \nonumber \\  \leq & C \left|\left|WQ_{C_1}\right|\right|^2 \left|\left|\hat{\beta}_\ell - \beta_0\right|\right|^2 \nonumber \\ & + C \left(\left|\left|\rho_{02}W^{\prime}QWQ_{C_1}\right|\right|^2 + \left|\left|WQ_{C_1}\right|\right|^2\right)\left|\left|\hat{\beta}_\ell - \beta_0\right|\right|^2  \nonumber \\ & + C \left|\left|C_1\right|\right|^2  \left|\left|W^{\prime}Q\left(Y - W\beta_0\right)\right|\right|^2 \left|\left|\hat{M}^{\dagger}_\ell - M^{\dagger}\right|\right|^2 \nonumber\\ & + C \left|\left|C_1\right|\right|^2 \left|\left|W^{\prime}QWQ_{C_1} \right|\right|^2 \left|\left|\hat{M}^{\dagger}_\ell - M^{\dagger}\right|\right|^2 \left|\left|\hat{\beta}_\ell - \beta_0\right|\right|^2 \nonumber \\ =  & O_p(1)\left|\left|\hat{\beta}_\ell - \beta_0\right|\right|^2 + O_p(1)\left|\left|\hat{\beta}_\ell - \beta_0\right|\right|^2  \label{opterm1}\\ & + O_p(1)\left|\left|\hat{M}^{\dagger}_\ell - M^{\dagger}\right|\right|^2 + O_p(1) \left|\left|\hat{M}^{\dagger}_\ell - M^{\dagger}\right|\right|^2 \left|\left|\hat{\beta}_\ell - \beta_0\right|\right|^2 \label{opterm2}\\  = & O_p(n^{-1/2}) \label{eqbound2},
	\end{align}
	where the results in \eqref{opterm1} and \eqref{opterm2} follow by \textit{ii)} and the Markov's inequality while \eqref{eqbound2} is implied by \textit{iii)}. Hence, by the triangle inequality $\left|\left|\hat{g}_{i\ell}\left(\hat{\psi}_\ell\right) - g_i(\psi_0)\right|\right|^2 = O_p\left(n^{-1/2}\right)$. Furthermore, observe that, again by the triangle inequality, 
	\begin{align}
		\left|\left|\bar{g}_\ell\left(\hat{\psi}_\ell\right) - \mathbb{E}\left[g_i\left(\psi_0\right)\right]\right|\right|^2 & \leq C \left|\left| \bar{g}_\ell\left(\hat{\psi}_\ell\right) - \mathbb{E}_{n}\left[g_i\left(\psi_0\right)\right] \right|\right|^2  + C \left|\left| \mathbb{E}_{n}\left[g_i\left(\psi_0\right)\right] - \mathbb{E}\left[g_i\left(\psi_0\right)\right] \right|\right|^2 \nonumber \\ & \leq C \frac{1}{n}\sum^L_{\ell = 1} \sum_{i \in I_\ell} \left|\left|\hat{g}_{i\ell}\left(\hat{\psi}_\ell\right) - g_i\left(\psi_0\right)\right|\right|^2 + O_p\left(n^{-1}\right) \label{term_a}\\ & = O_p\left(n^{-1/2}\right) + O_p\left(n^{-1}\right) \label{term_b}\\ & = O_p\left(n^{-1/2}\right), \nonumber
	\end{align}
	where \eqref{term_a} uses the Central Limit Theorem and \eqref{term_b} follows by the fact that $\left|\left|\hat{g}_{i\ell}\left(\hat{\psi}_\ell\right) - g_i(\psi_0)\right|\right|^2 = O_p\left(n^{-1/2}\right)$, as we have shown. Hence, $\left|\left|\hat{\tilde{g}}_{i\ell} - \tilde{g}_i\right|\right| = O_p\left(n^{-1/2}\right)$, by the triangle inequality, i.e., $\nu_n = n^{-1/2}$ in Assumption \ref{anormgconverg}. Lemma \ref{lWconv} then implies $\left|\left|\check{W}_n - W_g\right|\right| = O_p(n^{-1/2})$ and thus $\left|\left|\hat{W}^{\dagger}_n - W^{\dagger}_g\right|\right| = o_p(1)$ by Lemma \ref{lrankWhat}. The asymptotic properties of the test in terms of size and local asymptotic power follow by Propositions \ref{psizetest} and \ref{ppowertest}, respectively. The validity of the confidence regions (CRs) is obtained by Proposition \ref{pcivalidity}. These conclusions complete the proof. $\blacksquare$ \bigskip

		\begin{assumption}
			\label{regcondpanelalpha}
			Suppose that \textit{i)}  $\left|\left|\beta\right|\right| < C$, $\left|\left|\check{M}^{\dagger}\right|\right| < C$, for $\left|\left|\beta - \beta_0 \right|\right|$  and $\left|\left|\check{M}^{\dagger} - M^{\dagger} \right|\right|$ small enough;   \textit{ii)} $\left|\left|\hat{\beta}_\ell - \beta_0 \right|\right| = O_p\left(n^{-1/4}\right)$ and $\left|\left|\hat{M}^{\dagger}_\ell - M^{\dagger} \right|\right| = O_p\left(n^{-1/4}\right)$, for each $\ell = 1,\cdots,L$; \textit{iii)} $rank\left(W_g\right) >0$. 
		\end{assumption}

		\bigskip \noindent \textbf{Proof of Proposition \ref{validinferpanelalpha}:} As in the proof of Proposition \ref{validinferpanelbeta}, we verify the conditions of Lemmas \ref{lopmoment} and \ref{lrankWhat}. We start by verifying Assumptions \ref{aconvtozero}-\ref{adoublefrechet}. In this situation, we can write
		\begin{equation*}
			\begin{split}
				g\left(Z,\rho_1,\rho_2,\psi\right) & = \left(C^{\prime}_2H - \Gamma W^{\prime}Q\right)\left(Y - W\beta\right) - \psi \\ & = \phi_0\left(Z,\rho_1,\psi_0\right) + \phi_1\left(Z,\rho_1,\rho_2,\psi\right),
			\end{split}
		\end{equation*}
		where 
		\begin{equation*}
			\begin{split}
				\phi_0\left(Z,\rho_1,\psi\right) & = C^{\prime}_2H\left(Y - W\beta\right) - \psi, \\ 
				\phi_1\left(Z,\rho_{1},\rho_2, \psi\right) & \equiv   \phi_1\left(Z,\rho_{1},\rho_2\right) = -\Gamma W^{\prime}Q\left(Y - W\beta\right),
			\end{split}
		\end{equation*}
		with $\rho_1 = \beta$ and $\rho_2 = \Gamma$. Then, by our boundness assumptions, we obtain 
		\begin{equation*}
			\begin{split}
				\mathbb{E}\left[\left|\left|g\left(Z,\rho_{01},\rho_{02},\psi_0\right)\right|\right|^2\right] & \leq C \mathbb{E}\left[\left|\left|C^{\prime}_2H\left(Y - W\beta_0\right)\right|\right|^2\right] + C \left|\left|\Gamma_0\right|\right|^2 \mathbb{E}\left[\left|\left|W^{\prime}Q \right|\right|^2 \left|\left|\varepsilon\right|\right|^2\right] + C\left|\left|\psi_0\right|\right|^2 \\ & < \infty.
			\end{split}
		\end{equation*}
		By Assumption \ref{RankHDpanelalpha} \textit{i)}, we can also show that 
		\begin{equation*}
			\begin{split}
				\int \left|\left|\phi_0\left(z,\hat{\rho}_{1\ell},\psi_0\right) -  \phi_0\left(z,\rho_{01},\psi_0\right)\right|\right|^2 F_0(dz) & \leq \underbrace{\left|\left|\hat{\beta}_\ell - \beta_0 \right|\right|^2}_{o_p(1)} \underbrace{\mathbb{E}\left[\left|\left|C^{\prime}_2HW \right|\right|^2\right]}_{O_p(1)} \\ & = o_p(1).
			\end{split}
		\end{equation*}
		
		\begin{equation*}
			\begin{split}
				\int \left|\left|\phi_1\left(z,\hat{\rho}_{1\ell}, \rho_{02}\right) -  \phi_1\left(z,\rho_{01},\rho_{02}\right)\right|\right|^2 F_0(dz) & \leq \underbrace{\left|\left|\hat{\beta}_\ell - \beta_0 \right|\right|^2}_{o_p(1)} \underbrace{\left|\left|\Gamma_0 \right| \right|^2\mathbb{E}\left[\left|\left|W^{\prime}QW \right|\right|^2\right]}_{O_p(1)} \\ & = o_p(1).
			\end{split}
		\end{equation*}
		Additionally, 
		\begin{align}
			\int \left|\left|\phi_1\left(z,\rho_{01}, \hat{\rho}_{2\ell}\right) -  \phi_1\left(z,\rho_{01},\rho_{02}\right)\right|\right|^2 F_0(dz) & \leq \left|\left|\hat{\Gamma}_\ell - \Gamma_0\right|\right|^2 \mathbb{E}\left[\left|\left|W^{\prime}Q\right|\right|^2\left|\left|\varepsilon\right|\right|^2\right] \nonumber \\ & \leq C \left|\left|\hat{S}_\ell \hat{M}^{\dagger}_\ell - SM^{\dagger} \right|\right|^2 \mathbb{E}\left[\left|\left|W^{\prime}Q\right|\right|^2\left|\left|\varepsilon\right|\right|^2\right] \nonumber \\ & \leq O_p\left(\left|\left|\hat{S}_\ell \hat{M}^{\dagger}_\ell - SM^{\dagger} \right|\right|^2 \right) \label{bigoresult}.
		\end{align}
		Since $\left|\left|\hat{S}_\ell \hat{M}^{\dagger}_\ell - SM^{\dagger} \right|\right|^2 = O_p\left(\left|\left|\hat{M}^{\dagger}_\ell - M^{\dagger} \right|\right|^2\right) $, by Assumption \ref{RankHDpanelalpha} \textit{i)}, we know that \eqref{bigoresult} is $o_p(1)$. Thus, Assumption \ref{aconvtozero} holds. Next, 
		$$
		\hat{\Delta}_\ell(Z) = -C^{\prime}_2\left(\hat{S}_\ell\hat{M}^{\dagger}_\ell - SM^{\dagger}\right)W^{\prime}QW\left(\hat{\beta}_\ell - \beta_0\right).
		$$
		Assumption \ref{regcondpanelalpha} \textit{ii)} implies, for any $\ell$, 
		\begin{equation*}
			\begin{split}
				\left|\left| \sqrt{n} \int \hat{\Delta}_\ell(z) F_0(dz)\right|\right| & \leq \mathbb{E}\left[\left|\left|W^{\prime}QW \right|\right|\right] \sqrt{n} \left|\left|\hat{\Gamma}_\ell - \Gamma_0 \right|\right| \left|\left|\hat{\beta}_\ell - \beta_0 \right|\right| \\ & = O_p\left(\sqrt{n}\left|\left|\hat{M}_\ell - M \right|\right| \left|\left|\hat{\beta}_\ell - \beta_0 \right|\right|\right) \\ & \overset{p}{\rightarrow} 0.
			\end{split}
		\end{equation*}
		Similarly, 
		\begin{equation*}
			\begin{split}
				\int \left|\left|\Delta_\ell\left(z\right) \right|\right|^2 F_0(dz) & \leq \underbrace{ C \mathbb{E}\left[\left|\left|W^{\prime}QW \right|\right|^2\right]}_{O_p(1)} \underbrace{\left|\left|\hat{\Gamma}_\ell - \Gamma_0 \right|\right|^2}_{o_p(1)} \underbrace{\left|\left|\hat{\beta}_\ell - \beta_0 \right|\right|^2}_{o_p(1)} \\ & = o_p(1).
			\end{split}
		\end{equation*}
		As a result, Assumption \ref{adelta} \textit{i)} holds. By similar arguments as in the proof of Proposition \ref{validinferpanelbeta}, 
		\begin{equation*}
			\begin{split}
				\int \phi_1\left(z,\rho_{01},\hat{\rho}_{2\ell},\psi_0\right) F_0(dz) & =-\hat{\Gamma}_\ell \underbrace{\int W^{\prime}Q\left(Y - W\beta_0\right)F_0(dz)}_{= 0} \\ & = 0,
			\end{split}
		\end{equation*}
		and hence Assumption \ref{adoublefrechet} \textit{i)} holds. That $\mathbb{E}\left[\left|\left|\left(C^{\prime}_2H - \Gamma_0W^{\prime}Q\right)W \right|\right|^2\right]$ is finite imply that $\bar{g}\left(\cdot,\rho_{02}\psi_0\right)$ is twice Fréchet differentiable and thus by Proposition 7.3.3. of \cite{luenberger1997optimization}, there exists a $C>0$ such that $\left|\left|\bar{g}(\rho_{1},\rho_{02},\psi_0)\right|\right| \leq C \left|\left|\rho_{1} - \rho_{01}\right|\right|^2$ for all $\rho_1$ with $\left|\left|\rho_{1} - \rho_{01}\right|\right|^2$ small enough. Thus, \ref{adoublefrechet} \textit{ii)} is satisfied. We have therefore verified Assumptions \ref{aconvtozero}-\ref{adoublefrechet} and Lemma \ref{lopmoment} applies to this case.

		Next, we shall specify $\nu_n$ in Assumption \ref{anormgconverg}. Here, we have
		\begin{align}
			\left|\left|\hat{g}_{i\ell}\left(\hat{\psi}_\ell\right) - \hat{g}_i\left(\psi_0\right)\right|\right|^2 = &\left|\left|\hat{\psi}_\ell - \psi_0\right|\right|^2, \nonumber \\ \leq &  C\underbrace{\left|\left|C_2^{\prime}\mathbb{E}_n\left[H_i\left(Y_i - W_i\hat{\beta}_\ell\right)\right] - C^{\prime}_2\mathbb{E}\left[H\left(Y - W\beta_0\right)\right] \right|\right|^2}_{:=A_1} \label{A1} \\ & + C \underbrace{\left|\left|C^{\prime}_2\hat{\Gamma}_\ell W^{\prime}Q\left(Y - W\hat{\beta}_\ell\right) - C_2^{\prime}\mathbb{E}\left[\Gamma_0W^{\prime}Q\left(Y - W\beta_0\right)\right] \right|\right|^2}_{:=A_2} \label{A2} 
		\end{align}
		Simple algebra yields, by the Central Limit Theorem, our boundness assumptions, and Assumption \ref{regcondpanelalpha} \textit{ii)},
		\begin{align}
			A_1  \leq & C\left|\left|C^{\prime}_2\mathbb{E}_n[H_iY_i] - C^{\prime}_2\mathbb{E}\left[HY\right] \right|\right|^2 + C \left|\left|C^{\prime}_2\left(\mathbb{E}_n\left[H_iW_i\right] - \mathbb{E}\left[WH\right]\right)\hat{\beta}_\ell\right|\right|^2 \nonumber \\ & + C \left|\left|C^{\prime}_2\mathbb{E}\left[WH\right] \left(\hat{\beta}_\ell - \beta_0\right)\right|\right|^2 \nonumber\\  = & O_p\left(n^{-1}\right) + O_p(n^{-1} \left|\left| \hat{\beta}_\ell - \beta_0\right|\right|^2) + O_p(\left|\left| \hat{\beta}_\ell - \beta_0\right|\right|^2) \nonumber\\  = & O_p(\left|\left| \hat{\beta}_\ell - \beta_0\right|\right|^2) \nonumber\\  = & O_p(n^{-1/2}) \label{OpA1}.
		\end{align}
		
		\begin{align}
			A_2  \leq & C \left|\left|C^{\prime}_2\hat{\Gamma}_\ell \left(\mathbb{E}_n\left[W^{\prime}_iQ_iY_i\right] - \mathbb{E}\left[W^{\prime}QY\right]\right) \right|\right|^2 + C\left|\left|C^{\prime}_2\left(\hat{\Gamma}_\ell - \Gamma_0\right)\mathbb{E}\left[W^{\prime}QY\right] \right|\right|^2 \nonumber \\ & + C \left|\left|C^{\prime}_2\hat{\Gamma}_\ell \left(\mathbb{E}_n\left[W^{\prime}_iQ_iW_i\right] - \mathbb{E}\left[W^{\prime}QW\right]\right)\hat{\beta}_\ell \right|\right|^2 + C\left|\left|C^{\prime}_2\left(\hat{\Gamma}_\ell - \Gamma_0\right)\mathbb{E}\left[W^{\prime}QW\right]\left(\hat{\beta}_\ell - \beta_0\right)\right|\right|^2 \nonumber \\  = & O_p\left(n^{-1} \left|\left|\hat{M}^{\dagger}_\ell - M^{\dagger} \right|\right|^2\right) + O_p\left( \left|\left|\hat{M}^{\dagger}_\ell - M^{\dagger} \right|\right|^2\right) + O_p\left(n^{-1} \left|\left|\hat{M}^{\dagger}_\ell - M^{\dagger} \right|\right|^2 \left|\left|\hat{\beta}_\ell - \beta_0 \right|\right|^2\right) \nonumber \\ & + O_p\left(\left|\left|\hat{M}^{\dagger}_\ell - M^{\dagger} \right|\right|^2 \left|\left|\hat{\beta}_\ell - \beta_0 \right|\right|^2\right)\nonumber \\  = & O_p\left( \left|\left|\hat{M}^{\dagger}_\ell - M^{\dagger} \right|\right|^2\right) \nonumber \\  = & O_p\left(n^{-1/2}\right) \label{OpA2}.
		\end{align}
		Plugging \eqref{OpA1} and \eqref{OpA2} into \eqref{A1} and \eqref{A2}, respectively, we get $\left|\left|\hat{g}_{i\ell}\left(\hat{\psi}_\ell\right) - \hat{g}_{i\ell}\left(\psi_0\right)\right|\right|^2 = O_p\left(n^{-1/2}\right)$. In addition, notice that 
		\begin{align}
			\left|\left|\hat{g}_{i\ell}\left(\psi_0\right) - g_i\left(\psi_0\right)\right|\right|^2 \leq & C \left|\left|\phi_0\left(Z,\hat{\rho}_{1\ell},\psi_0\right) - \phi_0\left(Z,\rho_{01},\psi_0\right) \right|\right|^2 \nonumber \\ 
			& + C\left|\left|\phi_1\left(Z,\hat{\rho}_{1\ell},\rho_{02},\psi_0\right) - \phi_1\left(Z,\rho_{01},\rho_{02},\psi_0\right) \right|\right|^2 \nonumber \\ &  + C\left|\left|\phi_1\left(Z,\rho_{01}, \hat{\rho}_{2\ell}, \psi_0\right) - \phi_1\left(Z,\rho_{01},\rho_{02},\psi_0\right) \right|\right|^2 + C \left|\left|\hat{\Delta}_\ell(Z) \right|\right|^2 \nonumber \\  \leq & C \left|\left|C^{\prime}_2HW\right|\right|^2 \left|\left|\hat{\beta}_\ell - \beta_0\right|\right|^2 + C\left|\left|\Gamma_0\right|\right|^2 \left|\left|W^{\prime}QW\right|\right|^2\left|\left|\hat{\beta}_\ell - \beta_0\right|\right|^2 \nonumber \\ & +C\left|\left|\hat{\Gamma}_\ell - \Gamma_0\right|\right|^2 \left|\left|W^{\prime}Q\right|\right|^2\left|\left|\varepsilon\right|\right|^2 + C \left|\left|\hat{\Gamma}_\ell - \Gamma_0\right|\right|^2 \left|\left|W^{\prime}QW\right|\right|^2 \left|\left|\hat{\beta}_\ell - \beta_0\right|\right|^2 \nonumber + \\  = & O_p\left(\left|\left|\hat{\beta}_\ell - \beta_0\right|\right|^2 \right) + O_p\left(\left|\left|\hat{\beta}_\ell - \beta_0\right|\right|^2 \right) \label{bound11} \\ & + O_p\left(\left|\left|\hat{M}^{\dagger}_\ell - M\right|\right|^2 \right) + O_p\left(\left|\left|\hat{M}^{\dagger}_\ell - M\right|\right|^2 \left|\left|\hat{\beta}_\ell - \beta_0\right|\right|^2\right) \label{bound21} \\  = & O_p\left(n^{-1/2}\right), \label{eqbound21}
		\end{align}
		where the results in \eqref{bound11} and \eqref{bound21} follow by  the Markov's inequality while \eqref{eqbound21} is implied by Assumption \ref{regcondpanelalpha} \textit{ii)}. Then, by the triangle inequality, $\left|\left|\hat{g}_{i\ell}\left(\hat{\psi}_\ell\right) - g_i\left(\psi_0\right)\right|\right|^2 = O_p\left(n^{-1/2}\right)$. Similarly, by the triangle inequality, 
		\begin{align}
			\left|\left|\bar{g}_\ell\left(\hat{\psi}_\ell\right) - \mathbb{E}\left[g_i\left(\psi_0\right)\right]\right|\right|^2 & \leq C \left|\left| \bar{g}_\ell\left(\hat{\psi}_\ell\right) - \mathbb{E}_{n}\left[g_i\left(\psi_0\right)\right] \right|\right|^2  + C \left|\left| \mathbb{E}_{n}\left[g_i\left(\psi_0\right)\right] - \mathbb{E}\left[g_i\left(\psi_0\right)\right] \right|\right|^2 \nonumber \\ & \leq C \frac{1}{n}\sum^L_{\ell = 1} \sum_{i \in I_\ell} \left|\left|\hat{g}_{i\ell}\left(\hat{\psi}_\ell\right) - g_i\left(\psi_0\right)\right|\right|^2 + O_p\left(n^{-1}\right) \label{term_a1}\\ & = O_p\left(n^{-1/2}\right) + O_p\left(n^{-1}\right) \label{term_b1}\\ & = O_p\left(n^{-1/2}\right), \nonumber
		\end{align}
		where the results in \eqref{term_a1} and \eqref{term_b1} follow by the Central Limit Theorem and $\left|\left|\hat{g}_{i\ell}\left(\hat{\psi}_\ell\right) - g_i\left(\psi_0\right)\right|\right|^2 = O_p\left(n^{-1/2}\right)$, respectively. As a result,  $\nu_n = n^{-1/2}$ in Assumption \ref{anormgconverg}. Using Lemma \ref{lWconv}, \\ $\left|\left|\check{W}_n - W_g\right|\right| = O_p(n^{-1/2})$ and thus $\left|\left|\hat{W}^{\dagger}_n - W^{\dagger}_g\right|\right| = o_p(1)$ by  Lemma \ref{lrankWhat}. As before, the asymptotic properties of the test in terms of size and local asymptotic power are obtained by Propositions \ref{psizetest} and \ref{ppowertest}, respectively. Confidence regions (CRs) are valid,  by Proposition \ref{pcivalidity}. These observations show the results of the proposition and the proof is completed. $\blacksquare$
		
		\bigskip A key assumption in Propositions \ref{validinferpanelbeta} and \ref{validinferpanelalpha} states $\left|\left|\hat{\beta}_\ell - \beta_0 \right|\right| = O_p\left(n^{-1/4}\right)$ and $\left|\left|\hat{M}^{\dagger}_\ell - M\right|\right| = O_p\left(n^{-1/4}\right)$, for each $l = 1,\cdots, L$. We first work on studying the first condition in the next section. 

	\section{Asymptotic Properties and Implementation of the Lasso Program}
		
		\subsection{Asymptotic Properties}
		
		We first need to introduce some notation. Let $\tilde{W} = QW$. Let $\tilde{W}_t$ be its $t-th$ row. We use the same sub-index notation to refer to the $t-th$ entry or row of any vector or matrix, respectively. Let $\tilde{\varepsilon} = Q\varepsilon$. For an arbitrary vector $\beta \in \mathbb{R}^p$, let $\left|\left|\beta\right|\right|_0$ be the pseudo-norm that counts the number of non-zero entries in $\beta$. As in \cite{belloni2016inference}, let us define 
		$$
		\phi^2_{j,\ell} := \frac{1}{(n-n_\ell)T} \sum^L_{l \neq \ell}  \sum_{i \in I_\ell}\left(\sum^T_{t=1} \tilde{W}_{itj} \tilde{\varepsilon}_{it}\right)^2.
		$$
		Moreover, let
		$$
		\imath_T := T \underset{1 \leq j \leq p}{\operatorname{min}}\;\; \frac{\mathbb{E}\left[T^{-1} \sum^T_{t=1} \tilde{W}^2_{itj}\tilde{\varepsilon}^2_{it}\right]}{\mathbb{E}\left[T^{-1} \left(\sum^T_{t=1} \tilde{W}_{itj}\tilde{\varepsilon}_{it}\right)^2\right]},
		$$
		which is a measure of within-individual dependence. Let us define the minimal and maximal $m-$sparse eigenvalues of $\hat{M}$ as
		\begin{equation*}
			\begin{split}
				\phi_{min}(m)\left(\hat{M}_\ell\right) & := \underset{\delta \in \Delta(m)}{\operatorname{min}}\;\; \delta^{\prime}\hat{M}_\ell\delta, \\ 
				\phi_{max}(m)\left(\hat{M}_\ell\right) & := \underset{\delta \in \Delta(m)}{\operatorname{max}}\;\; \delta^{\prime}\hat{M}_\ell\delta,
			\end{split}
		\end{equation*}
		respectively, where $\Delta(m) = \left\{\delta \in \mathbb{R}^{p}: \left|\left|\delta\right|\right|_0,\; \left|\left|\delta\right|\right|_2 = 1\right\}$. Finally, we define 
		$$
		\varpi := \left(\mathbb{E}\left[\left|\frac{1}{\sqrt{T}} \sum^T_{t=1}\tilde{W}_{itj}\tilde{\varepsilon}_{it}\right|^3\right]\right)^{1/3}.
		$$
		
		\begin{lemma}
			\label{lassoresultbeta}
			Suppose 
			\begin{itemize}
				\item[i)] $\left|\left|\beta_0\right|\right|_0 \leq s = o\left(n\imath_T\right)$;
				\item[ii)] For any $C >0$, there exist constants $0 < \kappa^{\prime} < \kappa^{\prime \prime} < \infty$, not depending on $n$, such that with probability approaching one, $\kappa^{\prime} \leq  \phi_{min}(Cs)\left(\hat{M}_\ell\right) \leq  \phi_{max}(Cs)\left(\hat{M}_\ell\right) \leq \kappa^{\prime \prime}$, all $\ell = 1,\cdots, L$;
				\item[iii)] For all $\ell$, a) $\left(T^{-1}\sum^T_{t=1}\mathbb{E}\left[\tilde{W}^2_{itj}\tilde{\varepsilon}_{it}\right]\right) + \left(T^{-1}\sum^T_{t=1}\mathbb{E}\left[\tilde{W}^2_{itj}\tilde{\varepsilon}_{it}\right]\right)^{-1} = O(1)$, \\ b) $1 \leq \underset{1 \leq j \leq p}{\operatorname{max}}\;\; \phi_{j,\ell} /\underset{1 \leq j \leq p}{\operatorname{min}}\;\; \phi_{j,\ell} = O(1)$, c) $1 \leq \underset{1 \leq j \leq p}{\operatorname{max}}\;\; \varpi_j/\sqrt{\mathbb{E}\left[\phi^2_{j,\ell}\right]} = O(1)$, d) $\log^3(p) = o\left(nT\right)$ and $s \log\left(p \vee nT\right) = o\left(n \imath_T\right)$, e) $\underset{1 \leq j \leq p}{\operatorname{max}}\;\; \left|\phi_{j,\ell} -  \sqrt{\mathbb{E}\left[\phi^2_{j,\ell}\right]}\right|/\sqrt{\mathbb{E}\left[\phi^2_{j,\ell}\right]} = o(1)$.
			\end{itemize}
			Then, 
			\begin{equation}
				\label{resultbelloni}
				\left|\left|\hat{\beta}_\ell - \beta_0\right|\right| = O_p\left(\sqrt{s \log \left(p \vee nT\right)/ n\imath_T}\right),
			\end{equation}
			for each $\ell = 1,\cdots,L$. Under the additional assumption that $\sqrt{s \log \left(p \vee nT\right)/ n\imath_T} = O(n^{-1/4})$, it follows that $ \left|\left|\hat{\beta}_\ell - \beta_0\right|\right| = O_p\left(n^{-1/4}\right)$.
		\end{lemma}
		\bigskip \textbf{Proof of Lemma \ref{lassoresultbeta}:} The result follows by Theorem 1 of \cite{belloni2016inference}, accommodating for the fact that in our case the slopes of $\alpha$, the UH, is the multidimensional matrix $V$ and thus the within projection must be $Q$ instead of the usual within projection that removes fixed effects by within individual demeaning, which is typically used when the slope of $\alpha$ is one, as in \cite{belloni2016inference}. Specifically, we work with the model
		\begin{equation*}
			\begin{split}
				\tilde{Y}  := QY & = QW\beta_0 + Q\varepsilon \\ & = \tilde{W}\beta_0 + \tilde{\varepsilon},
			\end{split}
		\end{equation*}
		and thus we are able to write it in terms of that studied by \cite{belloni2016inference}.\footnote{Technically, the model considered by  \cite{belloni2016inference} is more general than the one here in the sense that they work with an approximately sparse model while we impose exact sparsity by imposing the mean independence condition of \cite{chamberlain1992efficiency} and the sparse structure of $\beta_0$ in \textit{i)}.} Our assumptions are the corresponding Conditions ASM, SE, and R, in \cite{belloni2016inference} and thus the result of the lemma is an application of Theorem 1 in the aforementioned paper, which shows \eqref{resultbelloni}. If the right-hand side in this expression is $O(n^{-1/4})$, we directly obtain $ \left|\left|\hat{\beta}_\ell - \beta_0\right|\right| = O_p\left(n^{-1/4}\right)$, which completes the proof. $\blacksquare$

		\subsection{Implementation}
		\label{implementationlasso}
		An important ingredient that is implicit in Lemma \ref{lassoresultbeta} is the appropiate determination of the tuning parameter $c_n$. As discussed by \cite{belloni2016inference}, we let
		$$
		c_n = \frac{c}{\sqrt{(n-n_\ell)T}} \Phi^{-1}\left(1-\gamma/2p\right),
		$$
		where $c > 1$ and $\gamma = o(1)$. We set $c=1.1$ and $\gamma = 0.1/\log(p \vee (n-n_\ell)T)$.\footnote{We take into account that in the notation of \cite{belloni2012sparse}, the tuning parameter is $c_n/nT$ instead of $c_n$, as in our notation.} Finally, we construct the estimated penalty loadings simply as 
		$$
		\hat{\phi}_{j,\ell}^2 = \frac{1}{(n-n_\ell)T} \sum^L_{l \neq \ell}  \sum_{i \in I_l}\left(\sum^T_{t=1} \tilde{W}_{itj} \hat{\varepsilon}_{it}\right)^2 = \frac{1}{(n-n_\ell)T} \sum^L_{l \neq \ell}  \sum_{i \in I_l}\sum^T_{t=1} \sum^T_{t^{'}=1}\tilde{W}_{itj}\tilde{W}_{it^{\prime}j} \hat{\varepsilon}_{it} \hat{\varepsilon}_{it^{\prime}},
		$$
		where $\hat{\varepsilon}_{it}$ is an estimate of $\tilde{\varepsilon}_{it}$. We implement the algorithm given in Appendix A of \cite{belloni2016inference}. In our scenario, this is implemented as follows. Let $K \geq 1$ be a bounded number of iterations. For each $j = 1,\cdots,p$ and $\ell = 1,\cdots,L$, let us denote feasible loadings
		\begin{align}
			\text{Initial:} \;\;\; \hat{\phi}_{j,\ell} & = \left[\frac{1}{(n-n_\ell)T} \sum^L_{l \neq \ell}  \sum_{i \in I_l}\sum^T_{t=1} \sum^T_{t^{'}=1}\tilde{W}_{itj}\tilde{W}_{it^{\prime}j} \tilde{Y}_{it} \tilde{Y}_{it^{\prime}}\right]^{1/2} , \label{initial} \\ \text{Refined:} \;\;\; \hat{\phi}_{j,\ell}& = \left[\frac{1}{(n-n_\ell)T} \sum^L_{l \neq \ell}  \sum_{i \in I_l}\sum^T_{t=1} \sum^T_{t^{'}=1}\tilde{W}_{itj}\tilde{W}_{it^{\prime}j} \hat{\varepsilon}_{it} \hat{\varepsilon}_{it^{\prime}}\right]^{1/2}. \label{refined}
		\end{align}

		\begin{algorithm}
			\caption{Implementation of our Lasso program in the High-Dimensional Random Coefficient Model}\label{alg:lassoimpl}
			\begin{algorithmic}
				\For{$l = 1:L$}
				
				\textbf{Step 1:} Set penalty loadings according to \eqref{initial}. Use them to solve the Lasso program and get an initial $\hat{\beta}^{init}_\ell$. Then compute residuals $\hat{\varepsilon}_{it} = \tilde{Y}_{it} - \tilde{W}_{it}^{\prime}\hat{\beta}^{init}_\ell$, for all $i \in I_l$, $l \neq \ell$  and $t = 1,\cdots,T$.
				
				\textbf{Step 2:} \If{$K >1$} 
				\State Update the penalty loadings using \eqref{refined} and update the Lasso estimator $\hat{\beta}_\ell$. Next, compute a new set of residuals suing the updated $\hat{\varepsilon}_{it} = \tilde{Y}_{it} - \tilde{W}_{it}^{\prime}\hat{\beta}_\ell$, for all $i \in I_l$, $\ell \neq \ell$  and $t = 1,\cdots,T$.
				\Else 
				\State Repeat Step 2, $K-2$ times.
				\EndIf
				\EndFor
			\end{algorithmic}
		\end{algorithm}
		
		The feasible loadings constructed as above are asymptotically valid in the sense that yields a Lasso estimator asymptotically consistent; see Appendix A of \cite{belloni2016inference}.
		
		Step 1 and Step 2 require computationally minimizing our Lasso program. There exist well-known approaches to find a solution that minimizes a generalized Lasso objective function. In particular, we recommend the so-called coordinate descent approach for Lasso; see \cite{friedman2007pathwise, friedman2010regularization, fu1998penalized,daubechies2004iterative}. This needs to be accommodated by the fact that penalty loadings appears in the penalization of our Lasso program. The validity of the solution follows by well-known differentiability and convex arguments; see \cite{friedman2007pathwise} and \cite{tseng2001convergence}.
		
		\subsection{Asymptotic Properties of a Moore-Penrose Estimator}
		\label{moorepenroseest}
		We next turn to the construction  of an estimator $\hat{M}^{\dagger}_\ell$ that satisfies $\left|\left|\hat{M}^{\dagger}_\ell - M\right|\right| = O_p\left(n^{-1/4}\right)$, for each $l = 1,\cdots, L$. Recall that 
		$$\check{M}_{\ell}=\frac{1}{n-n_{\ell}}%
		\sum_{l\neq\ell}^{L}\sum_{i\in I_{l}}W'_iQ_iW_i
		$$
		From $\check{M}_{\ell}$, we construct a suitable estimator $\hat{M}^{\dagger}_\ell$, similarly to our construction for the Moore-Penrose inverse of the variance-covariance matrix of Section \ref{sectiontest}. Specifically, we apply the spectral decomposition to $\check{M}_{\ell}$ so that $\check{M}_{\ell} = \check{U}_\ell \check{\Lambda}_\ell \check{U}^{\prime}_\ell,$, where $\check{\lambda}_{\ell,j}$ is the $j-th$ element of the diagonal matrix $\check{\Lambda}_\ell$ containing the eigenvalues of $\check{M}_{\ell}$. We define
		$$
		\check{M}_{\ell} := \check{U}_\ell \check{\Lambda}_\ell \left(\pi_n\right)\check{U}^{\prime}_\ell,
		$$
		where $\check{\Lambda}_\ell \left(\pi_n\right)$ is a diagonal matrix whose $j-th$ element is equal to $\check{\lambda}_{\ell,j}$ if $\check{\lambda}_{\ell,j} \geq \pi_n$ and zero otherwise, and $\pi_n$ is a non-negative sequence converging to zero to be determined below. We state the following proposition:
		\begin{proposition}
			\label{propresultestM}
			Let $\left|\left|\tilde{M}\right|\right| < C$ for $\left|\left|\tilde{M} - M\right|\right|$ small enough and $\text{rank}\left(M\right)>0$. For all $\ell$, if $\left|\left|\check{M}_\ell - M\right|\right| = O_p\left(\pi_n\right)$ and $\text{rank}\left(M\right) \pi_n = o(1)$, then $\hat{M}_\ell \xrightarrow[]{p} M$ and $\mathbb{P}\left(\text{rank}\left(\hat{M}_\ell\right) = \text{rank}\left(M\right)\right) \rightarrow 1$. As a result, $\left|\left|\hat{M}^{\dagger}_\ell - M\right|\right| = O_p\left(\text{rank}\left(M\right)\pi_n\right)$.
		\end{proposition}
		
		\bigskip \noindent \textbf{Proof of Proposition \ref{propresultestM}:} We first show that if $\left|\left|\check{M}_\ell - M\right|\right| = O_p\left(\pi_n\right)$ and $\text{rank}\left(M\right) \pi_n = o(1)$, then $\hat{M}_\ell \xrightarrow[]{p} M$ and $\mathbb{P}\left(\text{rank}\left(\hat{M}_\ell\right) = \text{rank}\left(M\right)\right) \rightarrow 1$. This follows by a straightforward extension of the proof of Proposition S1 in \cite{lee2024locally} to the case where the dimension $p$ (and $\text{rank}(M)$) grows with the sample size.  Let us borrow (most of) their notation here. Define $\hat{r}_\ell := \text{rank}\left(\hat{M}_\ell\right)$, $r:=\text{rank}\left(M\right)$, and the event $R_\ell : = \left\{\hat{r}_\ell = r\right\}$. In addition, let $\lambda_{j}$, $\check{\lambda}_{\ell,j}$, and $\hat{\lambda}_{\ell,j}$ be the $j-th$ largest eigenvalue of $M$, $\check{M}_\ell$, and $\hat{M}_\ell$, respectively. Note that we assume $r>0$.\footnote{The proof of \cite{lee2024locally} also allows for the possibility that $r=0$.}
		
		Define $\underline{\pi}:= \lambda_r/2 >0$. Since $\left|\left|\check{M}_\ell - M\right|\right| = O_p\left(\pi_n\right)$, the Weyl's Perturbation Theorem \citep[e.g.,][Corollary III]{bhatia1997} yields $\underset{j=1,\cdots,p}{\operatorname{max}} \left|\check{\lambda}_{\ell,j} - \lambda_j\right| \leq \left|\left|\check{M}_\ell - M \right|\right| = O_p(\pi_n) \rightarrow 0$. Let us define $E_\ell : = \left\{\check{\lambda}_{\ell,r} \geq \pi_n\right\}$. For $n$ large enough such that $\pi_n < \underline{\pi}$, note that 
		$$
		\mathbb{P}\left(E_\ell\right) = \mathbb{P}\left(\check{\lambda}_r \geq \pi_n\right) \geq \mathbb{P}\left(\check{\lambda}_{\ell,r} > \underline{\pi}\right) \geq \mathbb{P}\left(\left|\check{\lambda}_{\ell,r} - \lambda_r \right|< \underline{\pi}/2  \right) \rightarrow 1.
		$$
		If $r = p$ we have that $E_\ell \subset R_\ell$ and hence $\mathbb{P}\left(R_\ell\right) \rightarrow 1$. Moreover, when $\check{\lambda}_{\ell,p} \geq \pi_n$, we have $\hat{\lambda}_{\ell,j} = \check{\lambda}_{\ell,j}$ for each $j = 1,\cdots,p$, implying that $\hat{M}_\ell = \check{M}_\ell$. Thus, $\left|\left|\hat{M}_\ell - M \right|\right| = \left|\left|\check{M}_\ell - M \right|\right| = O_p(\pi_n) \rightarrow 0$.
		
		Now suppose instead that $r < p$. Define $F_\ell := \left\{\check{\lambda}_{\ell,r+1} < \pi_n\right\}$. By Weyl's Perturbation Theorem and the fact that $\lambda_j = 0$ for all $j >r$, as $n \rightarrow \infty$,
		$$
		\mathbb{P}\left(F_\ell\right) = \mathbb{P}\left(\check{\lambda}_{\ell,r+1} < \pi_n\right) \geq \mathbb{P}\left(\left|\left|\check{M}_\ell - M \right|\right| < \pi_n\right) \rightarrow 1.
		$$
		Since $R_\ell \supset E_\ell \cap F_\ell$, the previous display implies that $\mathbb{P}\left(F_\ell\right) \rightarrow 1$, as $n \rightarrow \infty$. Moreover, notice that in the case where $\check{\lambda}_{\ell,r} \geq \pi_n$, $\check{\lambda}_{\ell,r+1} < \pi_n$, and $\left|\left|\check{M}_\ell - M\right|\right| \geq C$, we have that $\hat{\lambda}_{\ell,j} = \check{\lambda}_{\ell,j}$, for any $j \geq r$, and  $\hat{\lambda}_{\ell,j} = 0 = \lambda_j$, for any $j >r$ and hence
		$$
		\left|\left|\Lambda_\ell\left(\pi_n\right) - \Lambda \right|\right| = \underset{j=1,\cdots,r}{\operatorname{max}} \left|\hat{\lambda}_{\ell,j} - \lambda_j\right| = \underset{j=1,\cdots,r}{\operatorname{max}} \left|\check{\lambda}_{\ell,j} - \lambda_j\right| \leq \left|\left|\check{\Lambda}_\ell - \Lambda \right|\right| \leq \left|\left|\check{M}_\ell - M \right|\right| \leq C,
		$$
		so $\left\{\left|\left|\check{M}_\ell - M \right|\right| \leq C \right\} \cap E_\ell \cap F_\ell \subset \left\{\left|\left|\Lambda_\ell\left(\pi_n\right) - \Lambda \right|\right| \leq C\right\}$. From here, we conclude that \\ $\left|\left|\Lambda_\ell\left(\pi_n\right) - \Lambda \right|\right| = O_p\left(\pi_n\right)$, as $\left|\left|\check{M}_\ell - M \right|\right| = O_p\left(\pi_n\right)$. Suppose that $\left(\lambda_1, \cdots, \lambda_r\right)$ consists of $s$ distinct eigenvalues: $\lambda^1 >  \lambda^2 > \cdots, \lambda^s$, and each one is repeated $\mathbf{m_1}, \mathbf{m_2}, \cdots, \mathbf{m_s}$, respectively, where $\mathbf{m_a} \geq 1$, $a = 1,\cdots,s$. Note that $\lambda^{s+1} = 0$ is an eigenvalue repeated $\mathbf{m_{s+1}} = p -r$ times. Let $\ell^{k}_{j}$, for $k = 1,\cdots,s+1$ and $j=1,\cdots,\mathbf{m_k}$ be the column indices of the eigenvectors of $U$ associated with each $\lambda^k$. For each $\lambda^k$, $\Pi_k:= \sum^{\mathbf{m_k}}_{j=1}u_{\ell^k_{j}}u^{\prime}_{\ell^k_{j}}$ is the total eigenprojection of $\lambda^k$. Recall that total eigenprojections are continuous.\footnote{See, e.g., Theorem 8.7 of \cite{magnus2019}.}. This implies that if we construct $\Pi_{\ell,k}$ in an analogous manner to $\Pi_k$, replacing columns of $U$ with columns of $\check{U}_\ell$, we have $\Pi_{\ell,k} \xrightarrow[]{p} \Pi_k$ for all $k = 1,\cdots,s+1$ because $\check{M}_\ell \xrightarrow[]{p} M$. Using the spectral decomposition of $M$, we shall write $M = \sum^s_{k=1} \lambda^k \Pi_k$. Observe that the sum runs to $s$ rather than $s+1$ since $\lambda^{s+1} = 0$. Notice then
		$$
		\hat{M}_\ell = \sum^{s+1}_{k=1}\sum^{\mathbf{m_k}}_{j=1}\hat{\lambda}_{\ell^k_{j}}\check{u}_{\ell^k_j}\check{u}^{\prime}_{\ell^k_j} = \sum^{s+1}_{k=1}\sum^{\mathbf{m_k}}_{j=1}\left(\hat{\lambda}_{\ell^k_j} - \lambda^k\right)\check{u}_{\ell^k_j}\check{u}^{\prime}_{\ell^k_j} + \sum^s_{k=1} \lambda^k \Pi_{\ell,k}.
		$$
		From the previous construction we obtain that, by the triangle inequality,
		\begin{equation*}
			\begin{split}
				\left|\left|\hat{M}_\ell - M \right|\right| & \leq \sum^{s+1}_{k=1}\sum^{\mathbf{m_k}}_{j=1} \left| \hat{\lambda}_{\ell^k_j} - \lambda^k \right|\left|\left|\check{u}_{\ell^k_j}\check{u}^{\prime}_{\ell^k_j} \right|\right| + \sum^s_{k=1} \left|\lambda^k \right| \left|\left|\Pi_{\ell,k} - \Pi_{k} \right|\right| \\ & \leq O_p\left(r\pi_n\right) + O_p(s\pi_n) \\ & = O_p\left(r\pi_n\right) \\ & \rightarrow 0,
			\end{split}
		\end{equation*}
		where the second inequality follows by $\hat{\Lambda}(\pi_n) \xrightarrow[]{p} \Lambda$,  continuity of total eigenprojections, and that $\left|\left|\check{u}_{\ell^k_j}\check{u}^{\prime}_{\ell^k_j} \right|\right| = 1$ for all $j,k$. The last displays and the fact $\mathbb{P}\left(R_\ell\right) \rightarrow 1$ along with Lemma 1 in \cite{andrews1987asymptotic} yield the desired results. $\blacksquare$
		
		\bigskip A key condition in Proposition \ref{propresultestM} states that $\left|\left|\check{M}_\ell - M \right|\right| =O_p\left(\pi_n\right)$, for a specific sequence $\pi_n$. This is determined by the following result.
		
		\begin{proposition}
			\label{proprateMmatrix}
			Suppose that there exists a $C>0$ such that $\left|\left|W^{\prime}_iQ_iW_i \right|\right| < C$, for all $i$, and $\mathbb{E}\left[\left|\left| W^{\prime}QW\right|\right|^2\right] < C$. Then, for each $\ell = 1,\cdots,L$,
			$$
			\left|\left| \check{M}_\ell - M\right|\right| = O_p\left(\left(\frac{\log p}{n}\right)^{1/2}\right).
			$$
			Under the additional assumption that $\log p = O(n^{3/2})$, it follows that $\left|\left| \check{M}_\ell - M\right|\right| = O_p\left(n^{-1/4}\right).$
		\end{proposition}
		
		\bigskip \noindent \textbf{Proof of Proposition \ref{proprateMmatrix}:} Define $B_i := W^{\prime}_iQ_iW_i - \mathbb{E}\left[W^{\prime}QW\right] = W^{\prime}_iQ_iW_i - M$. Note that the norm of this matrix is bounded, which follows by the triangle inequality: 
		$$
		\left|\left| B_i\right|\right| \leq \left|\left| W^{\prime}_iQ_iW_i\right|\right| + \left|\left| M\right|\right| \leq 2C.
		$$
		Furthermore, the variance of $B_i$ satisfies: 
		$$
		Var(B) = \mathbb{E}\left[\left(W^{\prime}QW\right)\left(W^{\prime}QW\right)\right] - MM.
		$$
		Hence, $\left|\left|Var(B) \right|\right| \leq  \mathbb{E}\left[\left|\left| W^{\prime}QW\right|\right|^2\right] + \left|\left| M\right|\right|^2 \leq 2C$. For a sufficiently small $\delta$, for any $\ell$, and universal constants $\mathcal{C}_1, \mathcal{C}_2$, and $\mathcal{C}_3$, by  Theorem 6.17 in \cite{wainwright2019}, 
		$$
		\mathbb{P}\left(\left|\left|\check{M}_\ell - M \right|\right| \geq \delta\right) \leq 2p \exp\left(-\frac{ \mathcal{C}_1n\delta^2}{2\mathcal{C}_2 \left(\mathcal{C}_3 + \delta\right)}\right).
		$$
		This bound implies 
		$$
		\left|\left|\check{M}_\ell - M \right|\right| = O_p\left(\left(\frac{\log p}{n}\right)^{1/2}\right).
		$$
		yielding the first conclusion of the proposition. Consequently, assuming that $\log p = O(n^{3/2})$, we have $\left|\left| \check{M}_\ell - M\right|\right| = O_p\left(n^{-1/4}\right)$, as desired. $\blacksquare$

		\section{Relation to Functional Differencing}
		
		\label{BonhommeFD}
		
		In a seminal paper, \cite{bonhomme2012functional} constructs moments that do not depend on UH through functional differencing in the conditional parametric setting. In this section, we will briefly outline his method and state conditions under which his approach and ours are equivalent for the setting Bonhomme considers.
		
		Let us review the model in \cite{bonhomme2012functional}. The data observation is $Z=(Y,X),$ where $Y$ is a vector of dependent
		variables, $X$ is a vector of covariates, and possibly other variables such as
		initial conditions. Let $\eta_{0}\left(  \alpha|x\right)  $ denote the
		conditional density of UH given covariates $X=x$, and let the density of the
		observed data be given by%
		\begin{equation*}
			f_{\lambda_0}\left( z\right)  =\int f_{Y|\alpha,X}\left(
			y|\alpha,x;\theta_{0}\right)  \eta_{0}\left(  \alpha|x\right)  d\alpha. 
		\end{equation*}
		In this specification, $f_{\lambda_0}$ is the density of
		$\mathbb{P}_{0}$ with respect to $\mu=\mu_{Y}\times v_{X},$ where $\mu_{Y}$ is
		a $\sigma$-finite measure and $v_{X}$ is the probability measure of $X.$ The
		conditional density of UH, $\eta_{0}\left(  \alpha|X\right)  $,
		depends on $X$ in an unrestricted way, as in a fixed effects
		approach. Functional differencing focuses on $\psi(\lambda_0) = \theta_0$, a special case of the functionals we consider in this paper. Recall that, in this case, $\tilde{r}_\psi = \left(I_p, 0\right)$. 
		
		Bonhomme defines the linear operator
		\[
		L_{\theta_{0},X}h\left(  y\right)  :=\int_{\mathcal{A}}f_{Y|\alpha,X}\left(
		y|\alpha,X;\theta_{0}\right)  h\left(  \alpha\right)  d
		\alpha  ,
		\]
		which is defined on $L_{2}(\pi_{\alpha}),$ for a user-specific weight function
		$\pi_{\alpha},$ and where $\mathcal{A}$ is a known set that contains the
		support of $\eta_{0}\left(  \alpha|x\right)  ,$ for all $x.$ A key assumption is that
		$\pi_{\alpha}(\alpha)>0$ for all $\alpha\in\mathcal{A}.$ 
		
		In contrast, we work with the score operator $S_{\eta}$
		that, in this case, is a mapping from $B(\eta_0)$ to $L_2$:
		\[
		S_{\eta}b  (y)=\frac{1}{f_{\lambda_0}(y|x)}%
		\int_{\mathcal{A}}f_{Y|\alpha,X}\left(  y|\alpha,x;\theta_0\right)
		b(\alpha)\eta_{0}(\alpha|x)d\alpha.
		\]
		To make our setting comparable to that in \cite{bonhomme2012functional}, we fix $X=x$ so that
		the score operator (and $L_{\theta_{0},X}$) should now be understood as a function of $y$ only.
		
		Even though $L_{\theta,X}$ can be defined for any $\theta$, we set
		$\theta=\theta_0$ as we are interested in inference. Moreover, the
		functions $h$ and $b$ might depend on $x$, however, we hereafter suppress this explicit
		dependence to simplify the notation. Bonhomme introduces the spaces of
		square integrable functions with domains $\mathcal{A}$ and $\mathcal{Y}$, where $\mathcal{Y}$ contains the support of $f_{\lambda_0}(.|x)$ for all $x$,
		respectively:
		\begin{align*}
			\mathcal{M}_{\alpha}  &  =\left\{  h:\mathcal{A}\mapsto\mathbb{R}%
			,\int_{\mathcal{A}}h^{2}(\alpha)\pi_{\alpha}(\alpha)d \alpha <\infty\right\}, \\ \mathcal{M}_{Y}    & =\left\{  q:\mathcal{Y}\mapsto\mathbb{R},\int%
			_{\mathcal{Y}}q^{2}(y)\pi_{Y}(y)d\mu_{Y}\left(  y\right)  <\infty\right\}, 
		\end{align*}
		where $\pi_Y$ is a positive function. Under some conditions,
		$L_{\theta_0,X}$ is an operator that maps $h\in\mathcal{M}_{\alpha}$ to
		$L_{\theta_0,x}h\in\mathcal{M}_{Y}$. Note that $\mathcal{M}_{\alpha}$ and
		$\mathcal{M}_{Y}$ are both Hilbert spaces, endowed with the scalar products
		$\left\langle h_{1},h_{2}\right\rangle_{\mathcal{M}_\alpha}  =\int_{\mathcal{A}}h_{1}(\alpha
		)h_{2}(\alpha)\pi_{\alpha}(\alpha)d\alpha$ and $\left\langle q_{1}%
		,q_{2}\right\rangle_{\mathcal{M}_Y}  =\int_{\mathcal{Y}}q_{1}(\alpha)q_{2}(\alpha)\pi_{Y}%
		(y)dy$, respectively. The range of
		$L_{\theta_0,x}$ is defined as follows:
		\[
		\mathcal{R}(L_{\theta_0,x}):=\left\{  L_{\theta_0,x}h\in\mathcal{M}%
		_{Y}:h\in\mathcal{M}_{\alpha}\right\}  .
		\]
		Then, non-surjectivity holds in Bonhomme's setting when $\overline
		{\mathcal{R}(L_{\theta,x})}\neq\mathcal{M}_{Y}$ $v_{X}-$a.s. for all $\theta\in\Theta.$
		Functional differencing constructs a moment that does not depend on the UH ($\eta_{0}%
		$) by finding a function $g_{FD}$ that is orthogonal to $\mathcal{R}%
		(L_{\theta_0,x})$. This can be accomplished by letting
		$$
		g_{FD}\left(  y,\theta_0\right)  =m(y)-\Pi_{\overline{\mathcal{R}(L_{\theta_0,x})}}m(y),\;\;\;m\in\mathcal{M}_{y},
		$$
		where $\Pi_{\overline{\mathcal{R}(L_{\theta_0,x})}}$ is the orthogonal projection operator onto $\overline{\mathcal{R}(L_{\theta_0,x})}$. If $f_{\theta_0,\eta_{0}}\in\mathcal{M}_{y}$, $v_{X}-$a.s.,  Theorem 2 of \cite{bonhomme2012functional} shows that $g_{FD}$ necessarily satisfies
		\[
		\mathbb{E}\left[  \left.  \pi_{Y}(Y)g_{FD}(Y,\theta_{0})\right\vert X\right]
		=0,\text{ }v_{X}-a.s.,
		\]
		This moment does not depend on
		$\eta_{0}$ and is trivially an LR moment.
		
		Next, we aim to find $L_{\theta_0,X}^{\ast}$, the adjoint of $L_{\theta_0,X}$. Simple algebra shows, by Fubini,%
		\begin{align*}
			\left\langle L_{\theta_0,X}h,g\right\rangle_{\mathcal{M}_Y}  &  =
			\int_{\mathcal{A}}\left(  \int_{\mathcal{Y}}f_{Y|\alpha,X}\left(
			y|\alpha,x;\theta_0\right)  \pi_{Y}(y)g(y,\theta_0)\frac{1}{\pi_{\alpha
				}(\alpha)}d\mu_{Y}(y)\right)  h(\alpha)\pi_{\alpha}(\alpha)d
			\alpha \\
			&  =\left\langle h,L_{\theta_0,x}^{\ast}g\right\rangle_{\mathcal{M}_\alpha} .
		\end{align*}
		Thus,
		\[
		L_{\theta_0,X}^{\ast}g=\frac{1}{\pi_{\alpha}(\alpha)}\int_{\mathcal{Y}%
		}f_{Y|\alpha,X}\left(  y|\alpha,X;\theta_0\right)  \pi_{Y}(y)g(y,\theta_0)d\mu_{Y}(y),
		\]
		and, by duality, 
		\begin{align*}
			\overline{\mathcal{R}(L_{\theta_0,X})}^{\perp}  &  =\mathcal{N}%
			(L_{\theta_0,X}^{\ast})\\
			&  =\left\{  g_{FD}(\cdot,\theta_0)\in\mathcal{M}_{Y}:\int_{\mathcal{Y}%
			}g_{FD}(y,\theta_0)\pi_{y}(y)f_{Y|\alpha,X}\left(  y|\alpha
			,X;\theta_0\right)  d\mu_{Y}(y)=0\right\} \\
			&  =\left\{  g_{FD}(\cdot,\theta_0)\in\mathcal{M}_{Y}:\mathbb{E}\left[
			\left.  \pi_{Y}(Y)g_{FD}(Y,\theta_0)\right\vert X,\alpha\right]
			=0,v_{X}\times\pi_{\alpha}-a.s.\right\}  .
		\end{align*}
		As $\tilde{r}_\psi = \left(I_p, 0\right)$, Proposition \ref{LRmomentUHresult}, and in particular, Equation \eqref{fde}, implies that $\pi_{y}(Y)g_{FD}(Y,\theta_0)$ is an LR moment, provided that the support of $v_{X}\times\pi_{\alpha}$ contains that of $v_{X}\times\eta_{0}$. This leads to the following result, whose proof is omitted as is obtained directly from our displays above and Proposition \ref{LRmomentUHresult}:
		\begin{proposition}
			Let the assumptions in Proposition \ref{LRmomentUHresult} hold.	If $g_{FD}$ is a function obtained by functional differencing, then $\pi_{Y}g_{FD}$ is an LR moment function provided that the
			support of $v_{X}\times\pi_{\alpha}$ contains that of $v_{X}\times\eta_{0}$.
		\end{proposition}
		
		\bigskip Proposition \ref{LRmomentUHresult} states that a debiased moment must satisfy\footnote{The restriction for $\theta$ in the proposition can be ignored as for any function satisfying \eqref{ourcondUH}, $$\mathbb{E}\left[s_\theta(Z,\lambda_0)g\left(Z,\theta_0\right)\right] = I_p,$$ which is the orthogonality restriction for $\theta$, as Remark \ref{remarkthetapsi} points out.} 
		\begin{equation}
			\label{ourcondUH}
			\mathbb{E}\left[\left. g(Y,\theta_0)\right|\alpha,X\right] = 0, \;\;\; v_X \times \eta_0\;\; a.s.
		\end{equation}
		It turns out that, under additional restrictions, \eqref{ourcondUH} makes $g_{FD} = g/\pi_Y$ a.s. a function constructed by functional differencing, depending only on $\theta_0$. This implies that our approach and that in \cite{bonhomme2012functional} are equivalent. 
		\begin{proposition}
			\label{fdourequiv}
			Let the assumptions in Proposition \ref{LRmomentUHresult} hold. If $\mathbb{E}\left[g\left(Y,\theta_0\right)|\alpha,X\right]$ is a real-analytic function of $\alpha$ in $\mathcal{A}$, $v_X-a.s.$, and the support of $\eta_0(\cdot|X)$ contains an open set contained in $\mathcal{A}$, $v_X-a.s.$, then if \eqref{ourcondUH} holds, then 
			\begin{equation}
				\mathbb{E}\left[  \left. g(Y,\theta_0)\right\vert X,\alpha\right]
				=0,\text{ }v_{X}-a.s.,\text{ \textit{all} }\alpha\in\mathcal{A},
				\label{orthog5}%
			\end{equation}
			and $g_{FD} = g/\pi_Y$, a.s., is a function obtained by functional differencing.
		\end{proposition}
		
		\bigskip \noindent \textbf{Proof of Proposition \ref{fdourequiv}:} By the the Identity Theorem for real-analytic functions \cite[see Corollary 1.2.6 in][]{krantzparks2002}, and the fact that $f_{Y|\alpha,X}(Y|\alpha,X;\theta_0)$ does not depend on $\eta_0$, \eqref{ourcondUH} implies \eqref{orthog5} when $g_{FD} = g/\pi_Y$ a.s. Hence, by Theorem 2 of \cite{bonhomme2012functional}, we conclude that $g_{FD} \in \overline{\mathcal{R}(L_{X,\theta_0})}^{\perp}$. $\blacksquare$
		
		\begin{remark}
			If the support conditions do not hold, the function satisfying \eqref{ourcondUH} might depend on $\eta_0$ (and thus on $\lambda_0$) through its support.   
		\end{remark}
		\bigskip This section concludes that commonly used sufficient conditions for obtaining functions $g(Y,\theta_0)$ in the
		functional differencing literature \citep[e.g.,][]{honore2024moment} such as (\ref{orthog5}), are also necessary for
		existence of LR moments under support conditions on UH and smoothness of the
		model. Thus, functional differencing is a special case of our
		characterization. Without these additional assumptions, the conditions of
		functional differencing are sufficient when $\mathcal{A}$ includes the support
		of UH, but they may not be necessary. 

  	\section{Equation of the First Kind, Ill-posed Problems, and Estimation of their Solutions: A Brief Review}
		\label{illposedsection}
		In this section, we discuss some theoretical and practical aspects regarding the computation of debiased functions, as implemented by our proposed algorithm. We will provide the reader with an overview. A detailed discussion can be found elsewhere \cite[][]{CARRASCO20075633, kress1989}. 
		
		Step 1 and Step 2 of our algorithm require solving an equation of the first kind. In generic terms, we need to find a solution to 
		\begin{equation}
			\label{firstkind}
			Sh = \phi,
		\end{equation}
		where $S: \mathcal{H}_1 \mapsto \mathcal{H}_2$ is a linear operator,  $\mathcal{H}_1$ and $\mathcal{H}_2$ are both Hilbert spaces with inner products $\left< \cdot, \cdot\right>_{\mathcal{H}_1}$ and $\left< \cdot, \cdot\right>_{\mathcal{H}_2}$, respectively. In particular, $S$ is an integral operator (specifically, the conditional expectation operator), which can be written as follows 
		$$
		Sh = \int h(v) K\left(v,t\right) d\mu_v(v),
		$$
		for some $\sigma-$finite measure $\mu_v$. For example, one will need to write
		\begin{equation}
			\label{adjth}
			Sh = \mathbb{E}\left[\left.h\left(Z,\theta_0\right)\right|\alpha,X\right] = \int h\left(Z,\theta_0\right) f_{Y|\alpha,X}\left(y|\alpha,x;\theta_0\right)\;d\mu(y),
		\end{equation}
		where $v \equiv z$, $t \equiv \left(\alpha,x\right)$, $K\left(v,t\right) \equiv  f_{Y|\alpha,X}\left(y|\alpha,x;\theta_0\right)$. We will impose some smoothness on the operator $S$. In particular, if we let $\mu_t$ be some $\sigma-$finite measure,\footnote{Whenever the support of $\alpha$ is involved, we understand that the support is $\mathcal{A}$ and hence $\mu_t$ is always positive over $\mathcal{A}$.}
		\begin{assumption}
			\label{boundedoperator}
			$\int \int \left|K\left(v,t\right) \right|^2 d\mu_t(t) d\mu_v(v) < \infty$.
		\end{assumption}
		The previous assumption implies that $S$ is a bounded operator. In addition, we can define its adjoint $S^{*}: \mathcal{H}_2 \mapsto \mathcal{H}_1$, which is also an integral (bounded) operator:
		$$
		S^{\ast} q := \int q(t) K^{*}\left(t,v\right) \; d\mu_t(t),
		$$
		and $K\left(v,t\right) = K^{*}\left(t,v\right)$. If \eqref{adjth} holds, its adjoint is the score operator (wrt $\eta_0$), as we discussed in the main text. An important observation is that Assumption \ref{boundedoperator} also implies that both $S$ and $S^{*}$ are compact operators. Hence, they can be approximated by finite dimensional operators, which will be crucial when it comes to estimation, as we will show below. 
		
		Equation \eqref{firstkind} will typically be ill-posed due to two reasons. First, the solution to \eqref{firstkind} may not be unique. Second, if a solution exists, it will not be stable with respect to small changes in $\phi$. In other words, the solution will not be continuous in $\phi$. Let us assume, 
		\begin{assumption}
			\label{minnormas}
			$\phi \in \mathcal{R}\left(S\right)$.
		\end{assumption}
		The previous assumption implies that \eqref{firstkind} admits a solution of minimum norm, denoted by $h^{\ast}$.  This is the solution we will be interested in computing. Notice that this ``addresses" the uniqueness problem. Under Assumption \ref{minnormas}, the minimum norm solution can be written as 
		$$
		h^{\ast} = S^{\dagger} \phi,
		$$
		where $S^{\dagger}$ is the Moore-Penrose inverse of $S$. The minimum norm solution can equivalently be written as 
		\begin{equation}
			\label{minnormsol}
			h^{\ast} = \left(S^{\ast}S\right)^{\dagger}S^{\ast} \phi,
		\end{equation}
		which is our preferred representation. 
		
		Since $S$ is, by assumption, a compact finite dimensional operator, $\mathcal{R}(S)$ is not closed, and $h^{\ast}$ is not continuous in $\phi$, unfortunately. This is a problem when $\phi$ is unknown as in practice we will be working with a suitable estimator of it, which will be the case in many applications. This stems form the singular values of $S$ converging to zero sufficiently fast \citep[see Section 3.1 of][]{CARRASCO20075633}. 
		
		To solve the lack of stability, one can use regularization schemes; see ch. 15 of \cite{kress1989}. The idea consists of approximating the true solution $h^{\ast}$ by one in which the influence of small singular values is dampened. This will be accomplished by computing 
		$$
		h^{\ast}_\gamma = R_\gamma \phi,
		$$
		where $R_\gamma $ is a bounded linear operator $R_\gamma: \mathcal{H}_2 \mapsto \mathcal{H}_1$, with regularization parameter $\gamma >0$, with the property of pointwise convergence: 
		$$
		\lim_{\gamma \rightarrow 0} R_\gamma \left(S^{\ast}S\right)S^{\ast}\phi = S^{\ast}\phi,
		$$
		for all $\phi \in \mathcal{H}_1$. A small $\gamma$ improves the accuracy of the solution whereas a large $\gamma$  improves its stability, imprlying that there is a trade-off in the choice of $\gamma$. There are three common regularization schemes that we can mention:\footnote{For a detailed discussion of regularization schemes, we refer the reader to Section 15.5 of \cite{kress1989}.} 
		\begin{enumerate}
			\item \textit{Landweber-Fridman:} In this case, the regularized operator and the solution are, respectively, 
			\begin{align*}
				R_\gamma & = c \sum^{\frac{1}{\gamma} - 1}_{\ell = 0} \left(I - cS^{\ast}S\right)^{\ell}  \\ 
				h^{\ast} & =  R_\gamma S^{\ast}\phi,
			\end{align*}
			where $0 < c \leq 1$.
			\item \textit{Tikhonov regularization:} This scheme is characterized by 
			\begin{align*}
				R_\gamma & =  \left(\gamma I + S^{*}S\right)^{-1}\\ 
				h^{\ast} & =  R_\gamma S^{\ast}\phi,
			\end{align*}
			\item \textit{Spectral cut-off} In this case, only singular values above a threshold depending on $\gamma$ is considered. We discuss an implementation of this scheme in our Monte Carlo simulations; see Section \ref{implementationMC}. 
		\end{enumerate}
		We next turn to estimation aspects. After all, $S$, $S^{\ast}$, and $\phi$ will often be unknown objects. Let $\hat{\phi}_n$ be a suitable estimator of $\phi$. Also, let us assume that we have some estimators of $S$ and $S^{\ast}$, denote by $\hat{S}_n$ and $\hat{S}^{\ast}_n$ such that these have finite range and satisfy 
		\begin{align*}
			\hat{S}_n h = \sum^{L_n}_{\ell = 1} d_\ell(h) \psi_\ell, \\
			\hat{S}^{\ast}_n q = \sum^{L_n}_{\ell = 1} b_\ell(q) \varphi_\ell,
		\end{align*}
		where $\psi_\ell \in \mathcal{H}_2$, $\varphi_\ell \in \mathcal{H}_1$, $d_\ell(h)$ is linear in $h$ and $b_\ell(q)$ is linear in $q$. It follows that 
		$$
		\hat{S}^{\ast}_n\hat{S}_n h = \sum^{L_n}_{\ell, \ell^{\prime}} d_{\ell^{\prime}}(h)b_\ell (\psi_{\ell^{\prime}})\varphi_\ell. 
		$$
		In the case of the \textit{Landweber-Fridman regularization}, the estimated regularized solution can be computed recursively by 
		$$
		\hat{h}^{*}_{\ell,n} = \left(I - c\hat{S}^{\ast}_n\hat{S}_n\right)\hat{h}^{*}_{\ell - 1,n} + c\hat{S}^{\ast}_n \hat{\phi}_n, \;\;\; \ell = 1,2,\cdots, 1/\gamma_n - 1,
		$$
		starting with $\hat{h}^{*}_{0,n} = c\hat{S}^{\ast}_n \hat{\phi}_n$, where $1/\gamma_n$ is an integer. On the other hand, the regularized estimated solution by the \textit{Tikhonov regularization} is 
		$$
		\hat{h}^{*}_n = \frac{1}{\gamma_n} \left[\sum^{L_n}_{\ell = 1} b_\ell \left(\hat{\phi}_n\right)\varphi_\ell - \sum^{L_n}_{\ell, \ell^{\prime}} d_{\ell^{\prime}}(h)b_\ell \left(\psi_{\ell^{\prime}}\right)\varphi_\ell\right],
		$$
		where to find $d_\ell(h)$, we need to solve 
		$$
		\left(\gamma_n I + A\right)\underline{d} = \underline{b},
		$$
		where $\underline{d} = \left[d_1(h), d_2(h), \cdots, d_{L_n}(h)\right]^{\prime}$, $A$ is the $L_n \times L_n$ matrix with generic element 
		$$
		A_{j,\ell^{\prime}} = \sum^{L_n}_{\ell = 1} b_\ell\left(\psi_{\ell^{\prime}}\right)d_j\left(\varphi_\ell\right),
		$$
		and 
		$$
		\underline{b} =  \begin{bmatrix} \sum_\ell b_\ell \left(\hat{\phi}_n \right)d_1\left(\varphi_\ell\right)\\ \cdots \\ \sum_\ell b_\ell \left(\hat{\phi}_n\right) d_{L_n}\left(\varphi_\ell\right) \end{bmatrix}.
		$$
		For details, we refer the reader to Section 3.4 of \cite{CARRASCO20075633}. The specifcation of $\gamma_n$, the tuning parameter, is beyon the scope of this paper; see Section 4.1 oc \cite{CARRASCO20075633} for optimal choices. 
		
		We have provided general guidance in implementing estimators of linear operators to solve ill-posed equations of the first-kind. In then next section, we illustrate how we can construct these esitmators in the specific setting of our Monte Carlo exercise. 
		
	\subsection{Implementation of the solution in our Monte Carlo simulations}
		\label{implementationMC}
		We now provide specific details regarding how we numerically compute $g=\Pi_{\mathcal{N}(S_{\eta}^{*})}m$ in practice. Recall that by duality, we know that $\overline{\mathcal{R}\left(S_{\eta}\right)}^{\perp} = \mathcal{N}(S_{\eta}^{*})$. Moreover, observe that the range of $S_{\eta}S^{*}_{\eta}$ is dense in $\mathcal{R}\left(S_{\eta}\right)$. Hence,\footnote{For simplicity, let us omit the dependence of $g$ and $m$ on the nuisance parameters.} 
		$$
		g\left(Z\right) = m\left(Z\right) - \Pi_{\mathcal{R}\left(S_{\eta}S^{*}_{\eta}\right)}m\left(Z\right).
		$$
		Provided that $ m \in \mathcal{R}\left(S\right) \oplus \mathcal{R}\left(S\right)^{\perp}$, where here $S\equiv S_\eta S^{*}_\eta$, we can approximate the orthogonal projection that appears in the previous display; see \cite{CARRASCO20075633}, Section 3.1.\footnote{Note that our previous discussion applies to this section as well, where $m$ plays the role of $\phi$.} 
		
		We need to construct an estimator of the operator $S$ with finite range.  A common strategy is to use the so-called collocation method; see \cite{kress1989}, Section 13.2. In this case,  we evaluate the problem in a finite number of points $L_n \equiv n_z$ that we generate from $\hat{f}_{\beta_0,\hat{\eta}_\ell}$, an estimator of $f_{\beta_0,\eta_0}$,\footnote{Note that the obtained evaluation points depend on $\ell$, and for simplicity, we maintain the same number $n_z$ for all folds.} such that we now want to solve 
		$$
		\min_{a \in \mathbb{R}^{n_z}} \sum^{n_z}_{j=1} \left( m(z_j) - \frac{1}{n_z}\sum^{n_z}_{j^{\prime}=1}a(z_{j^{\prime}}) \mu_{z_{j^{\prime}}}(z_j)\right)^2,
		$$
	    where $\mu_{z_{j^{\prime}}}(z_j)$ is the discretized estimated version of $S$. Hence, we obtain that the optimal discretized solution $\underline{a}^{*}$ is the vector
		$$
		\underline{a}^{*} = \left[\sum^{n_z}_{j=1}\mu_{z_{j^\prime}}(z_j)\mu_{z_{j^\prime}}(z_j)\right]^{\dagger}\left[\sum^{n_z}_{j=1}\mu_{z_{j^\prime}}(z_j)m(z_j)\right].
		$$
		This is the sample counterpart of the minimum norm solution \eqref{minnormsol}, using finite-dimensional operators. Thus, a finite-sample approximation to  the projection $\Pi_{\mathcal{R}\left(S_{\eta_{0}}S^{*}_{\eta_{0}}\right)}m\left(z\right)$ is given by\footnote{Our implementation here is similar to that in the functional differencing approach of \cite{bonhomme2012functional}, with the important difference that we now compute relevant LR moments.} 
		$$
	 \sum^{n_j}_{j=1}\underline{a}^{*}(z_j)\mu_{z_j}(z).
		$$
		Notice that the computation of $\mu_{z_j}$ involves an integral with respect to $\eta_0$. However, $\eta_0$ is unknown. Then, we need to estimate it to obtain $\hat{\eta}_\ell$, and then generate evaluation points and solutions for each fold $\ell$, i.e., we generate $n_z$ points using $f_{\beta_0,\hat{\eta}_\ell}$ and the integral in $\mu_{z_j}$ is approximated by resampling $n_\alpha$ points from $\hat{\eta}_\ell$.\footnote{A standard assumption is that this estimation is conducted over a set $\mathcal{A}$ and that it contains the true support of $\eta_0$ with probability approaching one; see \cite{fox2016simple} and references therein.}$^{,}$\footnote{To generate points based on $f_{\beta_0,\hat{\eta}_\ell}$, generate $n_z$ points  $\underline{\alpha}$ from $\hat{\eta}_\ell$, $n_z$ points $\underline{\varepsilon_1}$ and $\underline{\varepsilon_2}$ from a standard normal, and then use \eqref{eq:chaberlain_example}.} This process is repeated for each $\ell = 1,\cdots,L$. In addition, notice that the starting $m$ is also unknown. We compute it by replacing $\eta_0$ for an estimator, which we describe below.
		
		The solution $\underline{a}^{*}$ can be equivalently written in a more convenient fashion. Define the matrices:
		\begin{align*}
			\mathbf{m}_{\beta_0,\hat{\eta}_{\ell}}  &  :=\left[  m(\tilde{z},\hat{\eta
			}_{\ell},\beta_0)\right] _{n_{z}\times1},\\
			\mathbf{S}_{\beta_0,\hat{\eta}_\ell}  &  :=\left[  \frac{f_{z|\alpha}(\tilde{z}|\tilde{\alpha
				}_1;\theta)}{f_{\beta_0,\hat{\eta}_\ell}(\tilde{z})}\right]  _{n_{z}\times n_{\alpha}}, \\
			\mathbf{f}_{\beta_0,\hat{\eta}_\ell}(z)  &  :=\left[  \frac{f_{z|\alpha}(z|\tilde{\alpha
				}_1;\theta)}{f_{\beta_0,\hat{\eta}_\ell}(z)}\right]  _{1\times n_{\alpha}}.
		\end{align*}
		Then, using the fact that for any matrix $A$, $A^{\dagger} = \left(A^{\prime}A\right)^{\dagger}A^{\prime} = A^{\prime}\left(AA^{\prime}\right)^{\dagger}$, we have
		\begin{align*}
		 \sum^{n_j}_{j=1}\underline{a}^{*}(z_j)\mu_{z_j}(z) & = \mathbf{f}_{\beta_0,\hat{\eta}_\ell}(z) 	\mathbf{S}_{\beta_0,\hat{\eta}_\ell}^{\prime} \left(	\mathbf{S}_{\beta_0,\hat{\eta}_\ell}	\mathbf{S}_{\beta_0,\hat{\eta}_\ell}^{\prime}\mathbf{S}_{\beta_0,\hat{\eta}_\ell}	\mathbf{S}_{\beta_0,\hat{\eta}_\ell}^{\prime}\right)^{\dagger}\mathbf{S}_{\beta_0,\hat{\eta}_\ell}	\mathbf{S}_{\beta_0,\hat{\eta}_\ell}^{\prime}\mathbf{g}_{\beta_0,\hat{\eta}_{\ell}} \\ & = \mathbf{f}_{\beta_0,\hat{\eta}_\ell}(z)\mathbf{S}_{\beta_0,\hat{\eta}_\ell}^{\dagger}\mathbf{g}_{\beta_0,\hat{\eta}_{\ell}}.
		\end{align*}
		We compute a regularized version of $\mathbf{S}_{\beta_0,\hat{\eta}_\ell}^{\dagger}$, using the spectral cut-off regularization. To this end, we obtain the singular value decomposition of the matrix $\mathbf{S}_{\beta_0,\hat{\eta}_\ell}$ and consider the first $ks$ largest singular values, which we set equal to 10. 
		
		The estimator $\hat{\eta}_\ell$ is obtained by the approached introduced by \cite{efron2016empirical}, which can be implemented by the R package \textit{deconvolveR}; see \cite{GmodelinR}. This method uses an exponential
		series expansion to approximate the true distribution $\eta_{0}$ and thus delivers a regularized estimator.  Additionally, we set $n_z = 1,000$ and $n_\alpha = 100$ in our discretization strategy.  We always fix $L=4$ for cross-fitting.

		\section{Implications to event-study desings}
		\label{sectioneventstudy}
		\cite{borusyak2024revisiting} develop an estimator for target functionals of heterogeneous treatment effects, based on an ``imputation" strategy. More broadly, the authors develop a framework for difference-in-differences staggered rollout designs and heterogeneous causal effects.\footnote{In staggered rollout designs, being treated is an absorbing state. That is, there is a specific date in which the treatment of an individual is on and remains in that state forever.} We revisit such a framework, connect it to our linear random coefficient model, and show that our theory could be useful in this setting. 
		
		\cite{borusyak2024revisiting} consider estimation of causal effects of a binary treatment $D_{it}$ on an outcome $Y_{it}$. For each individual $i$, we have an event date $E_i$ when $D_{it}$ changes from 0 to 1 forever. That means, $D_{it}= \bm{1}\left\{K_{it} \geq 0\right\}$, where $K_{it} = t - E_i$.\footnote{Our discussion can be extended to non-binary treatment intensity; see \cite{borusyak2024revisiting}, p. 3261.} Accordingly, units that are never treated present $E_i = \infty$. 
		
		Let $Y_{it}(0)$ denote the potential outcome of unit $i$ at period $t$ if she is never treated. The individual treatment effect is defined as 
		\begin{equation}
			\label{indte}
			\tau_{it} := \mathbb{E}\left[\left.Y_{it} - Y_{it}(0)\right| \Lambda_i\right],
		\end{equation}
		where $\Lambda_i := \left(X_i,\alpha_i\right)$, for all $t \geq E_i$. Note that \cite{borusyak2024revisiting} view $\Lambda_i$ as fixed, i.e., the analysis conditions on it. In our case, $\Lambda_i$ is a random variable, and thus we straightforwardly accommodate for that in the discussion to follow.  In this framework, we assume that the researcher is interested in an aggregated measure of \eqref{indte}. More specifically, let $\omega_{1i} = \left(\omega_{it}\right)_{t=1,\cdots,T}$ be a vector of known weights, which can depend on $\Lambda_i$ and $t$, we might be interested in 
		$$
		\tau_\omega = \sum^T_{t=1} \mathbb{E}\left[\omega_{it} \tau_{it}\right] \equiv \mathbb{E}\left[\omega^{\prime}_{1i}\tau_i\right]. 
		$$
		See \cite{borusyak2024revisiting} for different target parameters that might fall into the previous expression. Note that by correctly specifying weights, we might let them be nonnegative, for each $t$, which leads to a weakly causal parameter \citep[][]{blandhollate} by construction. 
		
		Suppose that $W_{1i}$ is a sub-matrix of $W_{it}$ with $d_1$ columns. We next impose three assumptions, obtained directly from \cite{borusyak2024revisiting}: 
		
		\begin{assumption}[Parallel trends/General model of $Y(0)$]
			\label{ab1}
			For all $it$,
			$$
			\mathbb{E}\left[\left. Y_{i}(0)\right|\Lambda_i\right] = W_{1i}\beta_{d1} + V_{i}\alpha_i 
			$$
		\end{assumption}
		
		\begin{assumption}[No-anticipation effects]
			\label{ab2}
			$Y_{it} = Y_{it}(0)$ for all pairs $it$ such that $D_{it} = 0$.
		\end{assumption}
		
		\begin{assumption}[Model of causal effects]
			\label{ab3}
			$\tau_i = B_i\beta_{0,T - d_2}$, where $\beta_{0,T - d_2}$ is a $(T - d_2) \times 1$ vector of unknown parameters and $B_i$ is a known $T \times (T - d_2)$ matrix of full column rank, depending on $\Lambda_i$.
		\end{assumption}
		
		We refer to \cite{borusyak2024revisiting} for a discussion on these assumptions. However, we highlight that Assumption \ref{ab3} imposes a ``parametric" model of treatment effects. Suppose that the researcher is willing to assume homogeneous treatment effect. In this situation, say that $\tau_{it} \equiv \beta_{01}$, $T - d_2 = 1$ and $B_i= (1,\dots,1)^{\prime}$. Conversely, with fully heterogeneous treatment effects, $\tau_{it}$ is unrestricted and hence $d_{2} = 0$ and $B_i = I_{T}$, an $T \times T$ identity matrix. An implication of Assumptions \ref{ab1}-\ref{ab2} is that 
		$$
		Y_{i} = W_{1i}\beta_{d1} + V_{i}\alpha_i + W_{2i}\beta_{0,n-d_2} +\varepsilon_i,\;\;\; \mathbb{E}\left[\left. \varepsilon_i\right|\Lambda_i\right] = 0,
		$$
		where $W_{2i}$ is a $T \times (T- d_2)$ matrix of interactions between $D_{it}$ and the $t-$row of $B_i$---a sub-matrix of $W_i$ in our notation.  Note that 
		$$
		\tau_\omega = \mathbb{E}\left[\omega^{\prime}_1B\right] \beta_{0,T - d_2} \equiv C^{\prime}_1\beta_0,
		$$ 
		for a suitable matrix $C_1$, where $\beta_0 = \left(\beta^{\prime}_{d_1},\beta^{\prime}_{0,T-d_2}\right)^{\prime}$. 
		
		From the previous discussion we observe that our analysis of the high-dimensional linear random coefficient model, accommodating high-dimensional regressors and multivariate UH, might be useful in this setting since a high-dimensional $\beta_0$  allows for a high-dimensional $\beta_{0,d_{1}}$. That implies that we can potentially allow for a flexible model in $W_{1i}$ while modeling $Y(0)$ in Assumption \ref{ab1}, which makes parallel trends assumption more plausible.\footnote{\cite{goodman2021difference}, p. 268, observes ``the ability to control for covariates is a common motivation for regression  DD [Difference-in-differences] as it is though to make a `common trends' assumption more plausible".} Note that the fact that our results apply to multidimensional UH is also key regarding this point.

 \end{appendix}  
\end{document}